%% file: main.tex
\documentclass[onecolumn]{aastex63}
\usepackage{natbib}
\usepackage{enumitem}

\setlist[itemize]{noitemsep}
\begin{document}

\shorttitle{ALMA Distributed Peer Review}
\shortauthors{Donovan Meyer et al.}

\title{Analysis of the ALMA Cycle 8 Distributed Peer Review Process}

\author[0000-0002-3106-7676]{Jennifer Donovan Meyer}
\affiliation{National Radio Astronomy Observatory (NRAO), 520 Edgemont Road, Charlottesville, VA 22903, USA}

\author{Andrea Corvill\'on}
\affiliation{Joint ALMA Observatory, Avenida Alonso de C\'ordova 3107, Vitacura, Santiago, Chile}

\author[0000-0003-2251-0602]{John M. Carpenter}
\affiliation{Joint ALMA Observatory, Avenida Alonso de C\'ordova 3107, Vitacura, Santiago, Chile}

\author[0000-0002-9912-5705]{Adele L. Plunkett}
\affiliation{National Radio Astronomy Observatory (NRAO), 520 Edgemont Road, Charlottesville, VA 22903, USA}

\author{Robert Kurowski}
\affiliation{European Southern Observatory, Karl-Schwarzschild-Str. 2, 85748, Garching bei Munchen, Germany}

\author{Alex Chalevin}
\affiliation{European Southern Observatory, Karl-Schwarzschild-Str. 2, 85748, Garching bei Munchen, Germany}

\author{Jakob Bruenker}
\affiliation{European Southern Observatory, Karl-Schwarzschild-Str. 2, 85748, Garching bei Munchen, Germany}

\author{D.-C. Kim}
\affiliation{National Radio Astronomy Observatory (NRAO), 520 Edgemont Road, Charlottesville, VA 22903, USA}

\author[0000-0003-1283-6262]{Enrique Mac\'ias}
\affiliation{Joint ALMA Observatory, Avenida Alonso de C\'ordova 3107, Vitacura, Santiago, Chile}
\affiliation{European Southern Observatory, Karl-Schwarzschild-Str. 2, 85748, Garching bei Munchen, Germany}
\affiliation{European Southern Observatory, Avenida Alonso de Córdova 3107, Vitacura, Santiago, Chile}

\correspondingauthor{J. Donovan Meyer}
\email{jmeyer@nrao.edu}

\input{abstract}
\input{introduction}
\input{overview}
\input{inputs}
\input{assignments}
\input{output}
\input{ranks}
\input{surveys}
\input{summary}

\bibliography{references}

\input{appendices/rules}
\input{appendices/guidelines}
\input{appendices/keywords}
\input{appendices/survey_pi}
\input{appendices/ranks_comp}

\end{document}

%% file: abstract.tex
\begin{abstract}

In response to the challenges presented by high reviewer workloads in traditional panel reviews and increasing numbers of submitted proposals, ALMA implemented distributed peer review to assess the majority of proposals submitted to the Cycle 8 Main Call. In this paper, we present an analysis of this review process. Over 1000 reviewers participated in the process to review 1497 proposals, making it the largest implementation of distributed peer review to date in astronomy, and marking the first time this process has been used to award the majority of observing time at an observatory. We describe the process to assign proposals to reviewers, analyze the nearly 15,000 ranks and comments submitted by reviewers to identify any trends and systematics, and gather feedback on the process from reviewers and Principal Investigators (PIs) through surveys. Approximately 90\% of the proposal assignments were aligned with the expertise of the reviewer, as measured both by the expertise keywords provided by the reviewers and the reviewers' self-assessment of their expertise on their assigned proposals. PIs rated 73\% of the individual review comments as helpful, and even though the reviewers had a broad range of experience levels, PIs rated the quality of the comments received from students and senior researchers similarly. The primary concerns raised by PIs were the quality of some reviewer comments and high dispersions in the ranks. The ranks and comments are correlated with various demographics to identify the main areas in which the review process can be improved in future cycles.

\end{abstract}

%% file: introduction.tex
\section{Introduction}
\label{sec:intro}

Prior to Cycle 8\footnote{In this paper, we refer to ALMA Cycle 8 2021 as ``Cycle 8" for simplicity. Cycle 8 2021 is the name given to the ALMA Cycle which began observations in October 2021, following the postponement of the original Cycle 8 (which had been expected to begin observing in October 2020) due to the COVID-19 pandemic.}, proposals submitted to ALMA Main Calls for Proposals were peer reviewed using solely the traditional panel review process.\footnote{A detailed description of the structure of the panels and ALMA review process is presented in \citet{Carpenter20a}, while reports describing the results of each Main Call can be found at \url{https://almascience.org/documents-and-tools/alma-reports}.} In a traditional panel review, scientists from the community are invited to take part in topical review panels, with each panel devoted to reviewing proposals in an individual scientific category. The panelists discuss and score the proposals to establish a scientifically-ranked list of projects. They also write consensus reports summarizing the strengths and weaknesses of each proposal, which are provided to the Principal Investigator (PI) of each proposal. The order of the final, merged scientifically-ranked list is the main factor to determine which proposals are entered into the ALMA observing queue.

The effectiveness of the panel review process relies on a number of factors. Foremost, ALMA community members generously volunteer their time to review proposals and participate in a face-to-face discussion, which requires a significant commitment. Reviewers also need to be comfortable reviewing the relatively broad range of topics that will be discussed within a panel. The panels therefore need to be carefully constructed to cover the expertise within a given science category, while also reflecting the diversity of the community in terms of gender, geographic region, and level of experience. 

Over time, the number of proposals submitted to ALMA has increased steadily, where 1700-1800 proposals are now submitted in any given Cycle. The number of panels has also steadily increased such that in Cycle 7, 25 panels involving 158 reviewers were needed to review the proposals. One of the main concerns resulting from the large number of submitted proposals has been the workload on the panels. In Cycle 7, panels typically reviewed 75 proposals, and in previous cycles some panels had as many as 150 proposals to review. The high workload undoubtedly impacts the amount of time a reviewer can devote to a given proposal and the subsequent quality of the reviews. With the increasing number of proposals over time and the continued high workload, the traditional panel review process had become increasingly difficult to sustain. 

In response to these challenges, ALMA introduced a new review process to the Main Call for Proposals in Cycle 8 where each proposal team was required to designate one member to review 10 proposals. Often called ``distributed peer review'' \citep{Merrifield09}, the process aims to have a large number of reviewers each assess a small number of proposals, instead of a relatively small number of reviewers assessing a large number of proposals as in panel reviews. Like any review process, distributed peer review has both advantages and disadvantages to consider relative to the panel review process (see also the discussions in \citealt{Patat18} and \citealt{Kerzendorf20}). The primary motivation of distributed peer review is that the workload on any individual reviewer is substantially reduced, affording the reviewer the opportunity to read each of their assigned proposals more carefully. The process is scalable, relying on the participation of potentially hundreds of astronomers, which also makes it inherently more inclusive; a broader range of viewpoints can be expressed than is possible in a panel review, and in turn the overall transparency of the process is increased. Further, a greater number of participants enables better matching of the expertise of the reviewers to a given proposal, whereas in panel reviews a limited number of people (typically $\sim$ 6-8) are expected to be familiar with a broad range of topics. Finally, it is relatively straightforward to run the process with more reviewers for a given proposal, ostensibly providing a better sampling of the consensus opinion, than in panel reviews. As with any review process, distributed peer review also has potential drawbacks. One of the main concerns is that there is no straightforward approach to encourage discussion amongst the reviewers on a given proposal, as is customary for a panel, both to clarify any misunderstandings about the proposal and to hear the feedback from more expert reviewers. Further, there is no specific incentive for completely anonymous reviewers to do a good job. The process initially suggested by \citet{Merrifield09} in fact included a provision to boost reviewers' own proposals if the order of their ranked lists closely matched the order of those proposals according to the final consensus, though such adjustments though could potentially penalize reviewers for honest differences of opinion on the scientific merit of the proposals. As a relatively new process, distributed peer review needs to be more established before it is as widely accepted as panel reviews.

Among astronomical observatories, distributed peer review is in use for the Gemini Fast-Turnaround proposal system \citep{Andersen19}, and it was also implemented by the European Southern Observatory (ESO) on a trial basis, where volunteers reviewed 172 proposals as described in \citet[see also \citealt{Kerzendorf20}]{Patat19}. These proposals represented a subset of the total number submitted in response to the ESO Call for Proposals for Period 103, requesting time on all ESO facilities except ALMA (F. Patat, private communication). In addition, HST has been using a process analogous to distributed peer review, where external reviewers are drawn from the community instead of the proposing teams. Like distributed peer review, each reviewer is assigned a relatively small number of proposals, and reviewers do not interact with each other. This process has been employed for 6 years in order to review HST mid-cycle proposals and was applied to all Small ($<$16 orbits) GO proposals beginning in HST Cycle 28 (N. Reid, private communication).\footnote{\url{https://hst-docs.stsci.edu/hsp/hubble-space-telescope-call-for-proposals-for-cycle-30}} Finally, ALMA used the distributed peer review process to review 249 proposals in the Cycle 7 Supplemental Call as a pilot program to gauge whether the process would be suitable for a Main Call. \citet{Carpenter20b} presented an extensive analysis of the Supplemental Call and found that PIs gave similar marks to the overall clarity, accuracy, and helpfulness of the reviewer comments as in panel reviews. The majority of reviewers and PIs from the Supplemental Call supported using distributed peer review for smaller proposals. Based on the results of the pilot program, and with the implementation of significant improvements in the process and user-facing tool based on feedback from the Cycle 7 Supplemental Call, ALMA implemented distributed peer review for a subset of proposals in the Cycle 8 Main Call. This is the largest implementation of distributed peer review to date in astronomy, and the first time distributed peer review has been used to issue the majority of time for an observatory. 

Because distributed peer review is a new process, an extensive analysis of the results is required to evaluate the effectiveness of the review process. In this paper, we describe in detail the overall process as implemented in Cycle 8, both to inform the community of the results and to inform other observatories which may be considering implementing distributed peer review. The paper is organized as follows. Section~\ref{sec:overview} presents an overview of the Cycle 8 distributed peer review process, and Section~\ref{sec:inputs} describes the demographics of the inputs to the process (the proposals, the reviewers, and the reviewers' expertise). Section~\ref{sec:assign} presents the methodology to assign proposals to reviewers, and Section~\ref{sec:output} presents analyses of reviewers' individual ranks and comments. Section~\ref{sec:ranks} describes how the overall rankings of proposals were determined from the individual reviews. Section~\ref{sec:surveys} presents the results and analysis of the surveys that were conducted of the reviewers and the PIs to gain their feedback on the review process. Finally, Section~\ref{sec:summary} presents a summary of the conclusions.

%% file: overview.tex
\section{The Cycle 8 Distributed Peer Review Process}
\label{sec:overview}

In Cycle 8, ALMA used distributed peer review to assess proposals that requested less than 25 h on the ALMA 12-m Array or less than 150 h on the 7-m Array in ``standalone" mode. The remaining proposals were peer reviewed using the traditional panel system as described in \citet{Carpenter20a}. The decision to use distributed peer review for only relatively smaller proposals was based on input from advisory committees and surveys of the participants (both reviewers and PIs) in the Cycle 7 Supplemental Call \citep{Carpenter20b}. This section presents an overview of the overall review process as implemented in Cycle 8, including highlighting the significant changes introduced since the Cycle 7 Supplemental Call.

\subsection{Process}
\label{subsec:process}

Appendix~\ref{app:rules} presents the full set of rules for participation in the distributed peer review process that were presented on the ALMA Science Portal for the Cycle 8 proposal call. We present a brief description here.

When submitting a proposal, the PI designates one member of the proposal team to participate in the distributed peer review process. Nominally the reviewer is expected to have a PhD in astronomy or closely related field. However, if the PI does not have a PhD, they can serve as the reviewer by also specifying a mentor (with a PhD) who will assist them during the proposal review. The reviewer cannot be changed after the proposal deadline.

Each reviewer is assigned 10 proposals to review, which is called a proposal set. A reviewer receives a different proposal set for each proposal on whose behalf they are designated the reviewer. After the proposal deadline, the Joint ALMA Observatory (JAO) assigns the proposal set(s) to each reviewer based on their scientific expertise. To better inform the proposal assignment process, each reviewer was advised to specify at least three scientific keywords in which they are experts through their user profile on the ALMA Science Portal. These keywords are identical to the list of keywords used by PIs when they prepare and submit their proposals. The ability for reviewers to specify their own expertise keywords is a new capability added for Cycle 8 in order to better match reviewers to submitted proposals. If the reviewer does not specify any keywords, then the reviewer is assumed to be an expert in the keyword(s) of their submitted proposal, which was the assumption for all assignments in the Cycle 7 Supplemental Call. Section~\ref{sec:assign} describes the assignment procedure in detail, including a description of the automatic handling of conflicts of interest in Section~\ref{subsec:conflicts}.

After the JAO assigns the proposal sets to all reviewers, the process proceeds in two stages. To begin Stage 1, reviewers must first acknowledge that they will behave in an ethical manner, keep their assignments confidential, and delete their review materials after the process is completed. They may then proceed to identify any potential conflicts of interest with their assigned proposals in order for the JAO to replace them with new assignments. Once the proposal set is confirmed to be without conflicts, reviewers write a comment to the PI of each assigned proposal to summarize the strengths and weaknesses of their proposal. These comments will ultimately be sent to the PI of the proposal verbatim. In addition, reviewers rank the proposals within their proposal set from 1 to 10, where a rank of 1 indicates the strongest proposal. If a reviewer is assigned multiple proposal sets, they rank the proposals separately in each set. A reviewer is also able to write a comment to the JAO regarding any proposal. These comments to the JAO are often used, for example, to report suspected violations of dual anonymous practices or technical concerns. Stage 1 is a mandatory process. If the ranks and comments are not submitted by the deadline, then the reviewer's submitted proposal is cancelled by the JAO.

ALMA introduced Stage 2 in the Cycle 8 review process based on feedback received from the Cycle 7 Supplemental Call. In Stage 2, reviewers can read the anonymized comments (but not the ranks) from other reviewers assigned the same proposals. While not intended to represent a replacement for panel discussion, this stage allows reviewers to determine if they overlooked a significant strength or weakness in their own Stage 1 assessment. Reviewers can then modify their own ranks and comments as needed until the Stage 2 deadline. Unlike Stage 1, Stage 2 is optional for reviewers. If a reviewer does not complete Stage 2, their Stage 1 results are considered final. 

Two surveys were created and employed to capture the feedback (separately) from reviewers and PIs. Reviewers were provided with a survey upon submission of each proposal set in Stage 1. The survey allows the reviewers to indicate their self-assessed expertise on each proposal to help improve the proposal assignment process in future Cycles, and to optionally provide written feedback on the overall process. The PI survey was opened one week after the proposal results were announced. This survey gathered feedback on the quality of the comments and allowed PIs to provide feedback on the review process. Section~\ref{sec:surveys} presents the results of these surveys. 

Table~\ref{tab:timeline} summarizes the timeline of the distributed peer review process. Stage 1 began approximately two weeks after the proposal deadline and lasted for four weeks. After a couple of days to verify that the Stage 1 results were complete, Stage 2 was opened for one week.

\input{tables/table_timeline}

\subsection{Reviewer Tool}
\label{subsec:tool}

The Reviewer Tool, the user-facing tool used by reviewers in Cycle 8, enables reviewers to view and download their assigned proposals as well as submit their ranks, comments, and the reviewer survey.\footnote{\url{https://almascience.org/proposing/alma-proposal-review/how-to-use-the-reviewer-tool}} The tool was updated after the Cycle 7 Supplemental Call based on community feedback. Improvements made to the tool for Cycle 8 included the ability to download all assigned proposals in a single proposal set in a batch (previously proposals had to be downloaded one at a time), the robustness of the auto-saving functionality, and ingesting the reviewer survey into the tool for improved accessibility.

In addition, given the confusion experienced by a few reviewers in the Cycle 7 Supplemental Call with respect to the ordering of the ranks (and concern from PIs that reviewer rank orders had been reversed), a ``drag and drop" functionality was incorporated into the Reviewer Tool. This capability replaced the previous drop-down menus, which a user needed to use to select individual rankings. The new functionality is more modern, makes the ranking order more intuitive, and is easier to manipulate if a user wishes to move proposals around in the list after setting initial ranks. 

\subsubsection{Documentation}

In lieu of scheduling a plenary session, which is customary for panel reviews, extensive documentation was placed on the ALMA Science Portal to describe the review process and expectations of reviewers given the involvement of over 1000 participants. The documentation included criteria for judging the scientific merits of the proposals (see Appendix~\ref{app:guidelines}). Guidelines were also provided on how to write comments to the PI, with an emphasis on summarizing the strengths and weaknesses of the proposal written in a constructive and professional manner. An example comment text was provided to illustrate the contents of a good comment. 

Reviewers were provided further guidance concerning declaring conflicts, guidelines for mentors, unconscious bias, the code of conduct,\footnote{\url{https://almascience.org/proposing/alma-proposal-review/guidelines-for-reviewers}} dual-anonymous practices,\footnote{\url{https://almascience.org/proposing/alma-proposal-review/dual-anonymous}} and a page of frequently asked questions (FAQ).\footnote{\url{https://almascience.org/proposing/alma-proposal-review/frequently-asked-questions}} Reviewers were also provided with the contact email for the Proposal Handling Team (PHT) at the JAO to ask any further questions.

\subsection{Overall results of the process}
\label{sec:results}

Of the 1497 proposal sets assigned on behalf of the proposals submitted to the distributed peer review process, 1492 were submitted by the review deadline (99.7$\%$). Given that Cycle 8 was the first Main Call where distributed review was implemented, the PHT provided the 4 reviewers of the remaining 5 proposal sets with a short grace period. All but one reviewer responded within that window, and as a result, only one submitted proposal was rejected for a lack of submitted ranks and comments.

\subsection{User support}

The PHT responded to 105 requests from reviewers for user support during the distributed review process. The most common issues needing clarification were due to user confusion over the simultaneous review processes (i.e., the separation of proposals above and below 25 requested hours on the 12-m Array into two review processes), suspected dual-anonymous violations by PIs in their review assignments, and questions about declaring conflicts of interest. Table~\ref{tab:support} lists the issues encountered four or more times during the review process.

\input{tables/table_usersupport}

After the review process was completed and the results of the review process were distributed to PIs, the PHT received nine further user-support requests from PIs. In four cases, PIs suspected the reviewer inverted the rank order; upon review, all four cases were confirmed to be correct as submitted. Other reasons that PIs contacted the PHT included missing notification emails of the Cycle 8 results (2 cases), unavailable proposal reviews in the ALMA PI-facing web-based tool called SnooPI (2 cases), and one instance in which a PI received a review for the wrong proposal, which was confirmed to be a copy and paste error by the reviewer.

%% file: tables/table_timeline.tex
\begin{deluxetable*}{cl}
\tablecolumns{2}
\tablecaption{Timeline of the ALMA Cycle 8 Distributed Review Process\label{tab:timeline}}
\tablehead{\colhead{Date} & \colhead{Milestone}}

\startdata
17 December 2020 & Cycle 8 Pre-announcement \\
17 March 2021 & Cycle 8 Call for Proposals \\
21 April 2021 & Proposal submission deadline \\
06 May 2021 & Proposals released to reviewers; Stage 1 begins \\
03 June 2021 & Distributed peer review Stage 1 deadline \\
08 June 2021 & Distributed peer review Stage 2 begins \\
15 June 2021 & Distributed peer review Stage 2 deadline \\
06 August 2021 & Disposition letters sent to all PIs \\
\enddata
\end{deluxetable*}

%% file: tables/table_usersupport.tex
\begin{table}[ht]
    \centering
    \caption{Most common requests for help from distributed peer review participants (via email or helpdesk)}
    \begin{tabular}{cl}
    \hline \hline
    {Number of inquiries} & {Topics} \\
    \hline
    18 & Confusion about distinction between distributed and panel reviews \\
    17 & Reporting potential violation of dual-anonymous guidelines \footnote{Additional suspected violations were reported via the Reviewer Tool}\\
    15 & Questions about conflicts of interest, or new conflicts identified after the conflict submission deadline \\
    11 & Missing the Stage 2 deadline of distributed peer review\\
     8 & Reporting lack of expertise in a proposal assignment\\
     6 & Asking for an extension to the Stage 1 deadline for distributed peer review \\
     4 & Questions about student reviewers and the role of mentors\\
     4 & Request to change the reviewer \\ 
     4 & Issues with the Reviewer Tool \\
    \hline
    \end{tabular}
    \label{tab:support}
\end{table}

%% file: inputs.tex
\section{Inputs to the Distributed Review Process}
\label{sec:inputs}

This section describes the demographics of the three main inputs to the distributed review process: the submitted proposals, the designated reviewers, and the declared and assumed expertise of the designated reviewers. 

\subsection{Proposal submissions}
\label{subsec:submissions}
ALMA PIs submitted 1735 proposals to the Cycle 8 Main Call, of which 1497 were reviewed by distributed peer review. The remaining proposals were reviewed by panels. Tables~\ref{tab:submissions_reg} and \ref{tab:submissions_cat} present the proposal submission statistics grouped by regional affiliation and scientific category in the context of the total requested time in Cycle 8. Accepted proposals receive a priority grade of A, B, or C, and the acceptance statistics are presented in Tables~\ref{tab:acceptances_reg} and \ref{tab:acceptances_cat} grouped by regional affiliation and scientific category in the context of the total approved time in Cycle 8. Due to the oversubscription rate in Cycle 8, only 15\% of submitted proposals were accepted with high priority (Grade A and B). Of the accepted proposals from both distributed peer review and the panels, 87\% of the highest priority proposals (Grade A and B) went through the distributed peer review process to account for 59\% of the available observing time on the 12-m Array.

\input{tables/table_submissions_regions}
\input{tables/table_submissions_cat}
\input{tables/table_acceptance_reg}
\input{tables/table_acceptance_cat}

\subsection{Designated reviewers}
\label{subsec:reviewers}
After accounting for the 336 reviewers designated multiple times, 1016 people performed the reviews. For reference, the breakdown of unique PIs and reviewers by region is tabulated in Table~\ref{tab:revdemographics_reg}. Unless noted, all further analysis presented in this paper is expressed per submitted proposal or proposal set (rather than per reviewer).

\begin{figure}
\centering
\includegraphics[width=7in, trim=0.5in 0in 0.5in 0in, clip]{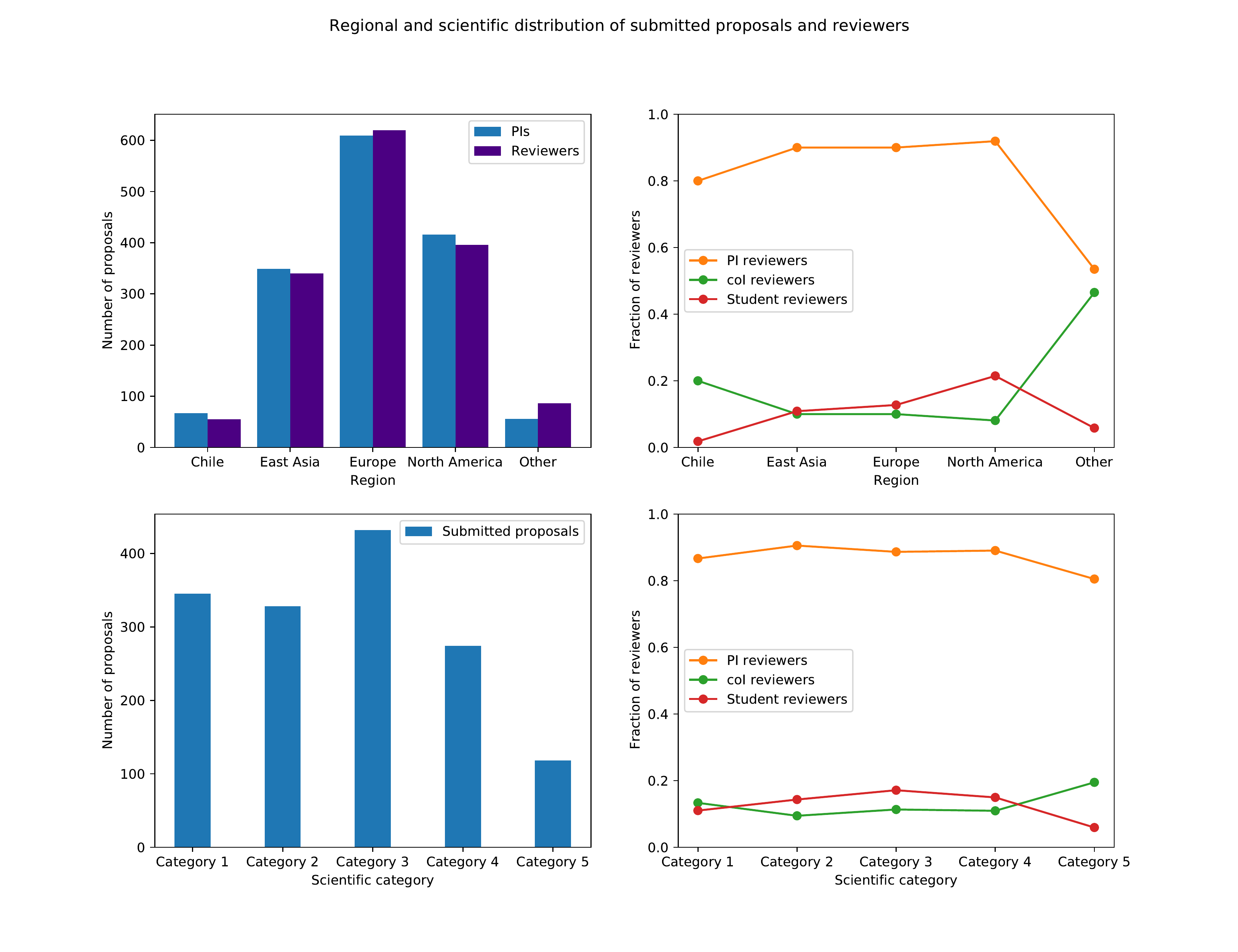}
\caption{(Top left) Regional distribution of PIs and reviewers. (Top right) Fraction of designated reviewers per region who were PIs, coIs, and students. The orange line represents all PIs, which includes student reviewers. (Bottom left) Scientific distribution of submitted proposals. (Bottom right) Fraction of designated reviewers per scientific category who were PIs, coIs, and students. The orange line represents all PIs, which includes student reviewers.}
\label{fig:regdist}
\end{figure}

PIs served as reviewers on behalf of 1318 submitted proposals (88\% of all proposals), and they designated coIs to serve in 179 cases (12\%). Figure~\ref{fig:regdist} shows the regional and scientific category distributions of PIs and reviewers, and it also displays the author status of the designated reviewers. In the right panels of the figure, the orange lines represent all proposal PIs, including reviewers with and without PhDs. The regional distributions of submitted proposals and reviewers were similar. Designating a coI was more common in the Open Skies region (47\% of proposal sets assigned on behalf of Open Skies proposals) than in the other regions. 

\input{tables/table_reviewerdemographics}

Reviewers without a PhD were designated in all regions and scientific categories; for simplicity, we will refer to this group as ``students" throughout this paper. After accounting for students assigned multiple proposal sets, 164 unique student reviewers participated in the process on behalf of 207 proposal sets (14\% of all proposal sets), with the help of 138 unique mentors. North American students were designated on the most proposal sets at 21\%, while Chile designated the fewest student reviewers (2\%), compared in each case to the numbers of reviewers designated from those regions. Proposals submitted to Category 3 designated the most student reviewers at 17\%, while those submitted to Category 5 designated the fewest at 6\%, compared in each case to the numbers of proposals submitted to each category.

\subsubsection{Assignment load}
\label{subsec:load}
\begin{figure}
\centering
\includegraphics[width=3.5in, trim=1in 0.5in 1in 0in, clip]{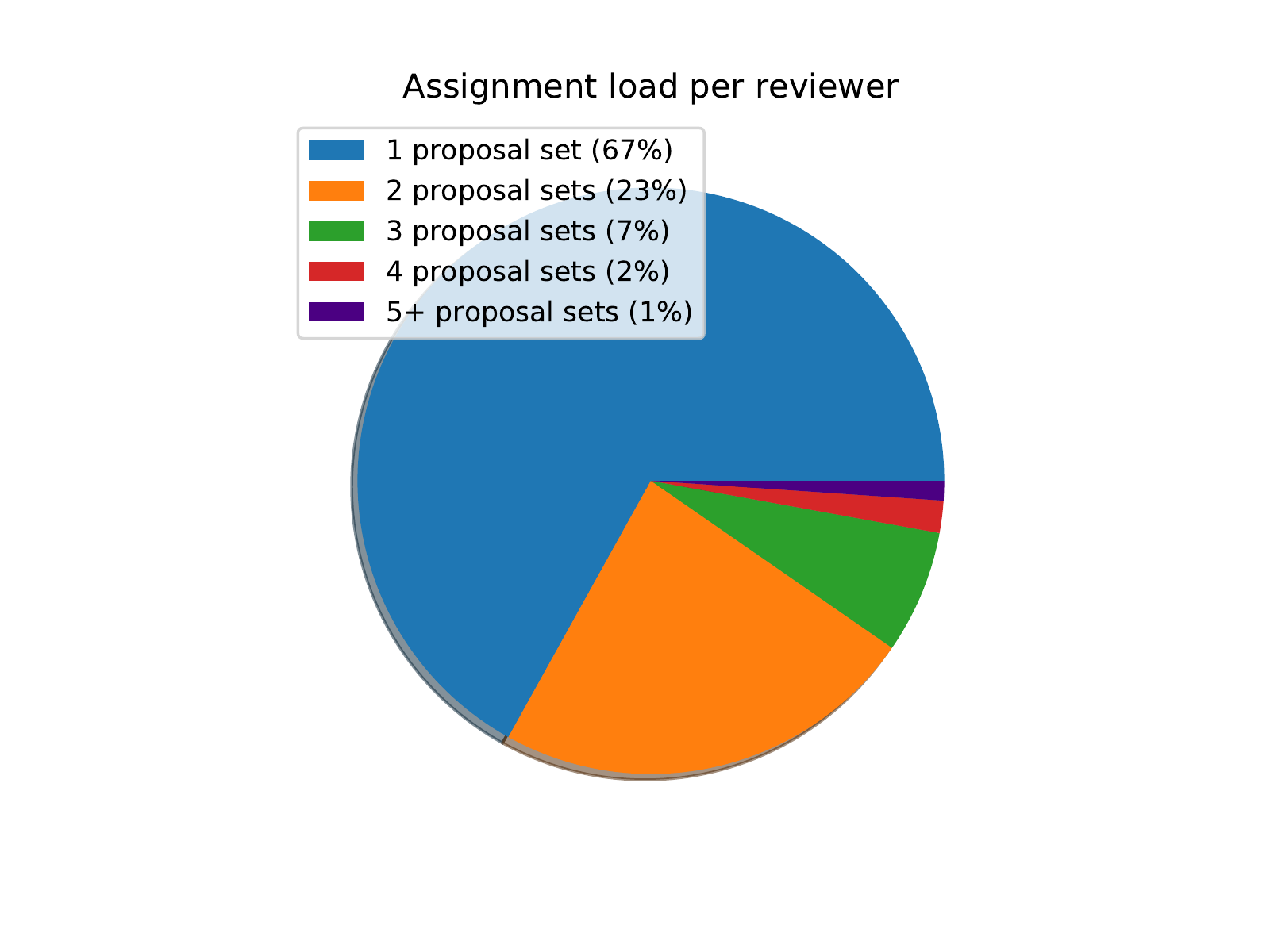}
\includegraphics[width=3.5in, trim=1in 0.5in 1in 0in, clip]{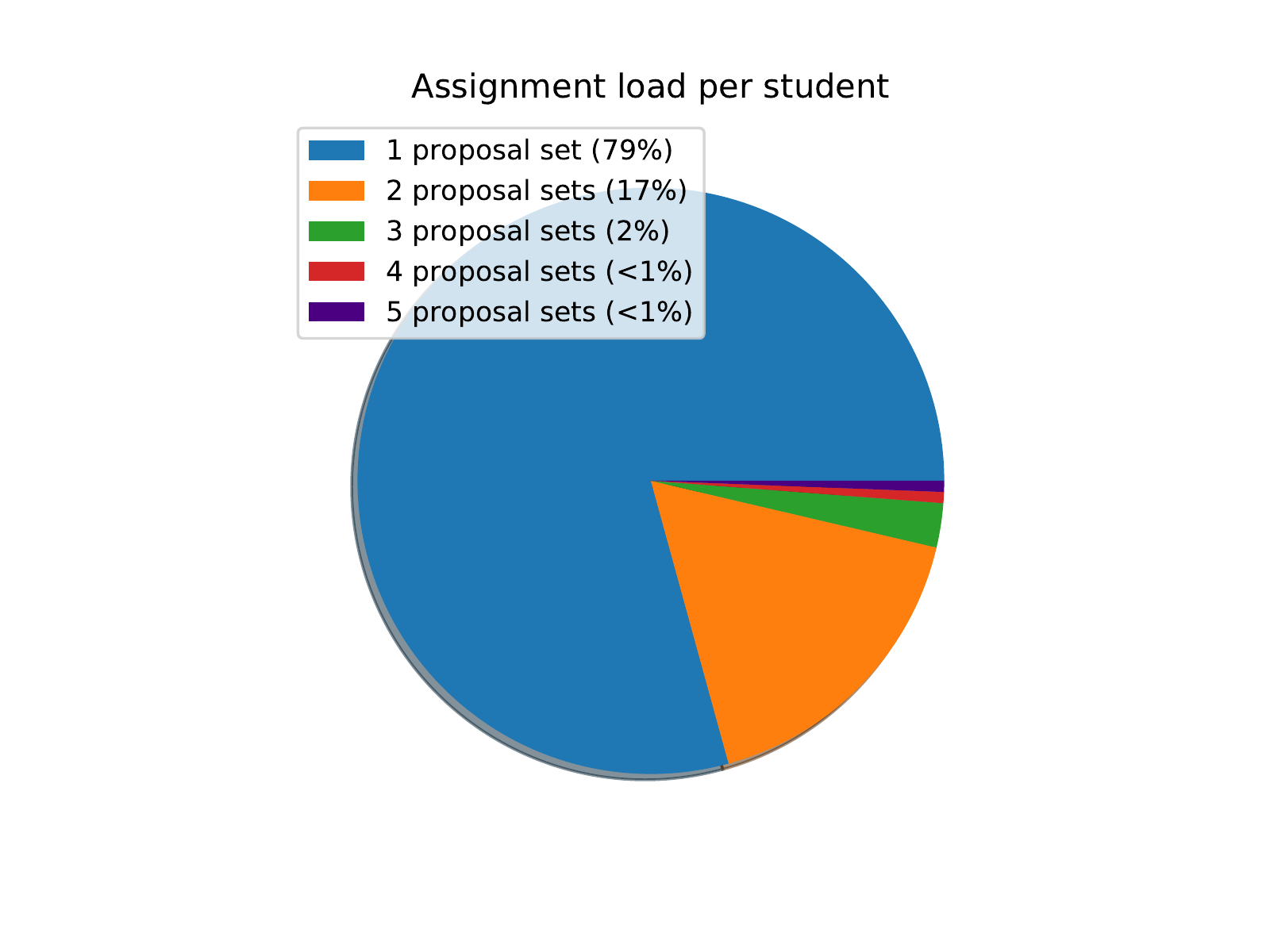}
\caption{(Left) Percentage of reviewers assigned a given number of proposal sets, shown for all reviewers. (Right) Percentage of student reviewers assigned a given number of proposal sets.} 
\label{fig:nsets}
\end{figure}

Figure~\ref{fig:nsets} shows the ``assignment load", or the number of proposals sets assigned, expressed as percentages of all reviewers (left panel) and student reviewers (right panel) that designated reviewers opted to review in Cycle 8. The majority of reviewers (90\%) were assigned either 1 (680 reviewers) or 2 (238 reviewers) proposal sets. A small number of reviewers (11) were responsible for 5 or more proposal sets, and the largest number of proposal sets assessed by a single reviewer was 9. As a result, 45\% of assignments were reviewed by a reviewer with a single proposal set, 32\% of assignments were reviewed by a reviewer with two proposal sets, and 23\% of assignments were reviewed by a reviewer with three or more proposal sets. Figure~\ref{fig:nsets} also indicates the assignment load for students, who were more likely to have been assigned a single proposal set compared to reviewers with a PhD.

\begin{figure}
\centering
\includegraphics[width=6.5in, trim=1in 0in 1in 0in, clip]{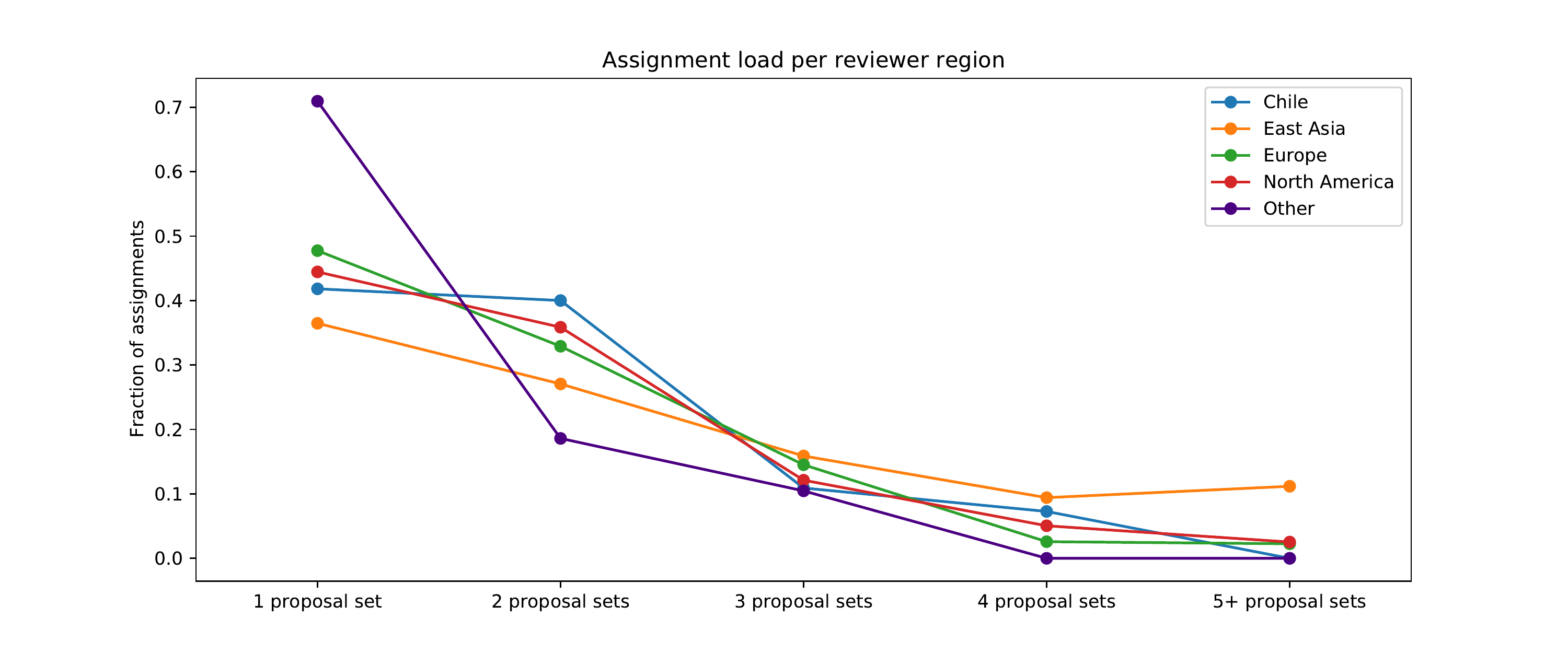}
\caption{Assignment loads assumed by reviewers, split by their region. Fractions are calculated relative to the total number of assignments made to reviewers from each region.}
\label{fig:nsets_cat}
\end{figure}

Reviewers took on a variety of assignment loads, as shown in Figure~\ref{fig:nsets_cat}. Reviewers from East Asia took on the smallest share of single-proposal set assignment loads (36\% of assignments to East Asian reviewers were in single-proposal set assignment loads) and the largest share of three-or-more proposal set assignment loads (when considering the 3, 4, and 5 proposal set bins, 36\% of assignments to East Asian reviewers were in three-or-more proposal set assignment loads). The other regions' reviewers took on single proposal set assignment loads more frequently -- closer to half of the time (42-48\% of assignments), and for Open Skies 71\% of the time -- and had less than 20\% of their assignments belonging to three-or-more proposal set assignment loads. Assignments to reviewers carrying the highest assignment loads (5+ proposal sets) comprised fully 11\% of the assignments made to East Asian reviewers, compared to less than 3\% in the other regions. In all cases, fractions are calculated relative to the total number of assignments made to reviewers from each region.

\subsection{Reviewer expertise}
As described in Section~\ref{subsec:process}, reviewers were advised to submit at least three keywords representing their expertise to aid matching of their assignments to the distribution of submitted proposals. In Cycle 8, 845 reviewers submitted their expertise on behalf of 1253 proposal sets (83\% and 84\% of the total number of reviewers and proposal sets, respectively). For students, 147 reviewers on behalf of 180 proposal sets submitted their expertise (90\% and 87\%, respectively). The expertise for the remaining reviewers was assumed from their submitted proposals. 

\subsubsection{Distribution of reviewers' submitted expertise}
\label{subsec:dist_expertise}
\begin{figure}
\centering
\includegraphics[scale=0.6]{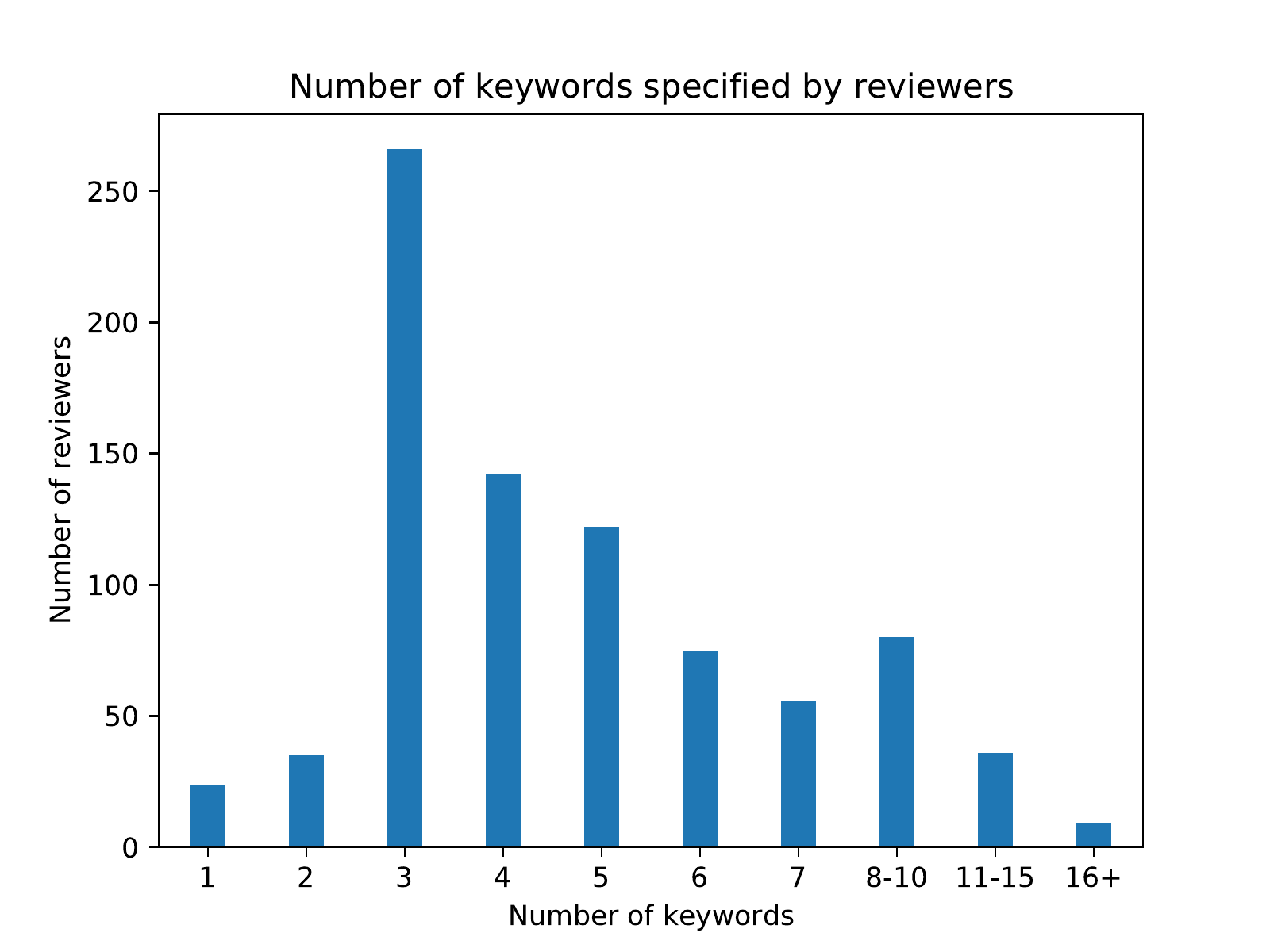} 
\includegraphics[width=\textwidth, trim=1in 0in 1in 0in, clip]{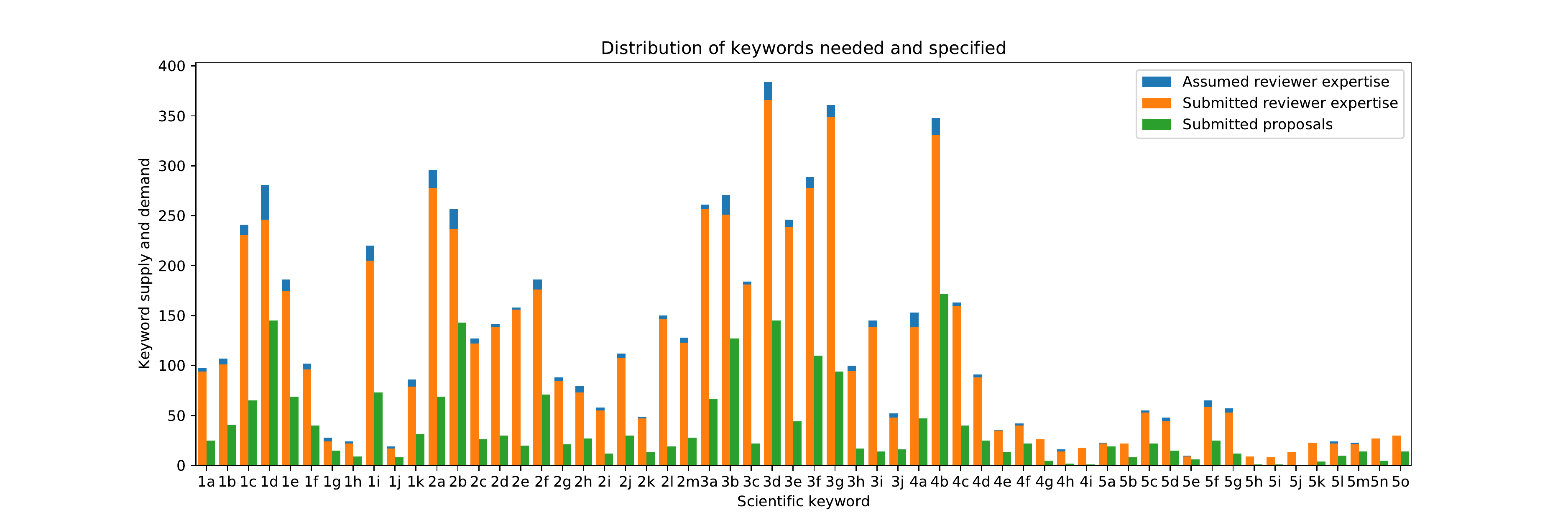}
\caption{(Top) Number of keywords specified by reviewers for use in the assignment process. (Bottom) Number of submitted proposals by scientific keyword (green) and the total number of reviewers with expertise in that keyword (orange and blue). If, for instance, a reviewer submitted 5 keywords in the top panel, they would appear in 5 different orange bars in the bottom panel. Appendix~\ref{app:keywords} lists the definitions of the keywords.}
\label{fig:pairs}
\end{figure}

Cycle 8 reviewers who specified their expertise submitted anywhere from 1 to 23 keywords for a total of 6,505 submitted expertise keywords. A histogram of the number of keywords submitted by reviewers is shown in the top panel of Figure~\ref{fig:pairs}. The most popular number of keywords submitted was three, likely due to the recommended minimum number of keywords to submit. No expertise keywords were submitted on behalf of 244 proposal sets. The bottom panel of Figure~\ref{fig:pairs} shows the histogram of keywords specified as expertise by reviewers, compared in each case to the number of proposals submitted with those keywords (i.e., ``supply and demand")\footnote{The definitions of the keywords, each denoted by a shorthand in the bottom panel of Figure~\ref{fig:pairs}, are listed in Appendix~\ref{app:keywords} and in the Cycle 8 Proposers Guide (https://almascience.org/documents-and-tools/cycle8/alma-proposers-guide, \citet{Braatz21}.}. In all keywords, the supply of reviewer expertise (orange/blue bar) is larger than the demand (green bar). However, a range of ``oversupply" can be seen in this figure, from keywords like 2e and 4i (``Surveys of galaxies" and ``Asteroids"), where 8-18 times as many reviewers claimed expertise as proposals were submitted) to keywords like 5a and 1g (``The Sun" and ``Damped Lyman Alpha systems"), where only 1.2-1.8 times as many reviewers claimed expertise as proposals were submitted. It is also important to keep in mind when interpreting this panel that a reviewer who, for instance, submitted 5 keywords in the top panel would appear in 5 different orange bars in the bottom panel. Therefore, a simple ``oversupply" in a given keyword does not mean that there will be enough exact matches between every reviewer and every submitted proposal after accounting for reviewers with expertise in many keywords and their conflicts of interest (see Section~\ref{sec:assign} for the details of the assignment process). 

\subsubsection{Diversity of reviewer expertise}
\begin{figure}
\begin{centering}
\includegraphics{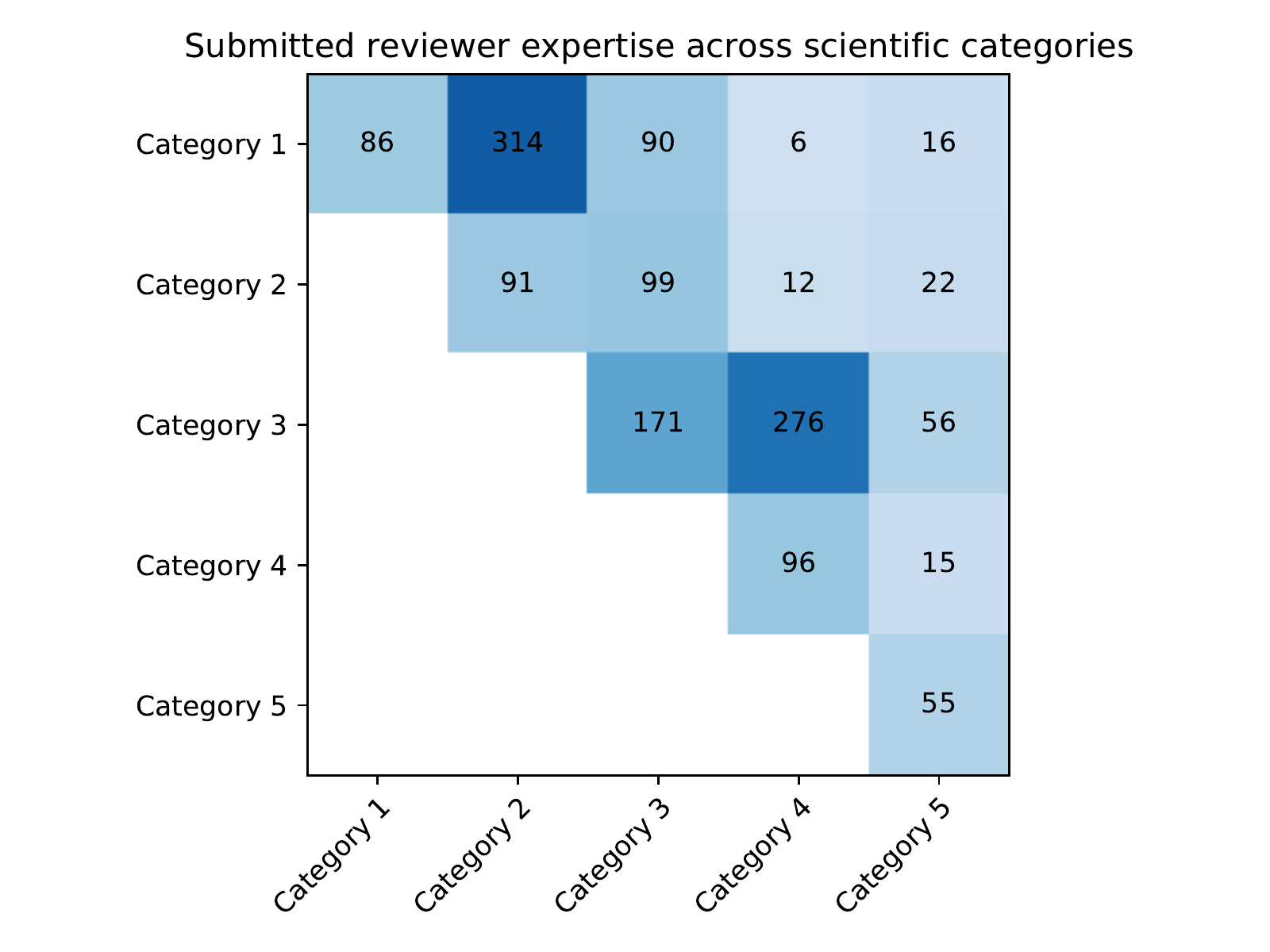}
\caption{Correlation of submitted reviewer expertise in pairs of scientific categories. Reviewers who submitted expertise in only one category appear in only the diagonal cells, and those who submitted expertise in more than one category appear in the off-diagonal cells. Reviewers who submitted expertise in 3 or 4 categories appear in multiple cells. The matrix is symmetric, and therefore only the upper triangle is shown.}
\label{fig:experts}
\end{centering}
\end{figure}

Almost half of all reviewers (488, or 48\%, and 58\% of reviewers who submitted any expertise) specified expertise in more than one scientific category on behalf of 754 proposal sets. This diversity of expertise was found in all scientific categories; the majority of reviewers who specified expertise in each category also selected expertise keywords in an additional category. The diversity of declared reviewer expertise, shown in terms of reviewers who submitted expertise in each scientific category correlated with themselves, can be seen in Figure~\ref{fig:experts}. Reviewers who submitted expertise in only one category appear in only the diagonal cells, and reviewers who submitted expertise in 3 or 4 categories (designated on behalf of 142 proposal sets) appear in multiple cells in this figure. This figure shows, for instance, that roughly 3.5 times as many reviewers specified both Category 1 and Category 2 expertise keywords (314 reviewers) compared to only Category 1 expertise (86 reviewers) or only Category 2 expertise (91 reviewers). Even for reviewers with expertise in Category 5, more reviewers specified expertise in both Category 3 and Category 5 (56 reviewers) than Category 5 alone (55 reviewers). 

In general, reviewers submitted expertise keywords which included the scientific category of their submitted proposal; in only 41 cases did reviewers specify expertise keywords that were completely out-of-category with respect to that reviewer's submitted proposal. In most scientific categories, such reviewers comprised a small fraction of proposal sets (2-3$\%$), but in Category 5, 8$\%$ of submitted proposals designated reviewers who specified only keywords outside of Category 5. 

%% file: tables/table_submissions_regions.tex
\begin{deluxetable*}{lcccccc}
\tablecolumns{7}
\tablecaption{Summary of submitted distributed peer reviewed (all reviewed) proposals, by region \label{tab:submissions_reg}}
\tablehead{\colhead{} & \colhead{Chile (CL)} & \colhead{East Asia (EA)} & \colhead{Europe (EU)} & \colhead{North America (NA)} & \colhead{Open Skies} &  \colhead{Total}}

\startdata
Number of proposals & 67 (76) & 349 (389) & 609 (721) & 416 (484) & 56 (65) & 	1497 (1735) \\
12-m Array time (hours) & 	676 (1308) & 3697 (5445) & 	6070 (11066) & 	4203 (7668) & 525 (838) & 	15170 (26325) \\
7-m Array time (hours) & 551 (862) & 3407 (4542) & 2435 (4447)	 & 2401 (4602) & 357 (394) & 	9151 (14846) \\
Total Power Array time (hours) & 132 (420) & 3895 (4537) & 1820 (3918) & 1756 (4822) & 106 (106) & 	7709 (13802) \\
\enddata

\end{deluxetable*}

%% file: tables/table_submissions_cat.tex
\begin{deluxetable*}{lcccccc}
\tablecolumns{7}
\tablecaption{Summary of submitted distributed peer reviewed (all reviewed) proposals, by category\tablenotemark{a} \label{tab:submissions_cat}}
\tablehead{\colhead{} & \colhead{Category 1} & \colhead{Category 2} & \colhead{Category 3} & \colhead{Category 4} & \colhead{Category 5} &  \colhead{Total}}

\startdata
Number of proposals & 345 (434) & 328 (386) & 432 (465) & 274 (320) & 118 (130) & 	1497 (1735) \\
12-m Array time (hours) & 	4175 (8570) & 3254 (5940) & 	3894 (5189) & 	2922 (5292) & 926 (1334) & 	15170 (26325) \\
7-m Array time (hours) & 1155 (1272) & 2918 (5931) & 3962 (6037)	 & 595 (971) & 522 (635) & 	9151 (14846) \\
Total Power Array time (hours) & 304 (304) & 1980 (5058) & 5140 (8155) & 31 (31) & 253 (253) & 	7709 (13802) \\
\enddata
\tablenotetext{a}{The scientific categories are provided in Appendix~\ref{app:keywords}.}
\end{deluxetable*}

%% file: tables/table_acceptance_reg.tex
\begin{deluxetable*}{lcccccc}
\tablecolumns{7}
\tablecaption{Summary of accepted distributed peer reviewed (all reviewed) proposals, by region \label{tab:acceptances_reg}}
\tablehead{\colhead{} & \colhead{Chile (CL)} & \colhead{East Asia (EA)} & \colhead{Europe (EU)} & \colhead{North America (NA)} & \colhead{Open Skies} &  \colhead{Total}}

\startdata
Grade A+B & & & & & & \\
\hline
Number of proposals & 26 (29) & 61 (67) & 58 (71) & 71 (83) & 3 (3) & 	219 (253) \\
12-m Array time (hours) & 	253 (383) & 607 (870) & 	700 (1310) & 	718 (1321) & 30 (30) & 	2309 (3914) \\
7-m Array time (hours) & 385 (385) & 398 (677) & 160 (518)	 & 664 (1002) & 0 (0) & 	1606 (2581) \\
Total Power Array time (hours) & 126 (126) & 280 (486) & 145 (462) & 535 (852) & 0 (0) & 	1086 (1927) \\
\hline
Grade C & & & & & & \\
\hline 
Number of proposals & 10 (13) & 47 (52) & 89 (94) & 58 (67) & 6 (7) & 	210 (233) \\
12-m Array time (hours) & 	125 (231) & 502 (658) & 	856 (1047) & 	664 (982) & 94 (126) & 	2241 (3044) \\
7-m Array time (hours) & 24 (49) & 192 (238) & 462 (494)	 & 156 (156) & 0 (0) &	833 (936) \\
Total Power Array time (hours) & 0 (0) & 286 (286) & 0 (0) & 37 (37) & 0 (0) & 323 (323) \\
\enddata

\end{deluxetable*}

%% file: tables/table_acceptance_cat.tex
\begin{deluxetable*}{lcccccc}
\tablecolumns{7}
\tablecaption{Summary of accepted distributed peer reviewed (all reviewed) proposals, by category \label{tab:acceptances_cat}}
\tablehead{\colhead{} & \colhead{Category 1} & \colhead{Category 2} & \colhead{Category 3} & \colhead{Category 4} & \colhead{Category 5} &  \colhead{Total}}

\startdata
Grade A+B & & & & & & \\
\hline
Number of proposals & 61 (74) & 49 (57) & 60 (66) & 34 (40) & 15 (16) & 	219 (253) \\
12-m Array time (hours) & 	765 (1308) & 478 (836) & 	541 (829) & 406 (787) & 119 (154) & 2309 (3914) \\
7-m Array time (hours) & 149 (149) & 684 (1126) & 507 (1002)	 & 104 (141) & 162 (162) & 	1606 (2581) \\
Total Power Array time (hours) & 0 (0) & 209 (335) & 824 (1538) & 0 (0) & 54 (54) & 	1086 (1927) \\
\hline
Grade C & & & & & & \\
\hline 
Number of proposals & 61 (69) & 47 (55) & 56 (60) & 36 (39) & 10 (10) & 	210 (233) \\
12-m Array time (hours) & 	776 (1060) & 496 (788) & 	503 (633) & 365 (461) & 101 (101) & 2241 (3044) \\
7-m Array time (hours) & 260 (260) & 271 (350) & 287 (312) & 0 (0) & 15 (15) &	833 (936) \\
Total Power Array time (hours) & 0 (0) & 149 (149) & 174 (174) & 0 (0) & 0 (0) & 323 (323) \\
\enddata

\end{deluxetable*}

%% file: tables/table_reviewerdemographics.tex
\begin{deluxetable*}{lcccccc}
\tablecolumns{7}
\tablecaption{Summary of reviewer demographics, by region \label{tab:revdemographics_reg}}
\tablehead{\colhead{} & \colhead{Chile (CL)} & \colhead{East Asia (EA)} & \colhead{Europe (EU)} & \colhead{North America (NA)} & \colhead{Open Skies} &  \colhead{Total}}

\startdata
PIs & 45 & 216 & 426 & 277 & 47 & 1011 \\
All reviewers & 37 & 203 & 434 & 270 & 72 & 1016 \\
PI-reviewers & 34 & 190 & 396 & 252 & 41 & 	913 \\
coI-reviewers & 9 & 26 & 58 & 28 & 37 & 158 \\
Student reviewers & 1 & 32 & 67 & 59 & 5 & 	164 \\
\enddata

\end{deluxetable*}

%% file: assignments.tex
\section{Assignment process}
\label{sec:assign}

The goal of the assignment exercise is to create proposal sets that closely align with the expertise of the reviewer and avoid potential conflicts of interest. In the following subsections, we describe the process of assigning reviews and the outcome of the distribution of assignments.

\subsection{Creating review assignments}
\label{subsec:rules}
Priority was placed on assigning proposals that aligned with the expertise specified by the reviewer. However, due to conflicts of interest and the fact that each reviewer was assigned 10 proposals and each proposal needed to be reviewed 10 times, this was not possible in all cases (recall the discussion in Section~\ref{subsec:dist_expertise}). In such cases, proposals were assigned within the same scientific category, or a similar scientific category, as the reviewer's expertise keywords. Given that the scientific topics within a category can be broad, for the first time in Cycle 8, the PHT pre-defined keywords within each category and even across categories that were similar to each other; these similar keywords were used to constrain the proposal assignments (e.g., 1a [Lyman Alpha Emitters/Blobs] and 1d [Sub-mm Galaxies], or 3d [Low-mass star formation] and 4b [Disks around low-mass stars]). Assignments which fell completely outside of the same or similar categories and keywords relative to a reviewer's expertise occurred in only a handful of cases. 

In practice, a list of 20 ``rules" were created that described the relationships between reviewers and each submitted proposal, where an increasing rule number corresponded to a bigger gap between a reviewer's submitted or assumed expertise and the content of the proposal to be reviewed. The matching algorithm assigned priority to low rule number matches. For all rule groups enumerated below, conflicts of interest were identified and avoided in the proposal assignments; conflicts with the PI (if they were not the reviewer) were permitted in some cases to allow a better match with a reviewer's expertise. Conflicts of interest are further described in Section~\ref{subsec:conflicts}. The groups of rules are summarized as follows: 

\textbf{Rule 1} assignments focus on three keywords from the reviewer's expertise keyword list which are selected from their best matching scientific category (or reference category) in order to provide a self-consistent proposal set. A reviewer's reference category is determined to be the one containing the most keywords in the their submitted expertise list or their submitted proposal category. \textbf{Rule 2} assignments expand the definition of the reviewer's expertise to all submitted keywords in the reviewer's reference category. Conflicts with the PI are not permitted in Rule 1-2 assignments.

\textbf{Rules 3-5} assignments expand the definition of the reviewer's expertise to their entire submitted keyword list if those keywords span multiple categories. Conflicts with PIs are only permitted if the PI is not the reviewer, and even then only as a last resort in order to enable a quality assignment.

Three groups of rules within \textbf{Rules 6-18} enable assignments using keywords deemed similar by the PHT to those included in a reviewer's expertise keywords list. The first group of rules (6-9) relies on similar keywords within the reference category, then expands to include other categories. The second group of rules (10-14) expands the possible assignments to any keyword within the reference category, any keyword within other categories containing reviewer-specified expertise, and then any category containing a keyword deemed similar to one specified by the reviewer. The third group of rules (15-18) applies the above logic to the reviewer's submitted proposal category and keyword (if not already used). In each group of rules, conflicts with PIs are only permitted if the PI is not the reviewer, and even then only as a last resort.

\textbf{Rules 19-20} assignments allow any available assignment to be made. Conflicts with PIs are only permitted if the PI is not the reviewer, and even then only as a last resort.

In summary, Rules 1-5 assign proposals that match the scientific keywords of the reviewer. Rules 6-18 assign proposals that are in the same or similar categories compared to the keywords specified in a reviewer's expertise. Rules 19-20 assign proposals in a category unrelated to the reviewer's expertise and their own submitted proposal.

\subsubsection{Conflicts of interest}
\label{subsec:conflicts}
For Cycle 8, the JAO implemented automated checks to avoid potential conflicts of interest in the proposal assignments and to enable reviews as unbiased as possible. The reviewer had a conflict of interest if the PI, designated reviewer, or mentor of their own submitted proposal was a PI, coPI, or coI on the proposal under consideration. The JAO also adopted automated checks to identify potential close collaborators, given that dual-anonymous review was used in Cycle 8 and the proposal team was not revealed to the reviewers. Close collaborators were identified if the PI, reviewer, or mentor of the submitted proposal and the PI of the proposal under consideration were collaborators on another proposal in Cycle 8, with one of the them serving as PI. An institutional conflict was identified if the PI, reviewer, or mentor the submitted proposal is at the same institution as the PI of the proposal under consideration.

Potential assignments which represented a conflict for the reviewer and/or mentor were removed from consideration in the assignment process. Assignments which represented a conflict for the submitted proposal PI, if the PI was not the reviewer, were avoided if possible but were permissible as noted in Section~\ref{subsec:rules} to allow a better match with the reviewer's expertise. 

After the assignments were sent to reviewers, each reviewer examined their assignments and submitted any conflicts to the JAO with a brief justification for the cause of the conflict. Of the 251 conflicts submitted (out of 14970 assignments, or 1.7$\%$), the PHT accepted 192 declared conflicts and assigned replacement proposals. In the other 59 cases, the PHT did not confirm the declared conflict and requested that the reviewer complete the review of the corresponding assignment. Typically, conflicts due to lack of expertise, similar but non-competing science, and suspected knowledge of the proposal team despite dual-anonymous practices were not accepted. No bias toward the best or worst ranks was detected in the 59 rejected conflicts.

\subsection{Assigned Proposal Sets}
\label{subsec:assignments}
\begin{figure}
\begin{centering}
\includegraphics[scale=0.8, trim=0.5in 0.5in 0.5in 0in, clip]{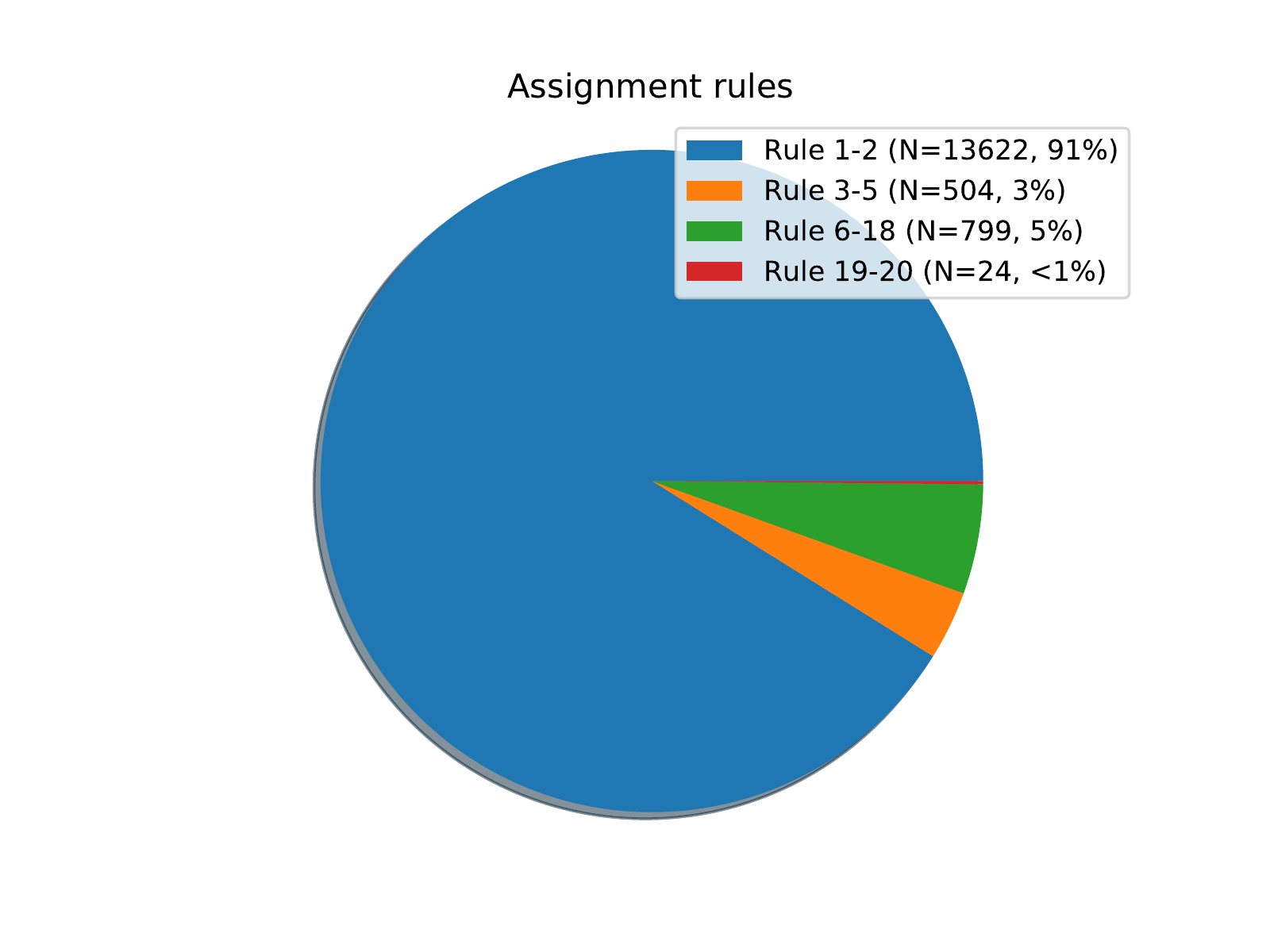}
\caption{Distribution of assignment rule groups in the Cycle 8 distributed review process. Out of 14,970 assignments, the assigned proposals matched the reviewer's specified expertise in 94\% of cases.}
\label{fig:rules}
\end{centering}
\end{figure}

Following the assignment and conflict rules described in the previous sections, the assignment of proposal sets resulted in primarily well-matched reviewers and proposals. As shown in Figure~\ref{fig:rules}, 91$\%$ of assignments were Rule 1-2 assignments (relying on keywords within a reviewer's reference category); the vast majority of these were Rule 1 assignments relying on up to three prioritized keywords. Fully 94$\%$ of assignments consisted of exact matches between a reviewer's specified expertise keyword and submitted proposal (Rules 1-5). Only 24 assignments fell into Rule 19. For most of the subsequent analysis in this paper, the assignment rules are grouped by their relevance to the reviewer's expertise as described in Section~\ref{subsec:rules}: Rules 1-5 match the scientific keywords of the reviewer, Rules 6-18 provide assignments related or similar to the reviewer's expertise, and Rules 19-20 provide assignments unrelated to the reviewer's expertise.

\begin{figure}
\begin{centering}
\includegraphics[scale=0.6]{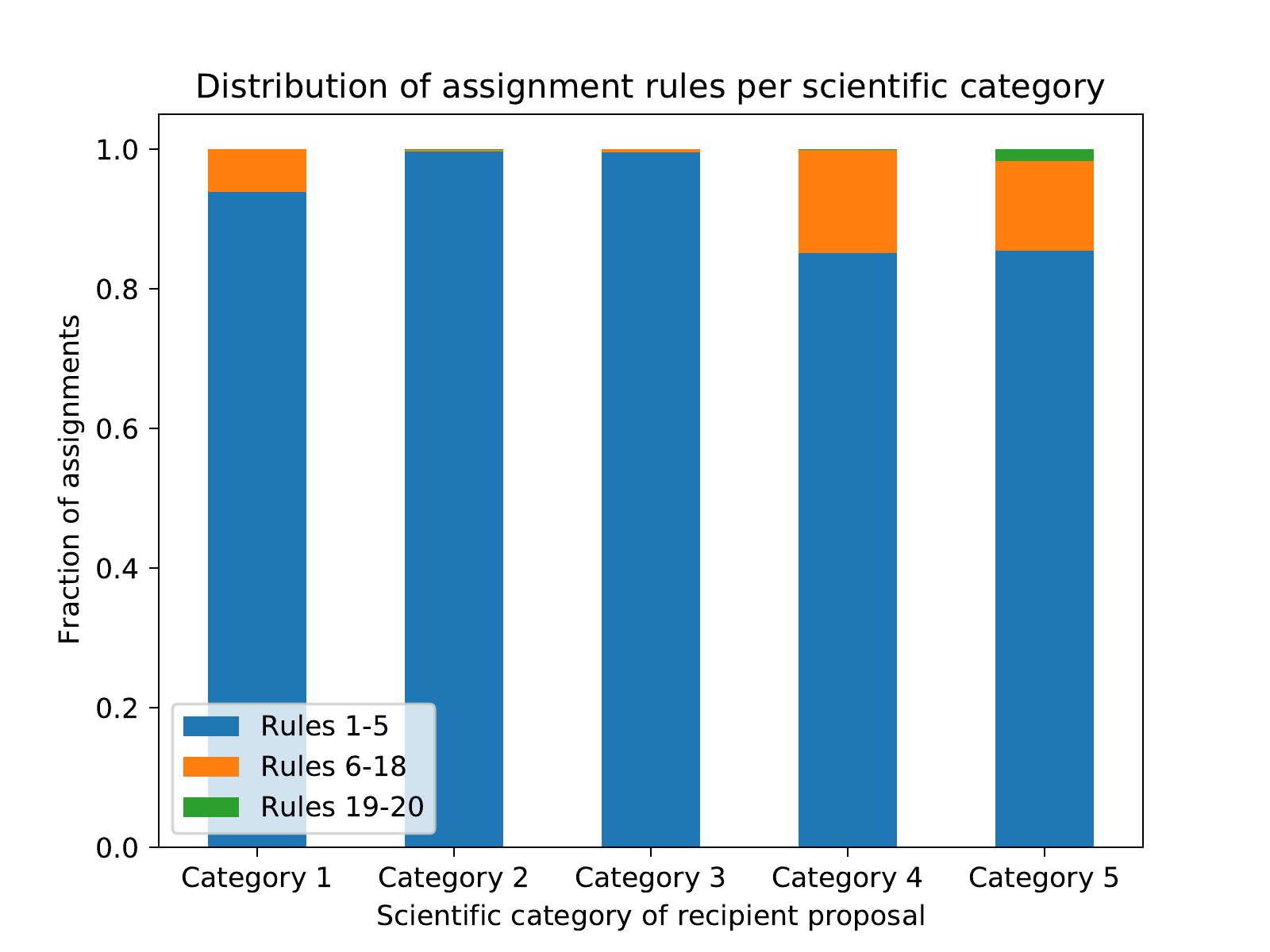} 
\includegraphics[width=\textwidth, trim=1in 0in 1in 0in, clip]{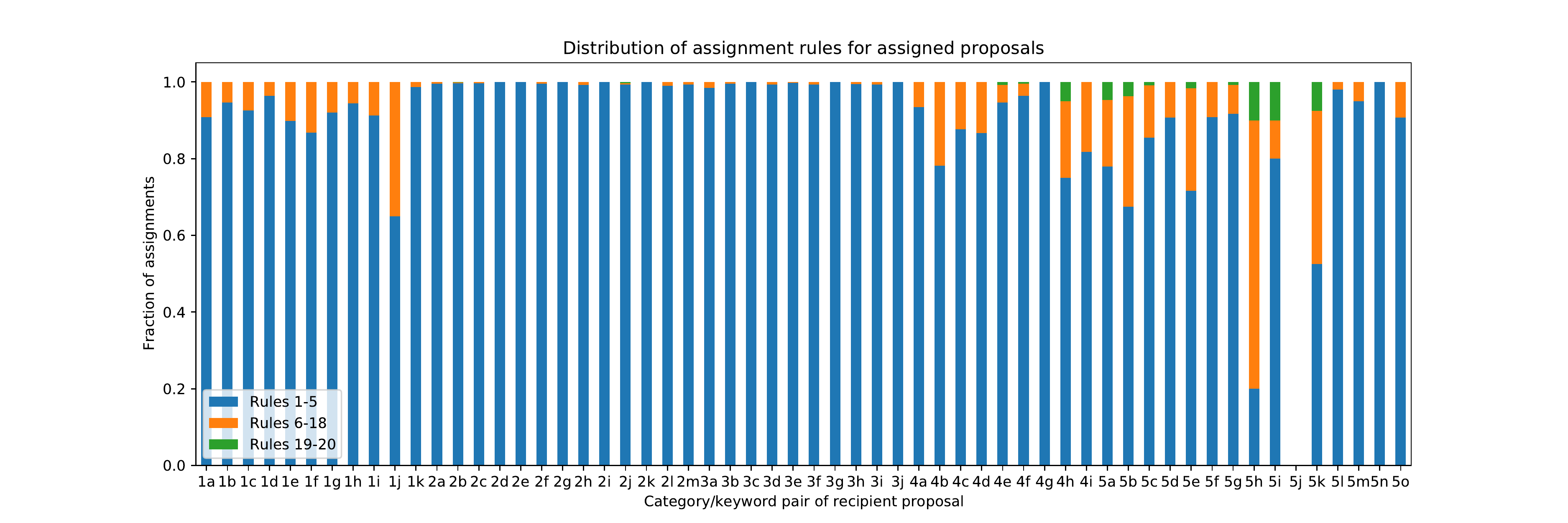}
\caption{(Top) Fraction of assignments per rule group, broken into scientific categories of the assigned proposal. Rules 1-5 match the scientific keywords of the reviewer, Rules 6-18 provide assignments related or similar to the reviewer's expertise, and Rules 19-20 provide assignments unrelated to the reviewer's expertise. (Bottom) Fraction of assignments, as in the top panel, now split by keyword.}
\label{fig:rules_scicat}
\end{centering}
\end{figure}

Some trends with assignment rule are apparent as a result of the availability of non-conflicted expert reviewers in each scientific category and keyword. In the top panel of Figure~\ref{fig:rules_scicat}, the assignment rule distribution for recipient proposals is shown broken down by scientific category. In each scientific category, the reviewer's specified expertise matched a keyword of the assigned proposal in at least 85\% of assignments, and this fraction is nearly 100\% in Categories 2 and 3. The bottom panel of Figure~\ref{fig:rules_scicat} breaks down the assignment rule distribution by individual keyword. Within each individual keyword, 65\% or more of the reviews were performed by reviewers who indicated expertise in that keyword (with exceptions in two undersubscribed Category 5 keywords, which represented 5 total submitted proposals). 

We can examine the assignment rule distributions even more finely (per reviewer's proposal set and per submitted proposal) to identify the limits of quality for individual sets of assignments. Nearly two-thirds of proposal sets (970) consisted entirely of assignments with matched keywords (i.e., Rule 1-5), and 91\% of proposal sets (1359) contained 9 out of 10 assignments with matched keywords. Submitted proposals received ten matched keyword reviewers 85\% of the time (1268 proposals), and they received 9 out of 10 matched keyword reviewers 91\% of the time (1364 proposals). In aggregate, when looking at individual proposal sets and submitted proposals, the vast majority of reviews were made by reviewers who specified expertise keywords which matched the keywords of their assigned proposals.

\subsubsection{Reviewer assignment load ``seen" by submitted proposals in each scientific category}

\begin{figure}
\plotone{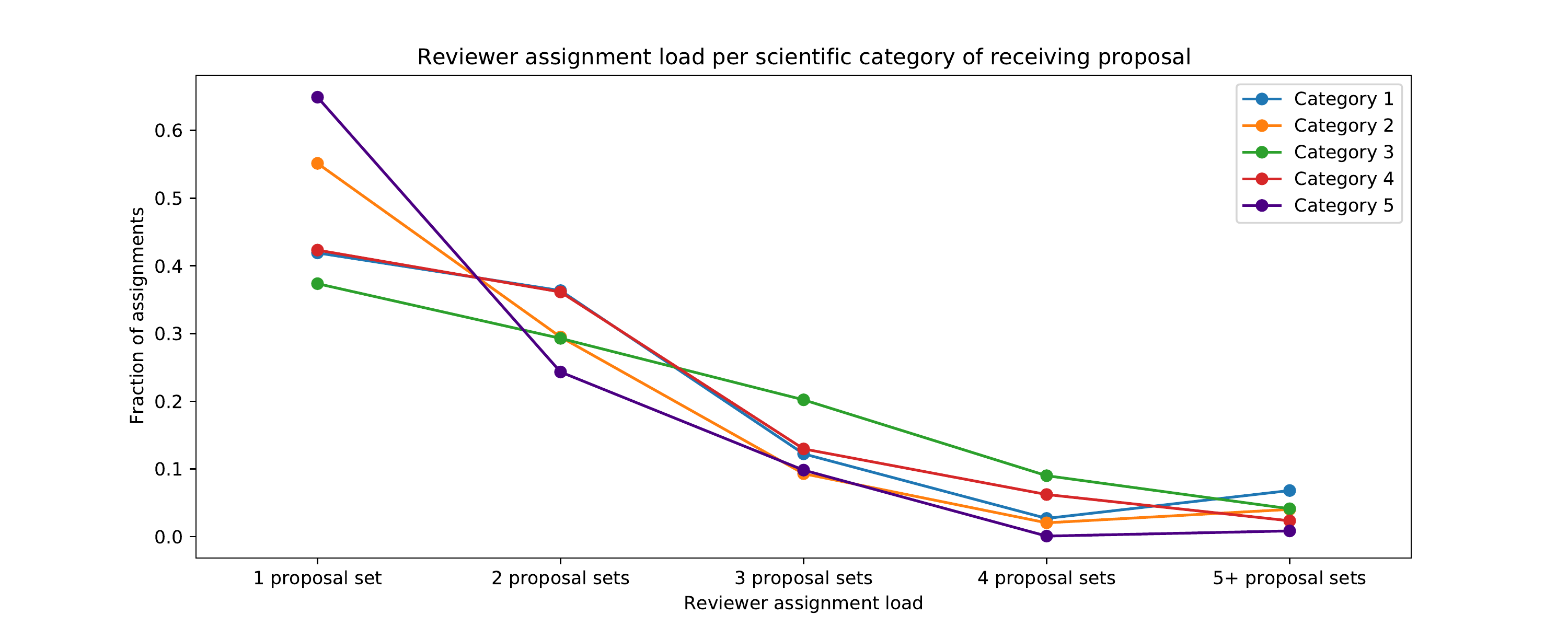}
\caption{The fraction of assignments to reviewers with a given assignment load is shown for the recipient proposals submitted to each scientific category.}
\label{fig:scicat_load}
\end{figure}

With respect to reviewer assignment loads, only minor variations with scientific category are seen in the assigned proposal sets. Figure~\ref{fig:scicat_load} indicates the reviewer workload ``seen" by each proposal receiving reviews, split by scientific category. The fraction of assignments in each category to reviewers carrying a given workload, relative to all assignments made in that category, are shown in the colored lines. For instance, assignments of proposals submitted to Category 5 were made to single proposal set reviewers 65\% of the time and to reviewers with 5 or more proposals sets less than 1\% of the time. While Category 3 proposals ``see" the largest share of reviewers with three-or-more proposal set assignment loads (at 33\%), no scientific category has more than 7\% of its reviews coming from reviewers with the highest (5 or more) assignment loads.
\\

%% file: output.tex
\section{Analysis of individual ranks and comments}
\label{sec:output}

Submitted proposal reviews consisted of both ranks and comments. In this section, we examine the changes that reviewers made to their ranks and comments during Stage 2, and we analyze the final ranks and comments.

\subsection{Changes made to the reviews during Stage 2}

In Stage 2, reviewers had the opportunity to read the anonymized comments written by other reviewers and modify their own Stage 1 ranks and comments. Though 842 reviewers opted to go through Stage 2 on behalf of 1252 proposal sets (84\% of all proposal sets), only 351 reviewers submitted any changes. Updates to the ranks were submitted on behalf of 1258 assignments, and updates to the comments were submitted on behalf of 1398 assignments (with 398 assignments recording a change to both the rank and comments). In the end, reviewers modified the ranks and/or comments in only 15\% of all assignments (18\% of the assignments that had been opted into Stage 2). Since most reviews remained unchanged in Stage 2, it appeared that most reviewers preferred to check the comments of other reviewers but did not find a reason to change their initial reviews.

\begin{figure}
\includegraphics[width=3.5in]{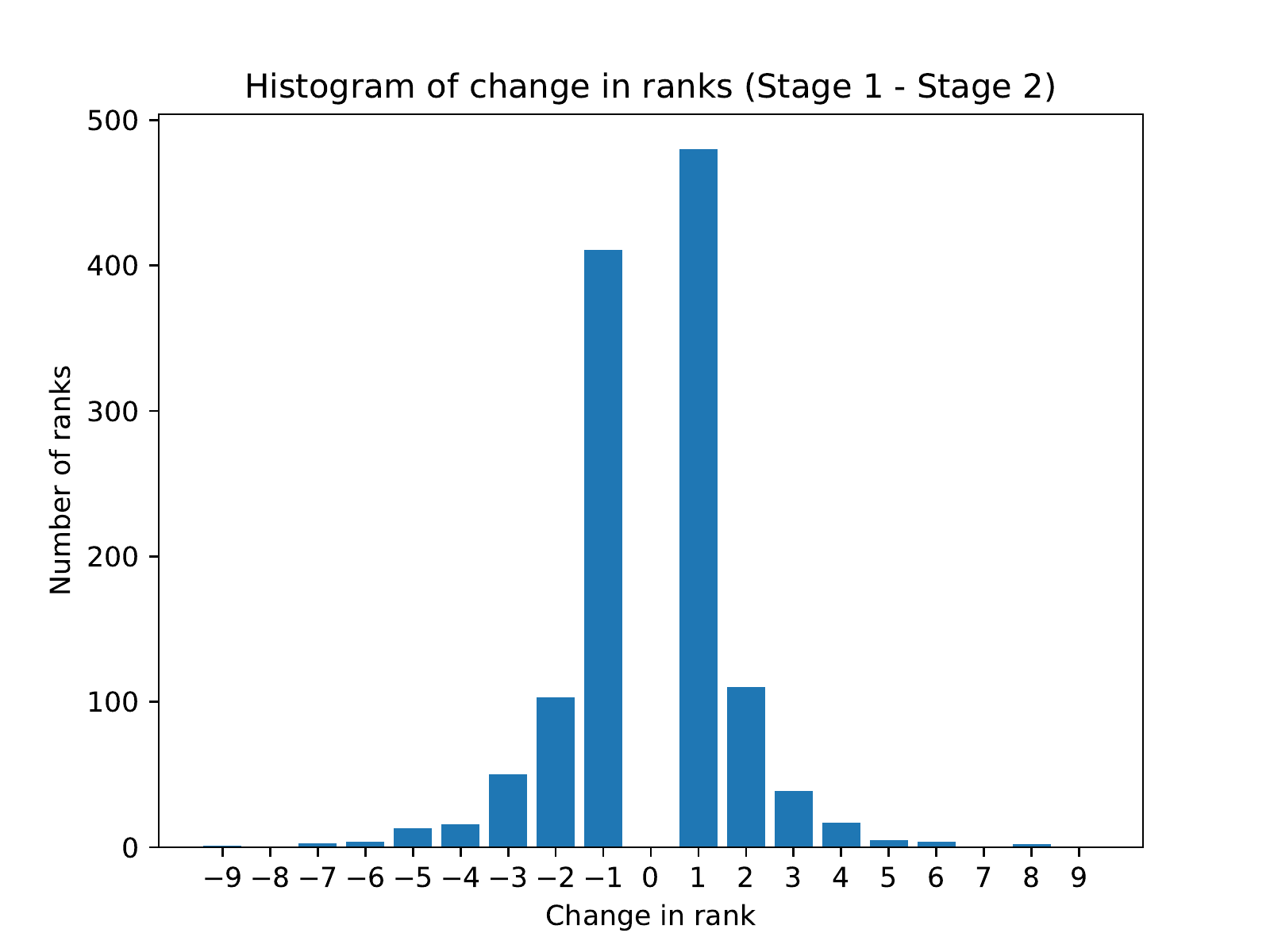}
\includegraphics[width=3.5in]{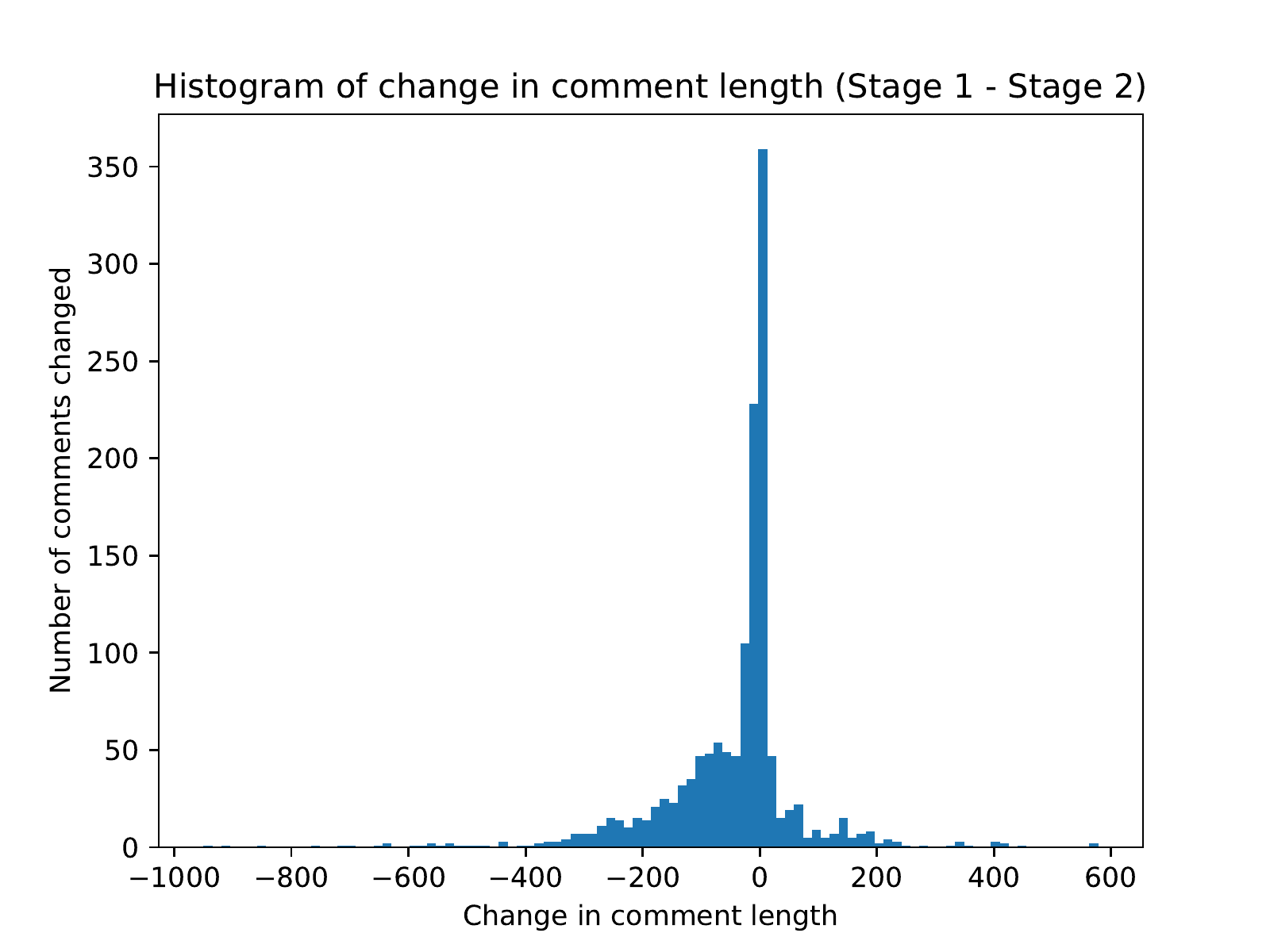}
\caption{(Left) Histogram of changes in ranks during Stage 2 (Stage 1 - Stage 2); i.e., a negative value indicates that the rank dropped in Stage 2, while a positive value indicates that the rank improved in Stage 2. (Right) Histogram of changes in comment length during Stage 2 (Stage 1 - Stage 2); i.e., a negative value indicates that the comment got longer in Stage 2, while a positive value indicates that the comment got shorter in Stage 2.}
\label{fig:stage2}
\end{figure}

Changes to both ranks and comments in Stage 2 tended to be small, as shown in Figure~\ref{fig:stage2}. Of the ranks that changed, most did so by only $\pm$1 (891 assignment ranks, or 71\%), with very few changing by 4 or more (65 assignment ranks, or 5\%). Of the comments that changed, the updated comment lengths were more evenly distributed between 1-10 characters (i.e., 1-2 words, in 506 assignments, or 36\%), 11-100 characters (i.e., 1-2 sentences, in 513 assignments, or 37\%), and more than 100 characters (i.e., a paragraph, in 379 assignments, or 27\%). 

The demographics of reviewers who submitted Stage 2 changes were fairly uniform across a number of axes. They submitted changes mostly uniformly with rank of the proposals, though with slightly higher instances for proposals ranked 2-5, and tended to be those who had written substantial comments already in Stage 1 (400 to 1200 characters). However, reviewers assigned a single proposal set were over-represented in the group of assignments with changes submitted in Stage 2 (they submitted 53\% of the changes, compared to 45\% of assignments which were assigned to this group), and reviewers with 4 or more proposal sets were under-represented (they submitted 4\% of the changes, compared to 9\% of assignments which were assigned to this group).

\subsection{Analysis of final individual ranks}
The final submitted rank of each assigned proposal represents a reviewer's quantitative, relative comparison of the merits of that proposal to the other proposals they reviewed. Since every reviewer was required to submit ten ranks per proposal set, but most reviewers were presented with proposal sets consisting of assignments with a mix of assignment rules, the PHT examined the distribution of ranks to look for trends with assignment quality, using the assignment rule as a proxy for quality. Recall that Rules 1-5 assign proposals matching the scientific keywords of the reviewer, Rules 6-18 assign proposals that are in the same or similar categories compared to a reviewer's expertise keywords, and Rules 19-20 assign proposals that are unrelated to the reviewer's expertise. Of the 1497 proposal sets assigned, 527 contained at least one assignment that fell into Rule 6-20, allowing a comparison of reviewer ranks within proposal sets that relied upon a range of levels of expertise. 

The main topics addressed in this analysis of the individual ranks are how reviewers rank proposals outside of their declared expertise and how they rank proposals in different categories against each other. Since reviewers are required to rank proposals from 1 to 10, the distribution of ranks should be uniform overall unless unintended systematics are present. In the following subsection, we look for potential systematics along a number of axes, including assignment rule, scientific category, student status, reviewer assignment load, reviewer region, and the number of expertise keywords submitted by a reviewer.

\subsubsection{Assignments outside of a reviewer's declared expertise}

In this subsection, we focus on the ranks given to the 799 assignments (5.3\% of all assignments) made using Rules 6-18, which are by definition outside of a reviewer's declared expertise. Though the impact of these assignments on the final ranked list is limited, analyzing them allows us to understand how reviewers ranked proposals outside of their declared expertise when presented with such assignments.

In the absence of biases, the expected distribution of ranks is a uniform distribution, given that every reviewer must submit each rank from 1 to 10 for each assigned proposal set. Any differences between the aggregated distributions were assessed using the Anderson-Darling $k$-sample function {\tt anderson$\_$ksamp} \citep{Scholz87} in {\tt scipy} \citep{Jones01}. This function also returns the probability ($p$, $0 \le p \le 1$) that the $k$ samples are drawn from the same (but unspecified) population. A low value of $p$ suggests that the $k$ samples are drawn from different distributions while a high value of $p$ suggests that the $k$ samples have similar distributions. Any differences in the cumulative ranks are  defined as ``significant" if the probability that the distributions are drawn from the same population is $p<0.01$ and ``marginally significant" if the probability is $0.01\le p \le 0.10$.

\begin{figure}
\begin{centering}
\includegraphics[width=7in, trim=1in 0in 1in 0in, clip]{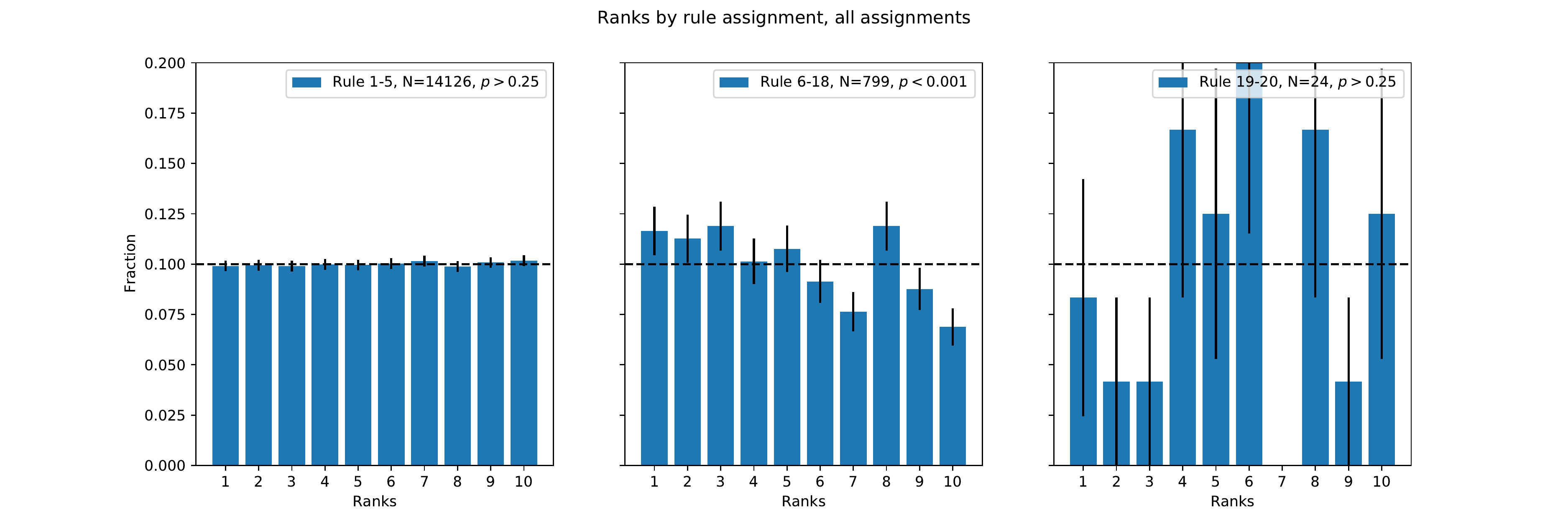}
\caption{Distribution of ranks with assignment rule. The ranks from all expert reviewers with Rule 1-5 assignments are combined into the left panel, the center panel shows the ranks cast by reviewers who were provided assignments similar to (but outside of) their declared expertise, and the right panel shows the ranks from reviewers with assignments unrelated to their expertise. The dotted line in each panel denotes the expectation of a uniform distribution among the ten rank options. The $\it{p}$-value in each panel indicates the probability that the observed distribution of ranks is consistent with a uniform distribution as computed from the Anderson-Darling test. Error bars in each panel reflect confidence intervals ($\sqrt N$) given the observed numbers of counts.}
\label{fig:ranks_agg_3panel}
\end{centering}
\end{figure}

The aggregate distributions of ranks given under each set of rules are shown in the three panels of Figure~\ref{fig:ranks_agg_3panel}. Assignments reviewed by experts (i.e., Rule 1-5; left panel) and those with unrelated expertise (i.e., Rule 19-20; right panel) reveal no trend and are consistent with a uniform distribution with $p>0.25$. This is expected. Since the overall ranks by design are uniformly distributed from 1 to 10, and the vast majority of assignments are in Rule 1-5, the ranks for the dominant subset shown in the left panel must also be largely uniform. On the other hand, only 24 assignments are in Rules 19-20 (as shown in the right panel), such that the statistics are too small to reveal any trends in all but the most extreme distributions. The consistency with a uniform distribution remains for expert (i.e., Rule 1-5) reviewers even when removing the 624 proposal sets consisting of only Rule 1 and 2 assignments (i.e., those who did not have to rank expert and non-expert assignments relative to each other, and those who were guaranteed to cast all ten ranks in the same category); this group is discussed further in the following subsection. However, for the 799 assignments (5.3\% of all assignments) in which reviewers were provided assignments related to, but outside of, their declared expertise (i.e., Rule 6-18, shown in the middle panel of Figure~\ref{fig:ranks_agg_3panel}), the distribution is inconsistent with a uniform distribution. Reviewers tended to give those assignments stronger ranks at a statistically significant level ($p<0.001$).

The same trend can be seen at a statistically significant level in some subsets of reviewers with relatively large instances of Rule 6-18 assignments (a few hundred or more). For instance, the trend visible in the center panel of Figure~\ref{fig:ranks_agg_3panel} is detected among reviewers with PhDs, but it is not detected in the reviewers without PhDs. Reviewers with PhDs produce a non-uniform distribution in the 686 ranks of Rule 6-18 assignments at a detectable level (with $p<0.001$). However, the distribution of the 113 student-assigned ranks in Rule 6-18 assignments is consistent with a uniform distribution at the level that we can detect. Among the Rule 1-5 and Rule 19-20 reviewers (both with and without PhDs), the consistency with a uniform distribution mirrors the behavior in Figure~\ref{fig:ranks_agg_3panel}. While the relatively large number (686) of Rule 6-18 assignments made to reviewers with PhDs may make this trend more detectable among that population, the student and PhD rank distributions are only marginally consistent with each other ($p=0.03$).

The second trend in ranks from Rule 6-18 assignments is seen with scientific category. Ranks cast by reviewers with similar knowledge (Rule 6-18) to Category 4 proposals show deviations from a uniform distribution with $p=0.002$, such that reviewers tend to give such proposals better ranks; it is the most significant detection among the categories and contains the most assignments of this rule group (404) among all of the scientific categories. Rule 6-18 reviewers of Category 1 proposals received 213 such assignments, and the significance of the trend in this rank distribution is more marginal (with $p=0.03$). The other three categories' rank distributions among Rule 6-18 assignments are even less numerous (and each is consistent with being uniform), so if this signal is present in these categories as well, we were not able to detect it. Given that only 5.3\% of review assignments fell into Rules 6-18, the impact on the final ranked list of these assignments is limited; specifically, in Category 1, the 213 assignments (6.2\% of the assignments in Category 1) were spread among 157 proposal sets, and in Category 4, the 404 assignments (14.8\% of the assignments in Category 4) were spread among 286 proposal sets. 

\subsubsection{Proposal sets that span multiple scientific categories}
\label{subsec:multcats}

\begin{figure}
\begin{centering}
\includegraphics[scale=0.6]{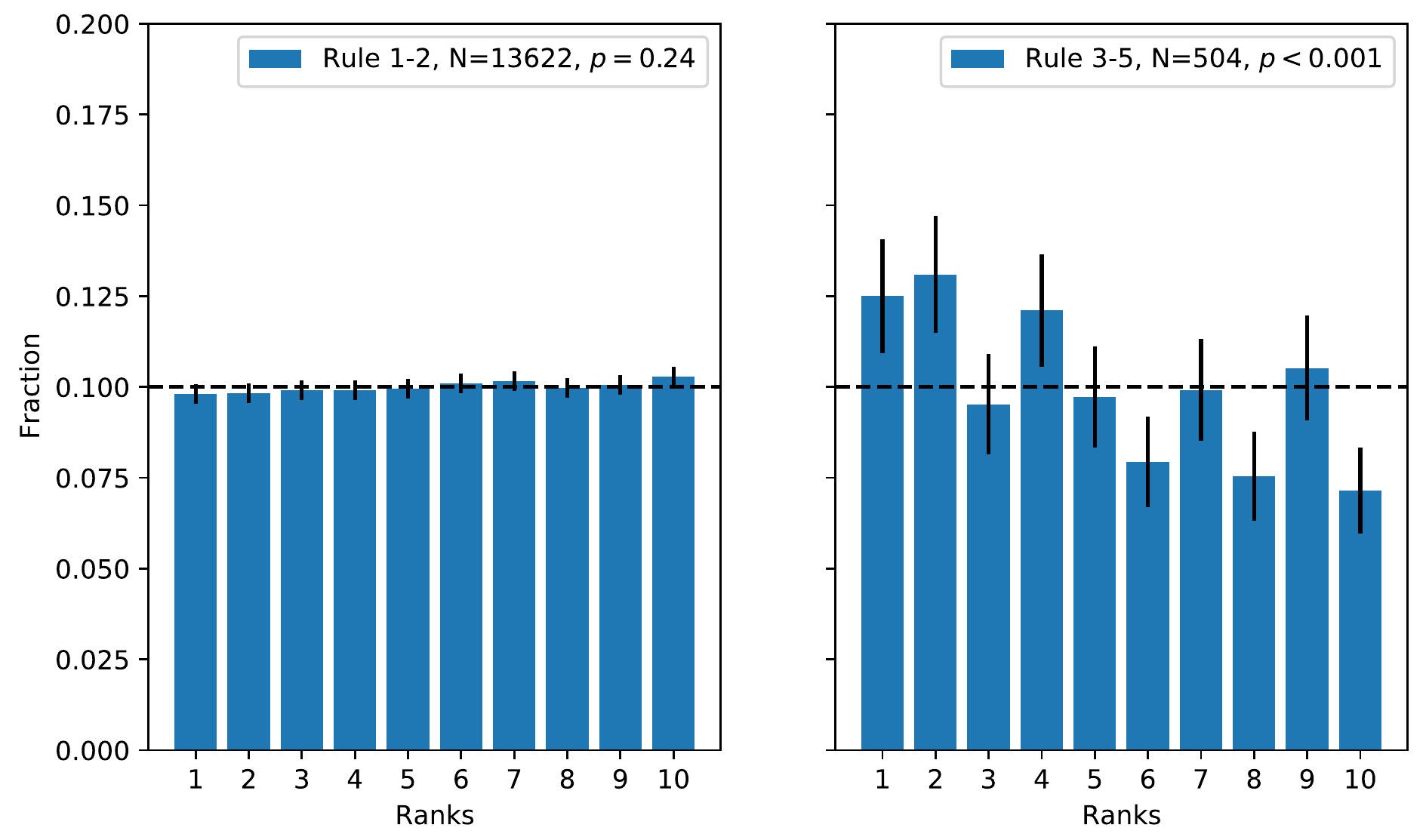}
\caption{Distribution of ranks with assignment rule, where the ranks from expert reviewers are split into Rule 1-2 (those who cast ranks in a single category, left panel) and Rule 3-5 (those who had to compare proposals across categories, right panel). The dotted line in each panel denotes the expectation of a uniform distribution among the ten rank options, and the $\it{p}$-value in each panel shows the result of the Anderson-Darling test. Error bars in each panel reflect confidence intervals ($\sqrt N$) given the observed numbers of counts.}
\label{fig:ranks_agg_4panel}
\end{centering}
\end{figure}

In this subsection, we describe the ranks given to the 8.9\% of assignments requiring reviewers to rank proposals in multiple categories relative to each other. The rules governing such assignments are Rules 3-5 (reviewers declaring expertise in the keyword of the proposal to be reviewed but outside of their JAO-defined ``reference category"), Rules 6-18 (reviewers with related expertise, discussed in the previous subsection), and Rules 19-20 (reviewers with unrelated expertise). The main result of this analysis is that reviewers with expertise who are required to rank proposals in more than one category against each other (i.e., those with Rule 3-5 assignments) behave similarly to those with similar expertise (i.e., those with Rule 6-18 assignments). The trend is apparent in Figure~\ref{fig:ranks_agg_4panel}, which splits the experts (the left panel of Figure~\ref{fig:ranks_agg_3panel}) into Rules 1-2 and Rules 3-5.

Similar to the previous subsection, this trend in ranks for Rule 3-5 assignments is detectable largely where the number of relevant samples is larger than a few hundred assignments. For instance, the trend is detected significantly (with a $\it{p}$-value of 0.002) among the ranks for the 454 Rule 3-5 assignments to reviewers with PhDs, but the ranks for the 50 such assignments to students are consistent with a uniform distribution (with a $\it{p}$-value of 0.22). With respect to scientific category, the trend is only significant ($p<0.001$) for the 222 Rule 3-5 assignments of Category 4 proposals. The other categories have very few Rule 3-5 assignments and/or the ranks are uniformly distributed. Here Category 1 has 206 such assignments but is consistent with a uniform distribution.

\begin{figure}
\includegraphics[width=\textwidth, trim=1in 0in 1in 0in, clip]{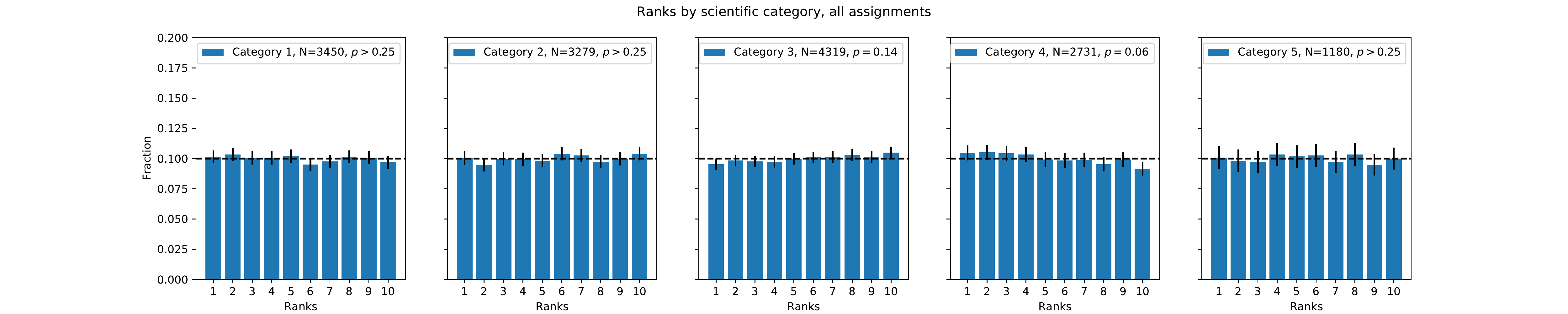}
\caption{Distribution of submitted ranks, broken down by scientific category of the receiving proposal. In Categories 3 and 4, slight deviations from uniform distributions in the ranks are detected but are not statistically significant. The dotted line in each panel denotes the expectation of a uniform distribution among the ten rank options, and the $\it{p}$-value in each panel shows the result of the Anderson-Darling test. Error bars in each panel reflect confidence intervals ($\sqrt N$) given the observed numbers of counts.}
\label{fig:totscicat}
\end{figure}

As part of this discussion with respect to scientific categories, it is relevant to recall Figures~\ref{fig:experts} and \ref{fig:rules_scicat}, which show the diversity of expertise submitted by reviewers and percentages of proposals in each category/keyword which receive expert and non-expert reviews. There are major overlaps in declared expertise between reviewers in Categories 1/2, Categories 3/4, and Categories 3/5, and proposals submitted to Categories 1, 4, and 5 received 6-15\% of Rule 6-20 reviews. Many proposal sets assigned to reviewers spanned more than one category (only 624 proposal sets consisted purely of Rule 1-2 assignments), so it is natural to look for links between these category pairs with respect to any detectable trends in the ranks. Given that reviewers with Category 3 expertise tended to also be experts in not only Category 4 but also Category 5, proposals submitted to Category 3 had the lowest percentage of ``perfect" (entirely Rule 1) proposal sets. The final effect on the trends with scientific category discussed in this subsection and the previous one can be best seen in Figure~\ref{fig:totscicat}, where the ranks submitted for all proposals are shown per scientific category. Even given the effects described here, only in Categories 3 and 4 are slight deviations from uniform distributions detectable in the ranks. However, these deviations are not statistically significant. 

In all cases, the trends explored in this section are suggestive and reported for completeness, but given that the number of assignments which fell into Rules 3-20 is less than 10\% of all assignments, the impact on the final rank-ordered list is not significant. However, these trends -- as well as the systematic trends examined in \citet{Carpenter22} -- will be monitored by the JAO in upcoming Cycles.

\subsection{Analysis of final Comments to the PI}
\label{subsec:comments}
Reviewer comments written to the PI represent their qualitative analysis of that proposal and, along with the rank, convey the merits of that proposal to the PI as well as to the JAO. High quality comments are critical for the PI to understand the context of the individual ranks that their proposal received and to enable adjustments to improve future proposals. Comment quality is difficult to quantify in an unbiased way, given that the same comment can be interpreted differently based on cultural and language variations between readers, but length is a quantitative measure that can be used for comparison. While the length of a comment does not necessarily translate directly to quality, we expect that a longer comment will typically contain more useful information than a very short comment. In this report, we analyze the comments written to PIs based on comment lengths as well as qualitative PI ratings for the helpfulness of individual comments (discussed in Section~\ref{sec:surveys}). 

\begin{figure}
\plotone{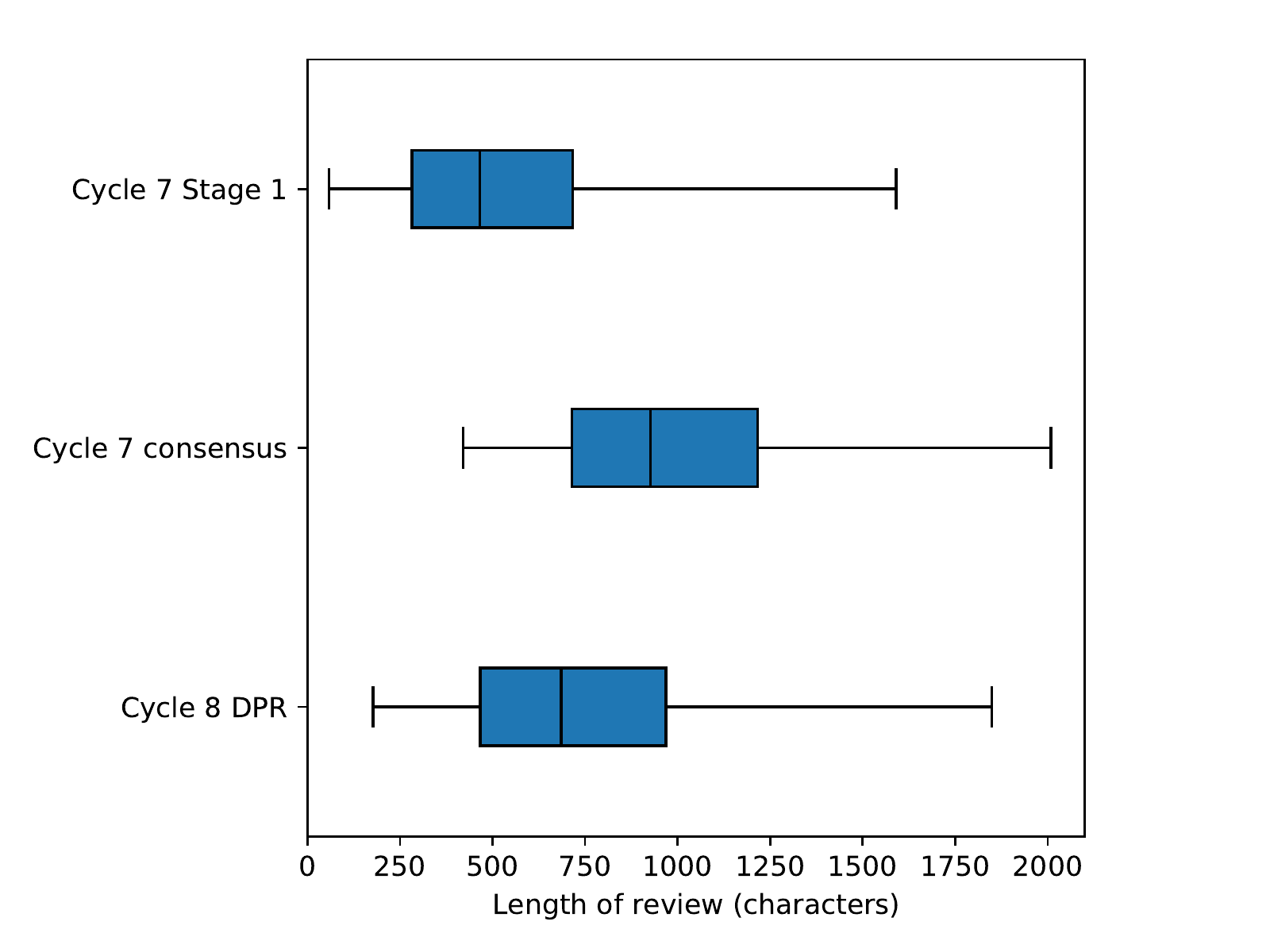}
\caption{Distribution of the length, in characters, of reviewers' comments to the PI for the Cycle 7 Stage 1 panel assessments (written by individual reviewers before the panel meetings), the Cycle 7 panel consensus reports (written post-discussion), and the Cycle 8 comments from distributed peer review. The box shows the 25-75 percentile distribution and the whiskers cover 95\% of the points. The median is shown by the vertical line within the box.}
\label{fig:comments}
\end{figure}

The Cycle 8 distributed reviewer comment lengths follow a lognormal distribution, resembling the behavior documented in at least one study of comment lengths in internet discussions \citep{Sobkowicz13}. The box plot in Figure~\ref{fig:comments} shows the range of comment lengths submitted in Cycle 8 in the bottom row. The median comment length is 685 characters, or roughly 5-6 sentences. For reference, this paragraph contains 680 characters. The Cycle 8 comment statistics are compared to the Cycle 7 panel review comments in the top two rows of the same Figure. The median length of the distributed reviewer comments in Cycle 8 falls between those of the two panel review distributions. 

\begin{figure}
\plotone{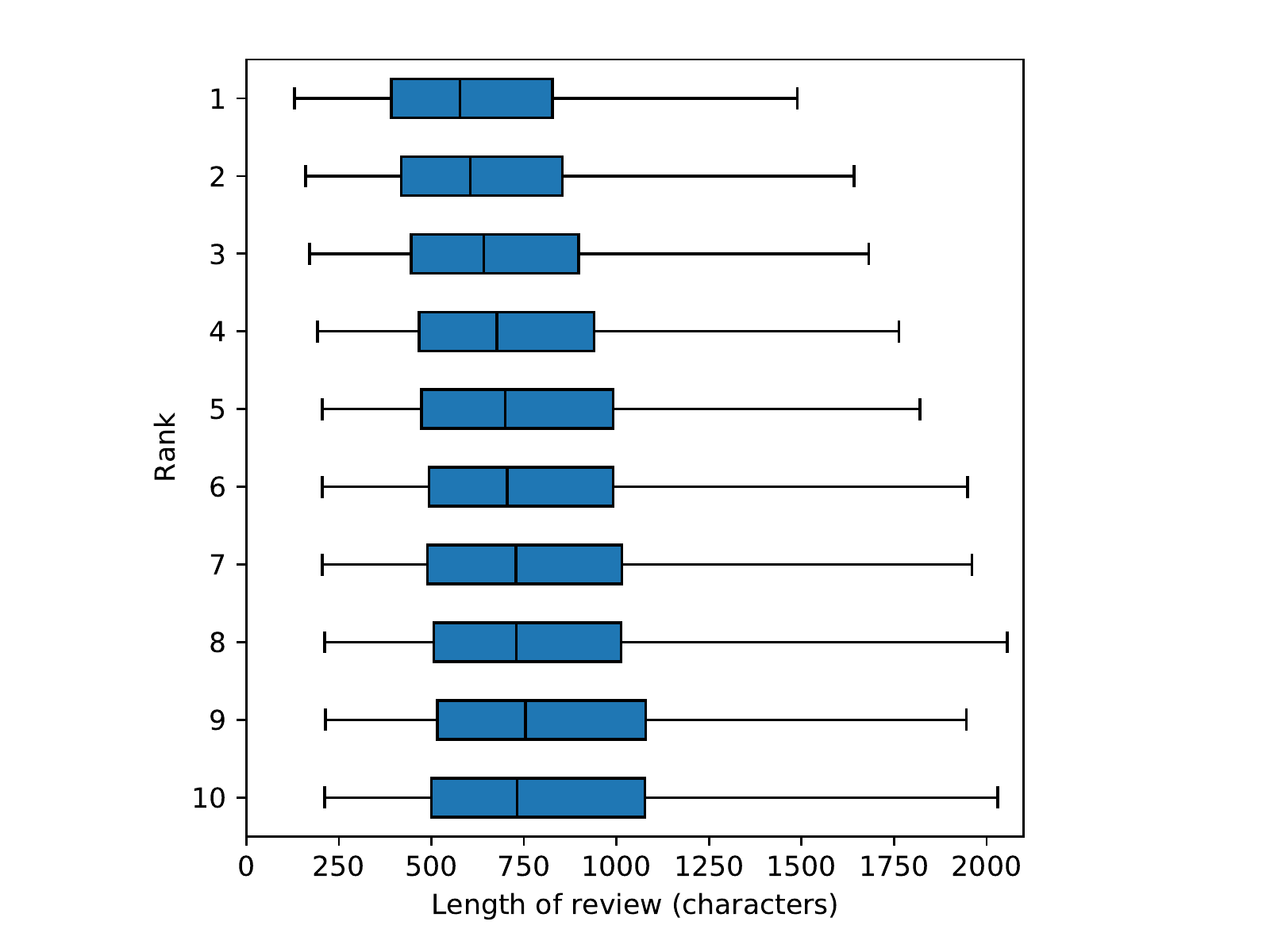}
\caption{Distribution of the length, in characters, of reviewers' comments to the PI as a function of rank given to that proposal. Reviewers generally wrote the shortest comments for the proposals they ranked best. The box shows the 25-75 percentile distribution and the whiskers cover 95\% of the points. The median is shown by the vertical line within the box.}
\label{fig:comments_ranks}
\end{figure}

\begin{figure}
\plotone{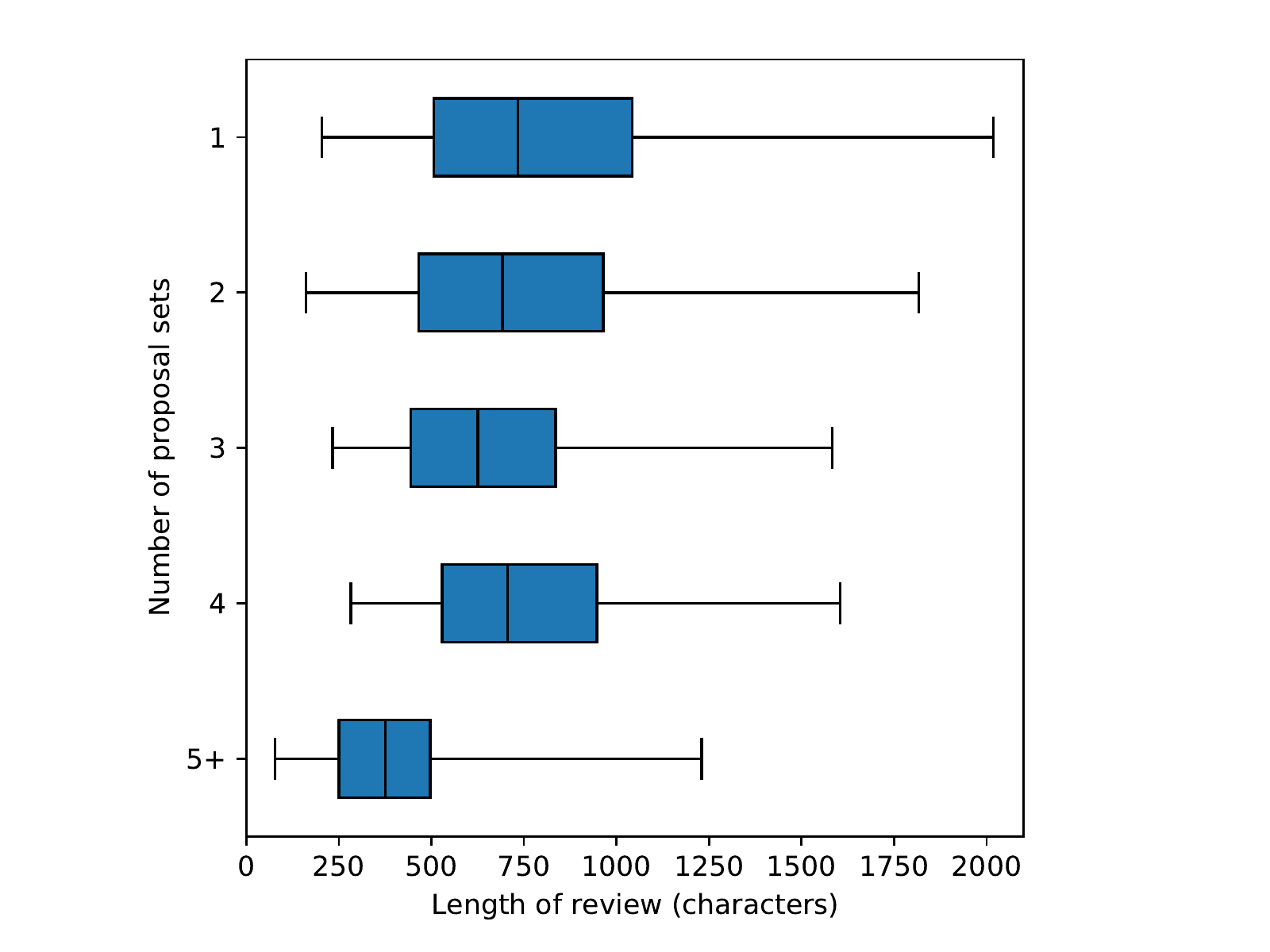}
\caption{Distribution of the length, in characters, of reviewers' comments to the PI as a function of reviewer assignment load. Reviewers with 5 or more proposal sets wrote significantly shorter comments than reviewers with 4 or fewer proposal sets. The box shows the 25-75 percentile distribution and the whiskers cover 95\% of the points. The median is shown by the vertical line within the box.}
\label{fig:comments_load}
\end{figure}

\begin{figure}
\plotone{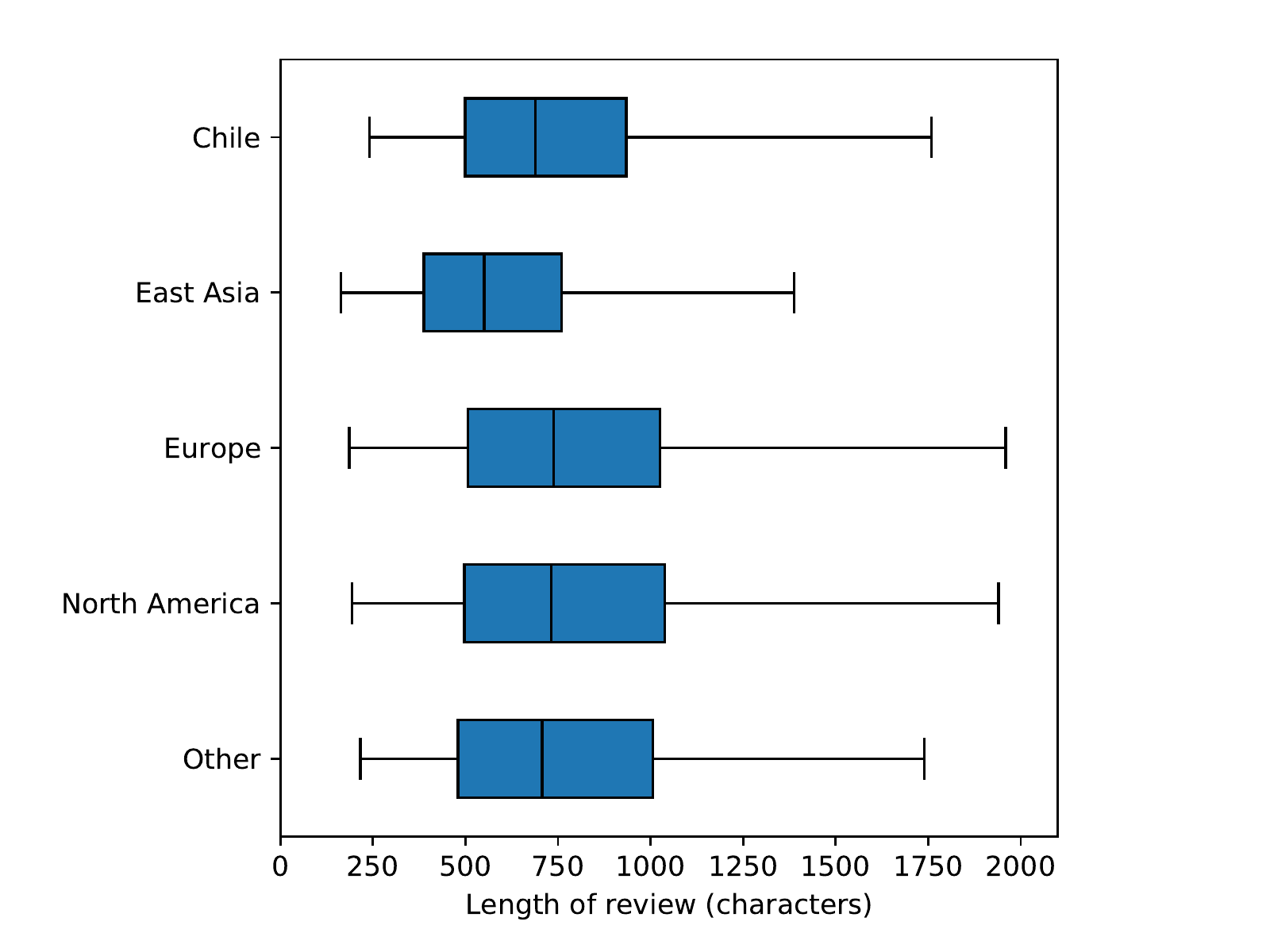}
\caption{Distribution of the length, in characters, of reviewers' comments to the PI split by reviewer region. The box shows the 25-75 percentile distribution and the whiskers cover 95\% of the points. The median is shown by the vertical line within the box.}
\label{fig:comments_region}
\end{figure}

Three global trends were detected in the distribution of comment lengths. First, as shown in Figure~\ref{fig:comments_ranks}, a trend in the comment length exists with the final rank given to a proposal by an individual reviewer. Broadly, reviewers wrote the shortest comments for their top-ranked proposals (median 577 characters for proposals with a rank of 1) and the longest comments for their lowest-ranked proposals (median 728 to 754 characters for proposals with ranks 7-10). Second, a trend in the comment length exists with the assignment load carried by the reviewer. Reviewers with 5 or more proposal sets wrote significantly shorter comments than reviewers with 4 or fewer proposal sets, as shown in Figure~\ref{fig:comments_load}. A third trend is that the length of the comments shows some regional dependencies as shown in  Figure~\ref{fig:comments_region}. East Asian reviewers tended to write shorter comments, with a median of 550 characters, compared to medians among the other regions of 689-738 characters. As described in Section~\ref{subsec:load}, East Asian reviewers tended to take high assignment loads more often than other regions, but the trend for shorter East Asian comments persists even when removing East Asian reviewers with 5 or more proposal sets; the median comment lengths for East Asian reviewers with 4 or fewer proposal sets was 574 characters, and the median for 5 or more proposal sets was 409 characters. 

\subsection{Comments to the JAO}
\begin{figure}
\plotone{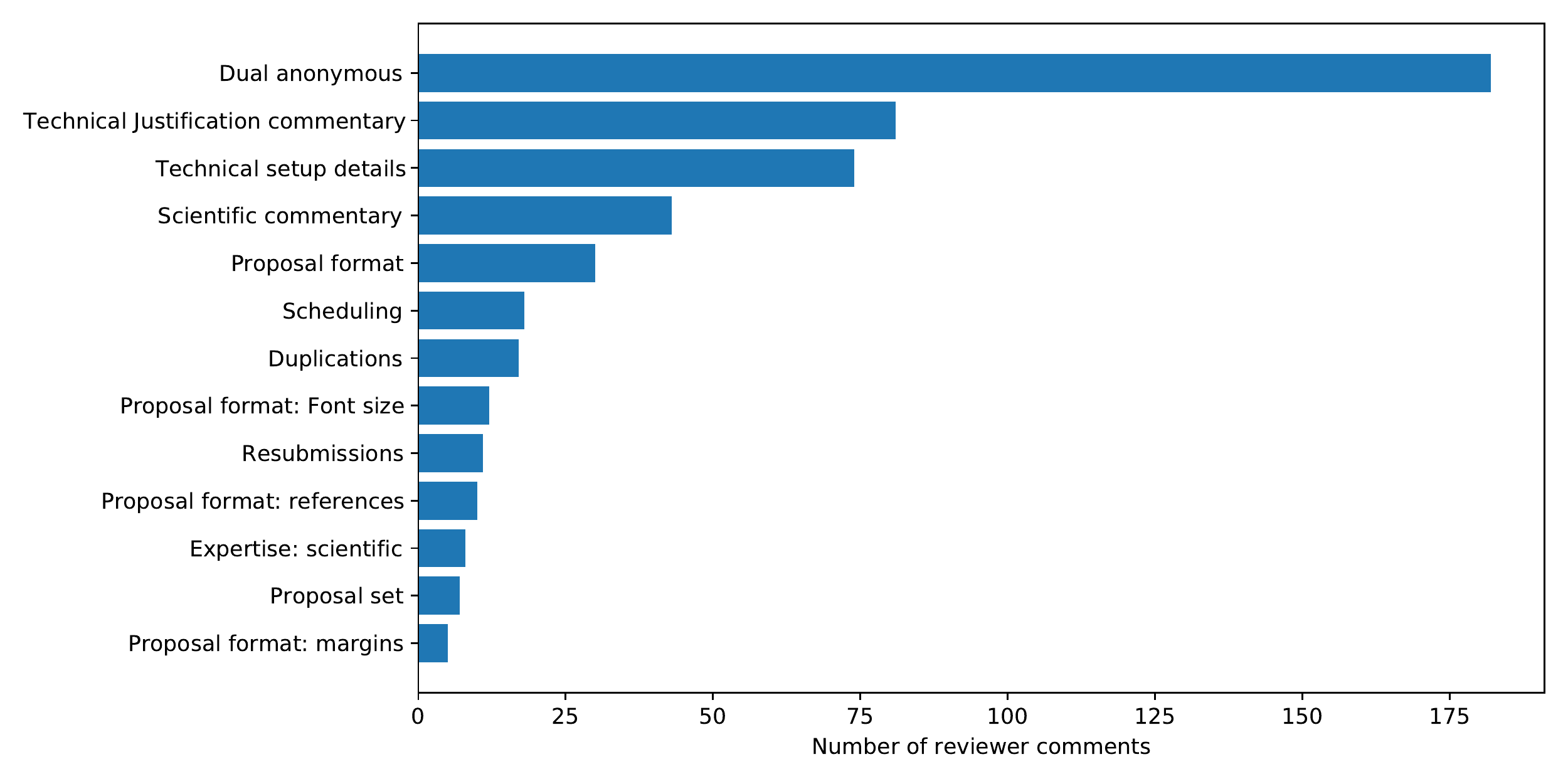}
\caption{Distribution of comments made to the JAO by the reviewers during the Stage 1 and Stage 2 distributed peer review process.}
\label{fig:comments_jao}
\end{figure}

When submitting ranks and comments, reviewers also had the opportunity to provide comments to the JAO on any assignment.  Figure~\ref{fig:comments_jao} shows the distribution of topics raised in the 627 comments sent by the reviewers. The three most common topics of the comments were suggested violations of the Cycle 8 dual anonymous practices and commentary on the Technical Justification and technical setup of the proposal. Several comments were received indicating that proposals did not follow the formatting guidelines.  Where appropriate, comments on technical issues were forwarded to the observatory for accepted proposals. The PHT examined each of the reported dual anonymous and format violations, and notified the PI where appropriate.

\subsection{Potential to reverse the ranks}
In the Cycle 7 Supplemental Call, at least two cases were identified where the reviewers reversed the rank scale by assigning rank 1 to the weakest proposal and 10 to the strongest proposal. Such cases were found by comparing the submitted ranks with the corresponding comments. In addition to producing errors in the overall rankings, such mistakes are disconcerting given that it is not practical to revise the outcomes of the review process if any ranks were entered incorrectly. For Cycle 8, considerable effort was placed in the re-design of the Reviewer Tool (see Section~\ref{subsec:tool}) to reduce the likelihood that such mistakes could be made. Four PIs did in fact contact the PHT to suggest that the ranks assigned to their proposals by some reviewers had been reversed (since the ranks and comments appeared to exhibit some inconsistency). The PHT examined each of these cases, and based on the comments the reviewers provided for all 10 proposals in their sets, the ranks appeared to be consistent with the comments.

To address more systematically if any ranks have been reversed, the PHT conducted a blind test to assess the order of the rankings based only on the comments. The test proceeded in two steps. In the first step, a proposal set was selected at random, and the reviewer comments for the strongest and weakest proposals were displayed in a random order. The objective was to guess which comment was associated with the rank 1 proposal and which was associated with the rank 10 proposal. If the guess was incorrect, the test proceeded to a second step where the comments for proposals ranked 2 through 9 were displayed in either ascending or descending order; again, the order was guessed based on the nature of the comments. Of the 400 proposal sets assessed in this manner, the correct order was discerned 96\% of the time in the first step. The correct order for the remaining 4\% of proposal sets was determined correctly in the second step. While the order was assessed correctly 100\% of the time, an element of luck was involved; in a small number of cases, all of the comments contained a balance of strengths and weaknesses for all proposals. Further, in one proposal set, the comments for proposals 1 to 9 aligned with the rankings, but the comment for the proposal ranked 10 seemed out of place. It is possible that the reviewer made an error in assigning this one proposal, but the overall rankings were nonetheless not reversed. Given that no instances of reversed rankings were identified in 400 proposal sets, we can place an upper limit of 1.3\% (99\% confidence) to the number of proposal sets where the ranks were reversed, or less than 20 proposal sets total.

%% file: ranks.tex
\section{Analysis of the overall rank-ordered list}
\label{sec:ranks}

The individual ranks submitted by reviewers were averaged and then sorted to determine the overall proposal ranks. Appendix~\ref{app:ranks} discusses in more detail the procedure and alternative approaches to determining the overall ranks. In this section, we examine the systematics in the rankings depending on the science category and popularity of the keywords, investigate the dispersion in the ranks, address the impact of outlier rejection, and compare the Cycle 8 results to those from the Cycle 7 panel reviews. Systematics in the rankings related to gender, regional affiliation of the PI and reviewer, and experience level of the PI are examined in \citet{Carpenter22}.

\subsection{Keywords and Categories}

The popularity of a proposal keyword (represented by the absolute sizes of the bars in Figure~\ref{fig:pairs}) could potentially lead to a bias in the final rankings if reviewers favor proposals from their own subfield. We used the proposal keywords selected by the PI to examine whether any systematics in the overall rankings are present with the overall popularity of the keyword.

\begin{figure}
\centering
\includegraphics[width=7in]{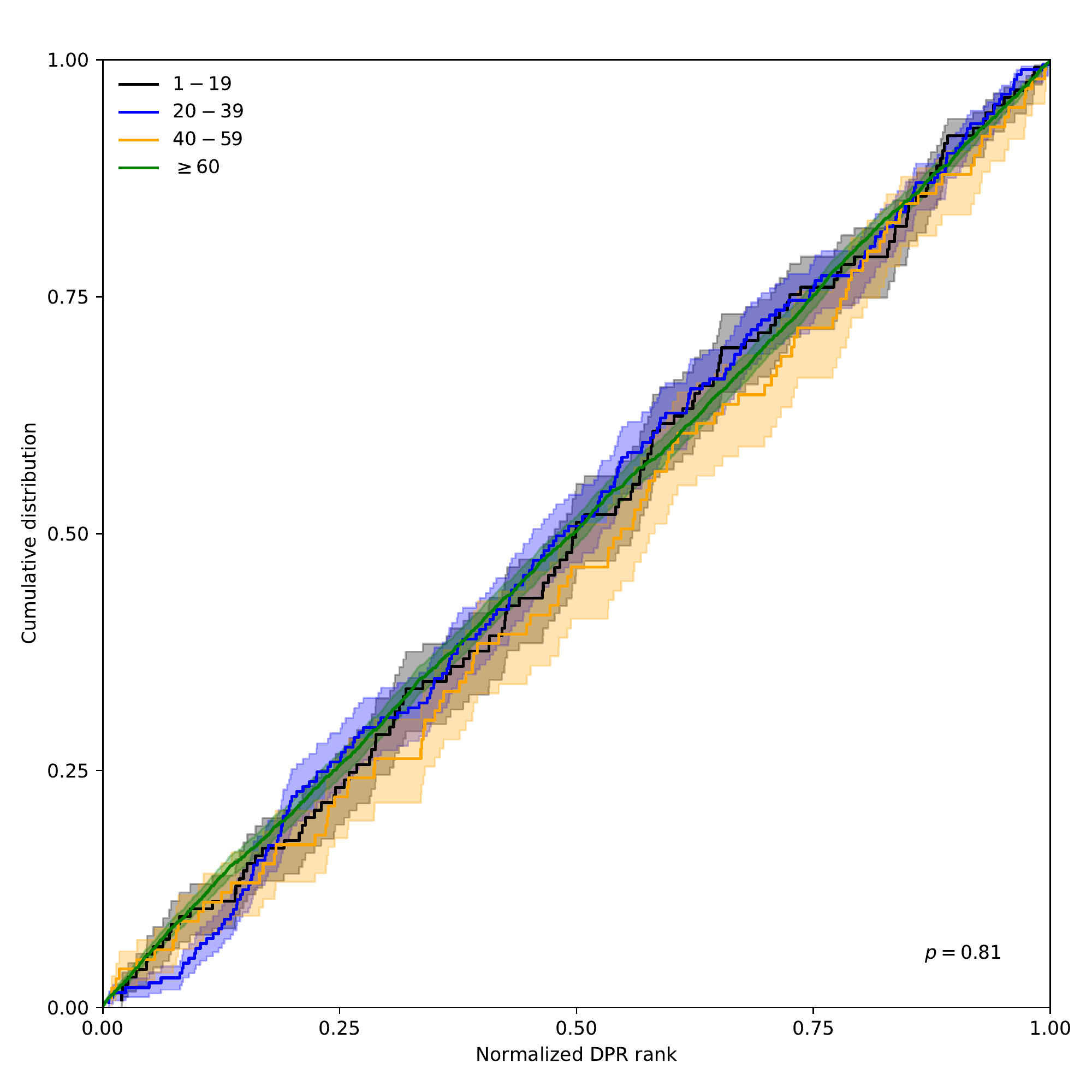}
\caption{Cumulative distribution of the overall proposal rankings as a function of the number of proposals submitted with a given keyword. The rankings have been normalized between 0 (best) and 1 (poorest). The shaded region indicates the 68.3\% confidence interval computed using the beta function. The probability ($p$), computed from the Anderson-Darling $k$-sample test \citep{Scholz19}, that the distributions within a cycle are drawn from the same population, and is indicated in the lower right corner.}
\label{fig:ad_keywords}
\end{figure}

Figure~\ref{fig:ad_keywords} shows the cumulative distribution of the overall proposal ranks, grouped by the number of proposals submitted with a given keyword. If a PI selected two keywords, then the more popular keyword was used for that proposal. The ranks shown in Figure~\ref{fig:ad_keywords} are normalized such that a rank of 0 is the best proposal rank and 1 is the poorest. Distributions shifted to the upper left therefore have better overall proposal ranks than distributions shifted to the lower right. The probability that the distributions are drawn from the same population was computed using the Anderson-Darling $k$-sample test and is indicated in the lower right corner of the plot. The results show that keywords that appear in less than 20 proposals (black curve) have similar distributions of ranks as keywords that appear in 60 or more proposals (green curve). While keywords that appear a total of 20-39 times (the blue curve in Figure~\ref{fig:ad_keywords}) are not as prevalent among the top-rated proposals, no meaningful differences ($p=0.81$) are found between the overall distributions. These results indicate that distributed peer review in Cycle 8 did not lead to significant systematics in the proposal rankings with the number of submitted proposals in a given keyword.

\begin{figure}
\centering
\includegraphics[width=7in]{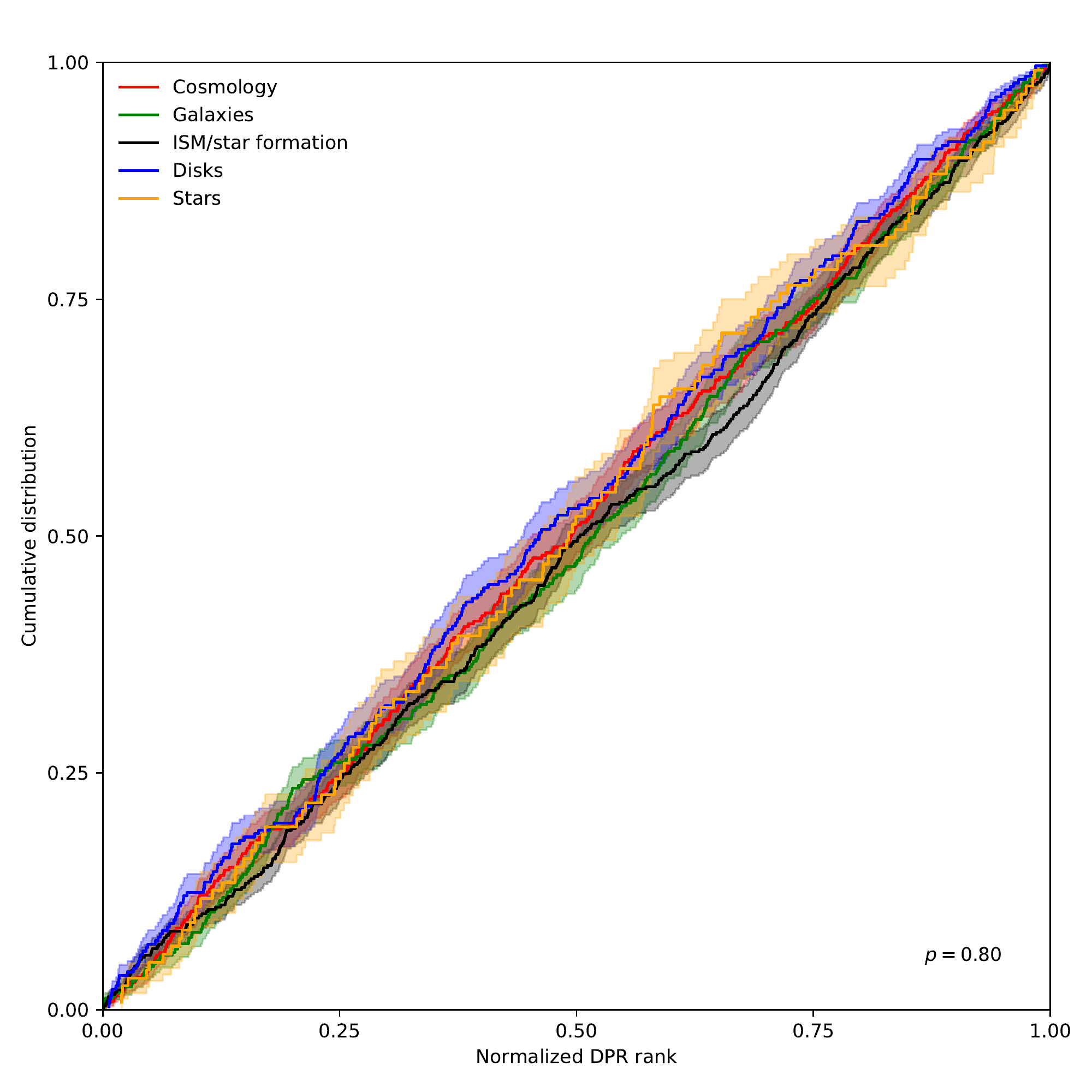}
\caption{Same as Figure~\ref{fig:ad_keywords}, but for the scientific categories.}
\label{fig:ad_categories}
\end{figure}

In the panel review process in previous cycles, the proposal ranks from the five science categories were normalized and combined to ensure that the overall rank distributions between the categories were the same. No explicit normalization by category was performed in the distributed peer review process. To check for any such systematics in the Cycle 8 proposal ranks, Figure~\ref{fig:ad_categories} shows the cumulative distribution of proposal ranks by scientific category. As with the keywords, the overall ranks have similar distributions for each proposal category. The $p$-value of 0.80 confirms that the distributions are indistinguishable. 

\subsection{Scatter in the ranks}
\label{subsec:scatter}

\begin{figure}
\centering
\includegraphics[width=\textwidth, trim=0 1.5in 0 1.5in, clip]{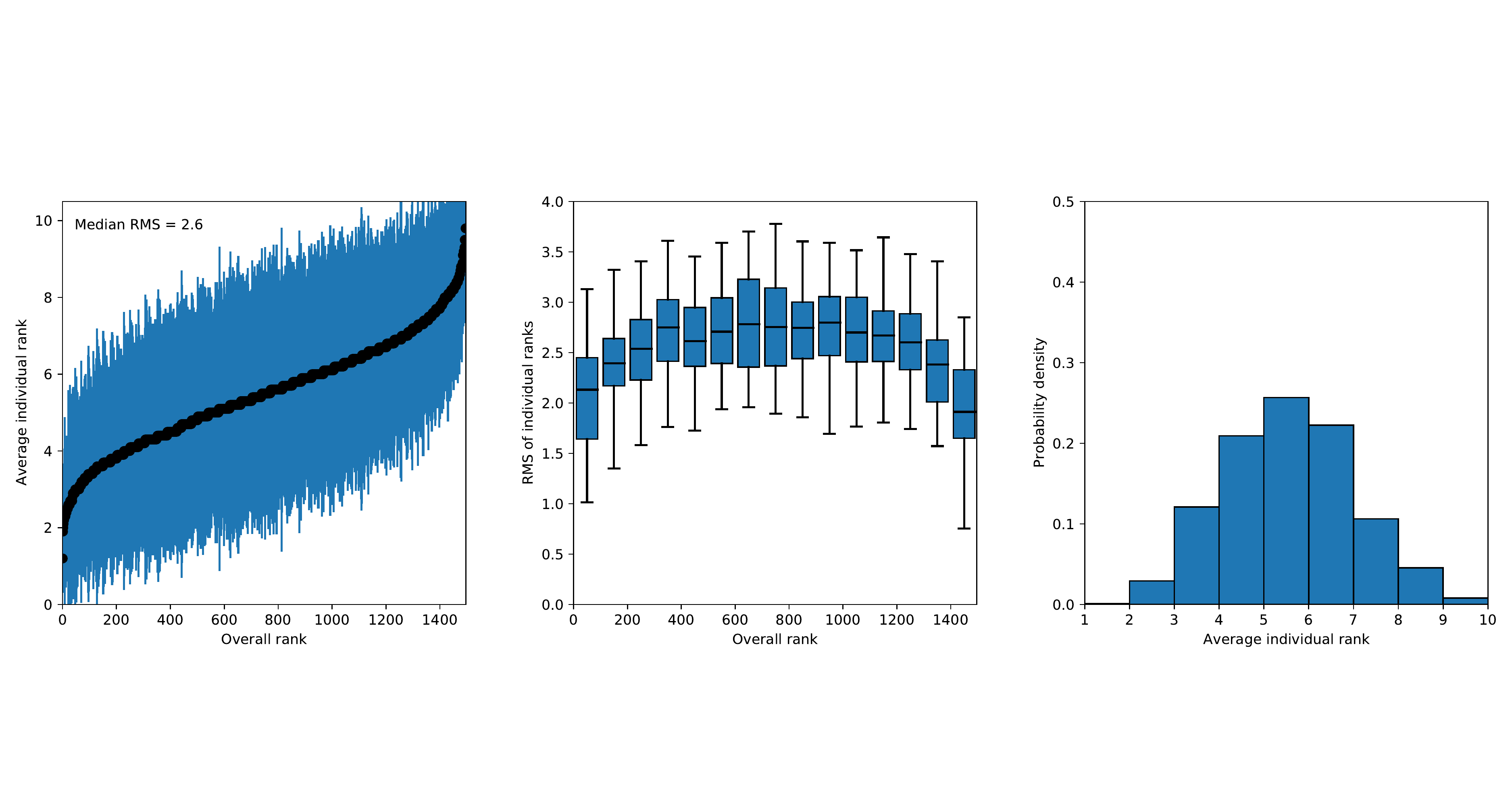}
\caption{(Left) The average rank (black points) from individual reviewers plotted as a function of the overall rank for proposals submitted to the Cycle 8 main call. The overall rank was determined by sorting the average individual ranks. The blue vertical lines for each point show the standard deviation of individual ranks given to each proposal. (Center) Boxplot showing the distribution of standard deviations as a function of the overall proposal rank. The box shows the 25-75 percentile distribution and the whiskers cover 95\% of the points. The median is shown by the horizontal line with the box. (Right) Histogram of the average individual ranks for the proposals.}
\label{fig:scatter_cycle8}
\end{figure}

The left panel of Figure~\ref{fig:scatter_cycle8} shows the mean and standard deviation of the individual proposal ranks in Cycle 8, shown as a function of the overall proposal rank. The median standard deviation of the ranks for all proposals is 2.6. As shown in the center panel of Figure~\ref{fig:scatter_cycle8}, the rms is largely independent of the overall rank of a proposal, with a slight decline toward the strongest and weakest ranked proposals. This result is qualitatively similar to the behavior extensively modeled and analyzed by \citet{Patat18} for a large sample of ESO panel reviews over 8 years; in that study, the agreement between individual referees and panels is shown to be significantly higher in the first and fourth quartiles than the second and third quartiles. While not shown, no significant difference in the median rms per overall rank bin is detected when comparing proposals reviewed with purely Rule 1-2 assignments and those reviewed with all assignment rules. The right panel of Figure~\ref{fig:scatter_cycle8} shows the distribution of the average individual ranks. The distribution is a Gaussian-like distribution with a mean value of 5.5.

\begin{figure}
\centering
\includegraphics[width=\textwidth, trim=0 1.5in 0 1.5in, clip]{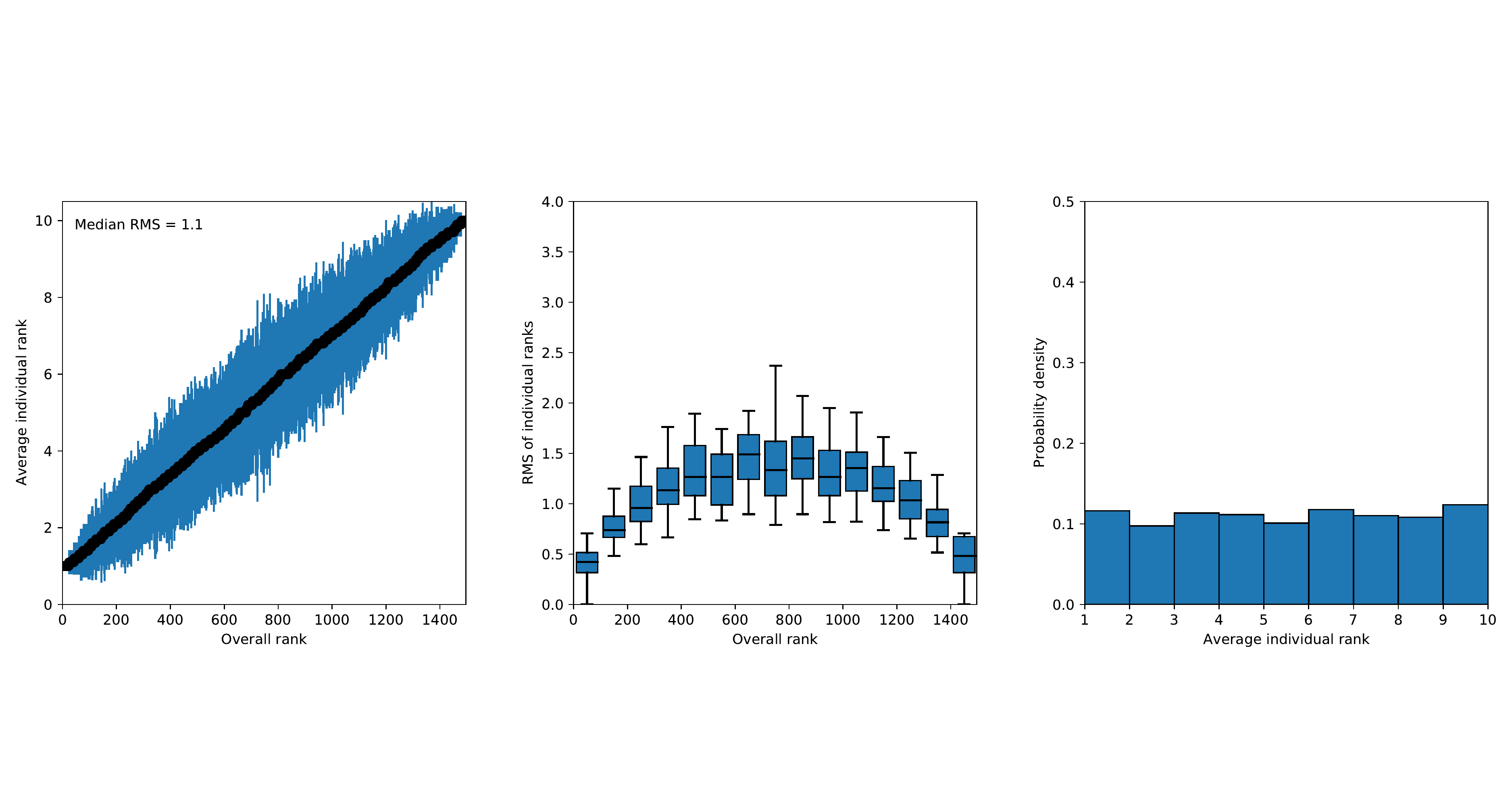}
\caption{Same as Figure~\ref{fig:scatter_cycle8}, but for a Monte Carlo simulation where all reviewers had the perfect ability to rank proposals by scientific merit.}
\label{fig:scatter_ideal}
\end{figure}

Part of the scatter in the ranks for a given proposal is caused by the finite number of proposals that are assigned to a reviewer and requiring reviewers to rank their proposals from 1 to 10. We used the Monte Carlo simulations in Appendix~\ref{subsec:ranks_sim} to assess this intrinsic scatter due to the finite size of the proposal sets. Figure~\ref{fig:scatter_ideal} shows one realization of the Monte Carlo simulation where reviewers can rank the proposals by their true scientific merit perfectly. The median rms in the proposals ranks for a given proposal is 1.1, but for intermediately ranked proposals, the median rms is about 1.5. Unlike the distribution of average ranks for the actual Cycle 8 proposals, the overall distribution of average ranks is uniform between 1 and 10. 

\begin{figure}
\centering
\includegraphics[width=\textwidth, trim=0 1.5in 0 1.5in, clip]{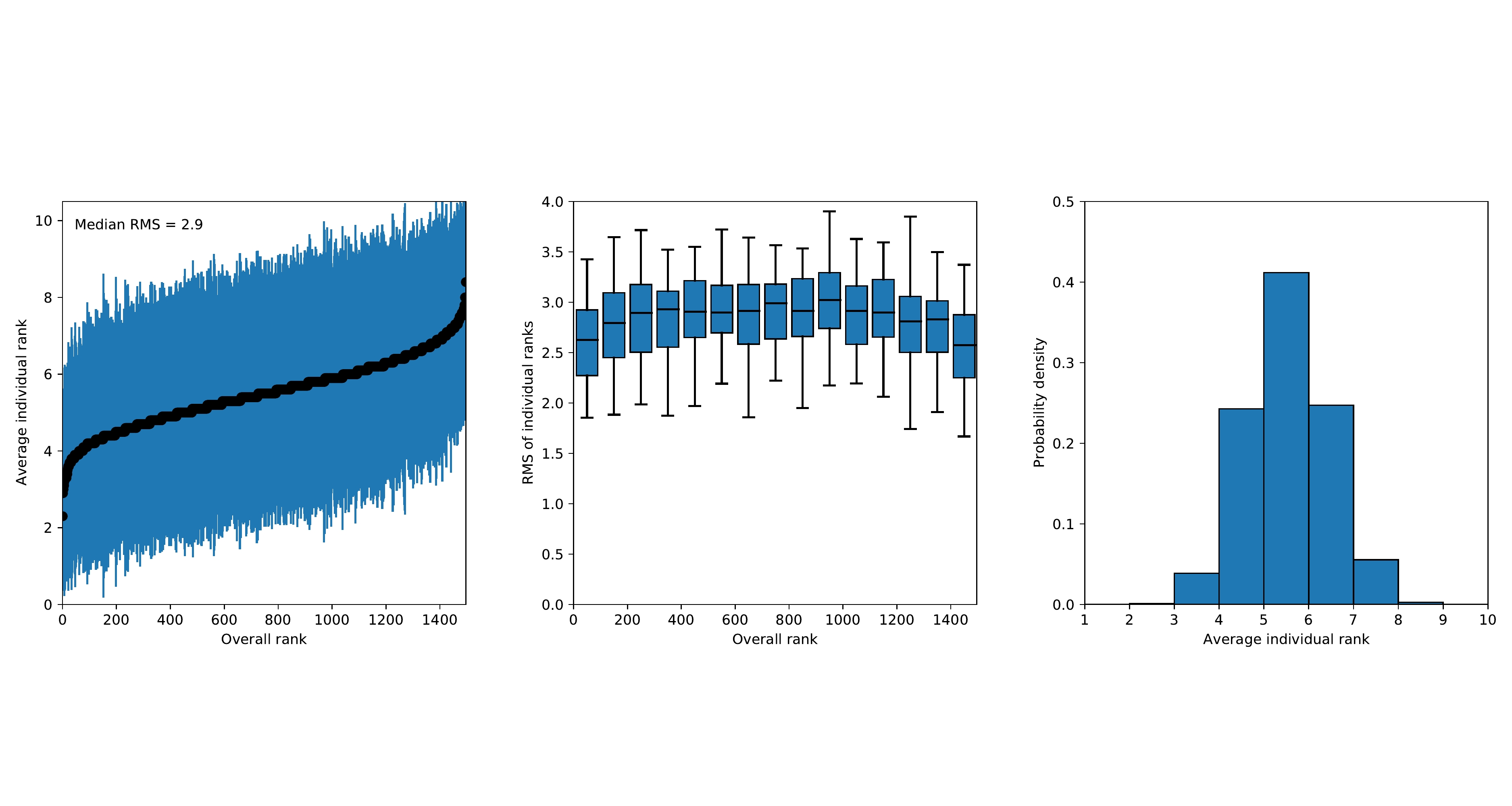}
\caption{Same as Figure~\ref{fig:scatter_cycle8}, but for a Monte Carlo simulation where there is no correlation in the ranks between reviewers.}
\label{fig:scatter_random}
\end{figure}

The opposite extreme to assuming reviewers have perfect ability to rank the proposals by scientific merit is to assume the ranks are random. Figure~\ref{fig:scatter_random} shows the expected distribution of proposals ranks in this limit. Compared to the actual Cycle 8 ranks, the median standard deviation is higher (2.9) with less variation with overall proposal rank. The probability distribution function is also Gaussian in shape, but the distribution is more sharply peaked than the distribution from Cycle 8.

Comparison between the Cycle 8 ranks and the simulations indicates that the dispersion in the individual ranks for a given proposal is mainly due to differences in the opinion of the reviewers on the scientific merit of the proposal (similar to the conclusion of \citealt{Patat18}) and not in reviewing a subset of proposals. Nonetheless, the ranks are not completely random. The median standard deviation of individual ranks for actual Cycle 8 proposals is smaller than that for a random distribution of individual ranks, indicating that reviewers have reached some consensus on the strongest and weakest proposals. 

\subsection{Impact of outlier rejection}

With 10 ranks per proposal, it is possible to identify individual ranks that differ greatly from the consensus. Most likely this will be a result of a non-consensus opinion of an individual reviewer, but it could also occur if a reviewer receives an unusually strong or weak proposal set. These outliers are mitigated in that an individual rank contributes only 10\% toward the final rank, since each proposal nominally receives 10 ranks. Nonetheless, allowing for outlier rejection may improve the accuracy of the overall rank of a proposal.

We investigated two possible outlier rejection algorithms. In the first approach, two ranks were removed before computing the average: the lowest rank received and the highest rank. The advantage of this approach is that it is both straightforward to compute and explain. However, it automatically removes two ranks even if they are not true outliers, and therefore the accuracy of the true rank will decrease by $\sim$12\% since only 8 ranks are used in the average instead of 10. The second approach is to identify outliers using sigma clipping. The rms ($\sigma$) of the individual ranks are computed, and any individual ranks that deviate more than $m\sigma$ from the median are rejected. The average and rms are then recomputed, and the computation is repeated until no more points are rejected or until a maximum number of allowed iterations is reached. The advantage of this approach is that outliers are removed only if they differ significantly from the other scores. However, the value of $m$ and the maximum number of iterations is subjective.  

\begin{figure}
\centering
\plotone{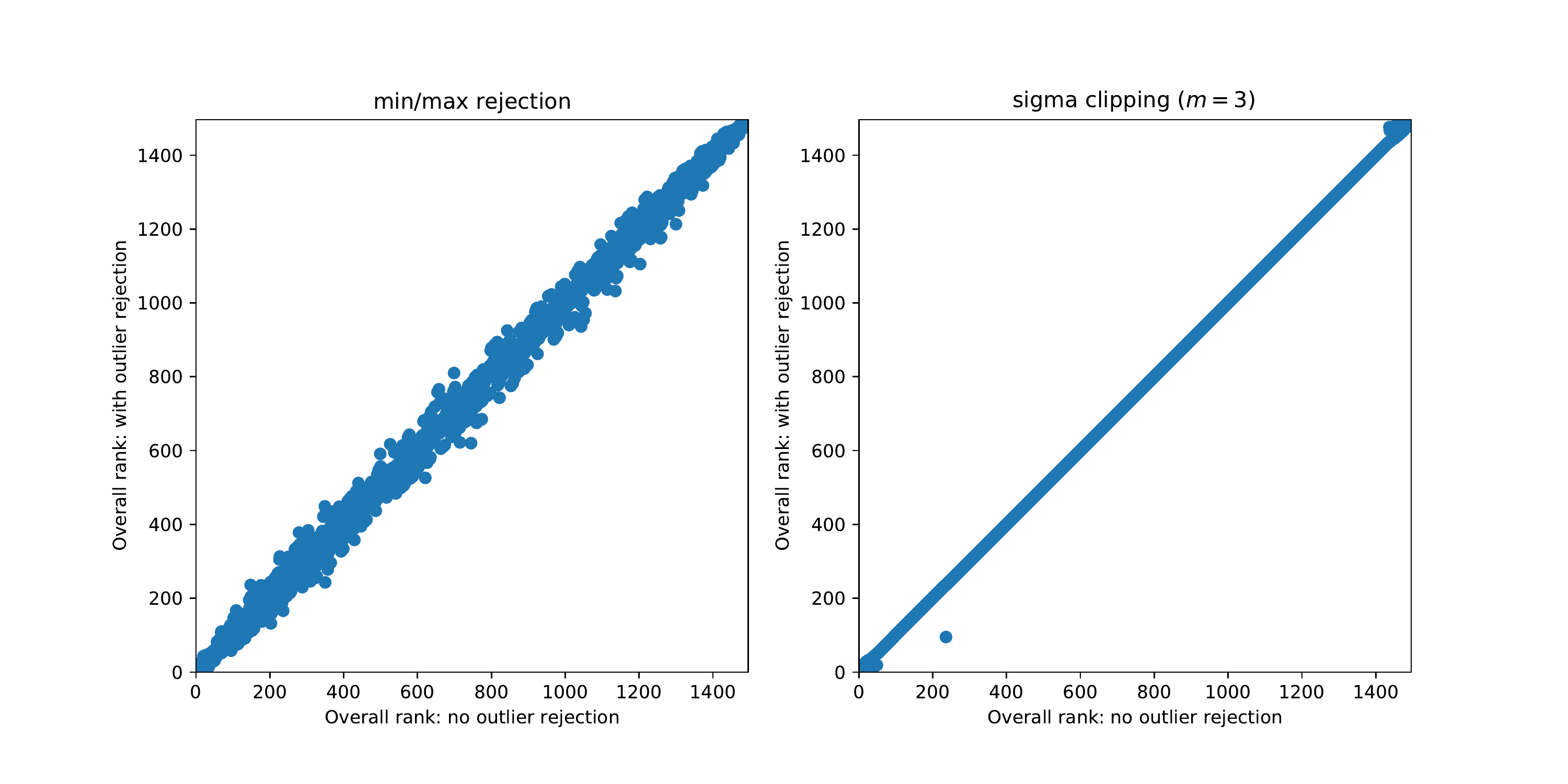}
\caption{Comparison of the overall proposals rankings from Cycle 8 distributed peer review with (ordinate) and without (abscissa) outlier rejection. The overall proposal rankings were determined from the average of the individual review ranks. The left panel shows the results by removing the strongest and weakest ranks, and the right panel shows the results with sigma clipping with $m=3$.}
\label{fig:cycle8_outliers}
\end{figure}

Figure~\ref{fig:cycle8_outliers} compares the overall ranks with and without outlier rejection for both min/max rejection (left panel) and a single iteration of sigma clipping with $m=3$ (right panel). A strong correlation is seen in the overall ranks with and without outlier rejection. After removing a best-fit linear relationship between the ranks with and without outlier rejection, the RMS of the residuals is 32 for min/max rejection. Given there are 1497 proposals, min/max outlier rejection has a relatively minor impact on the overall rankings. Nonetheless, eight proposals changed ranks by more than 100 places in the overall ranked list. The impact of sigma clipping with $m=3$ was less than that for min/max rejection, as only one proposal changed ranks by more than 100 places. In the end, no outlier rejection was performed in Cycle 8 given the overall small impact on the rankings and that the dispersion in the ranks appears to be generally dominated by a broad range of reviewer opinions rather than distinct outliers. However, outlier rejection is still under consideration for future cycles in order to mitigate the small number of cases where a single review drives the high dispersion.

\subsection{Comparison to panel reviews}

The significant scatter in the ranks from individual reviewers in distributed peer review raises the question if similar scatter is seen in the panel reviews. Given that reviewers are invited to participate in the panel based on their expertise and read all proposals in their panel, the scatter in the individual ranks could conceivably be reduced since the panel members have a broad perspective on the overall science that has been proposed (see \citealt{Patat18}, and references therein, for a thorough quantitative characterization of this inherently subjective process).

The results from the ALMA panels cannot be directly compared to distributed peer review since, in the panels, proposals are scored on a scale of 1 (strongest) to 10 (weakest) rather than being ranked. To allow for a comparison between the two systems, a simulation was run in which 10 proposals were selected at random from actual previous ALMA panel reviewers, and their ranks were set by sorting the proposals by the assigned scores. The simulation was run 100 times for each of the 158 reviewers that reviewed a total of 1759 proposals (excluding Large Programs) in Cycle 7. The assigned ranks were then averaged to produce the results as with distributed peer review. Of the 25 panels in Cycle 7, 21 panels contained 6 panelists and four panels with a broad range of proposals contained 8 panelists to have additional expertise. 

\begin{figure}
\centering
\includegraphics[width=\textwidth, trim=0 1.5in 0 2in, clip]{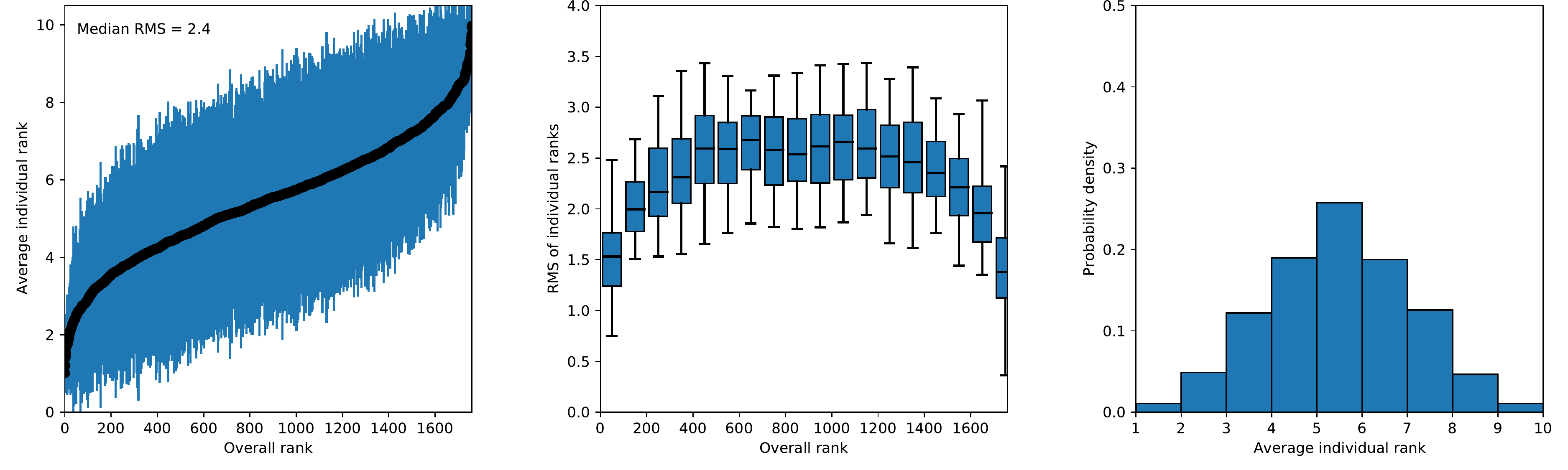}
\caption{Same as Figure~\ref{fig:scatter_cycle8}, but for the Cycle 7 panel reviewers, where the Stage 1 scores assigned to the proposals were converted into ranks (see text).}
\label{fig:scatter_cycle7_stage1}
\end{figure}

Figure~\ref{fig:scatter_cycle7_stage1} shows the distribution of ranks and scatter inferred from the Cycle 7 Stage 1 scores. As with distributed peer review, the range of ranks inferred from the Cycle 7 scores is considerable. The median rms of the individual ranks is 2.4 versus the 2.6 for Cycle 8 distributed peer review, and is considerably higher than the scenario in which all reviewers can rank proposals perfectly by the ``true" scientific merit (see Figure~\ref{fig:scatter_ideal}). Thus panel reviewers have similar scatter compared to reviewers in distributed peer review in how they prioritize the proposals.

\begin{figure}
\centering
\includegraphics[width=\textwidth, trim=0 1.5in 0 2in, clip]{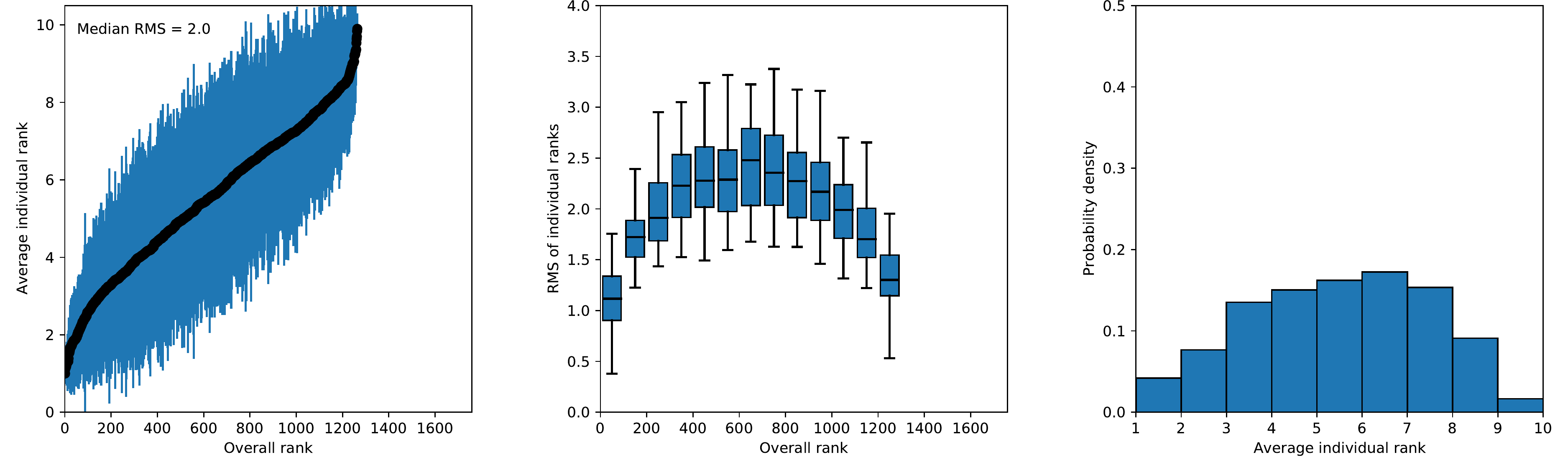}
\caption{Same as Figure~\ref{fig:scatter_cycle8}, but for the Cycle 7 panel reviewers, where the Stage 2 scores assigned to the proposals were converted into ranks (see text).}
\label{fig:scatter_cycle7_stage2}
\end{figure}

We also analyzed the Cycle 7 results after Stage 2, when the panel members meet to discuss and re-score the proposals. For Cycle 7, 1264 of the 1759 proposals were discussed in Stage 2, where the remaining proposals were ``triaged" and not further discussed because of their poor scores in the Stage 1 process. In addition to the discussion in Stage 2, reviewers can see the Stage 1 scores from the other reviewers. Therefore, the Stage 2 scores are no longer independent as they are in Stage 1. Figure~\ref{fig:scatter_cycle7_stage2} shows the results of the simulation on the Stage 2 scores. The overall rms declines to 2.0 due to reduced scatter for the the best and weakest proposals, but the RMS in the intermediate proposals remains comparable to distributed peer review even after the panel discussions.

%% file: surveys.tex
\section{Reviewer and PI surveys}
\label{sec:surveys}

ALMA conducted a survey of the reviewers after they completed the Stage 1 review process and another survey of the PIs after they received the results of the review process. This section analyzes the results from these surveys.

\subsection{Reviewer survey}
\label{subsec:survey_rev}

After reviewers submitted the ranks for a proposal set, they were invited to complete a survey and to provide feedback on the distributed peer review process. The reviewers were asked to provide their career status, to rate their expertise on each proposal they reviewed, and to provide comments on any aspect of the reviewer process. Of the 1016 unique reviewers, 942 (93\% of all reviewers) on behalf of 1354 proposal sets (90\% of all proposal sets) responded to the survey, and 195 reviewers provided free-form comments.

\subsubsection{Self-assessment of expertise by reviewers}
\begin{figure}
\centering
\includegraphics[width=4in]{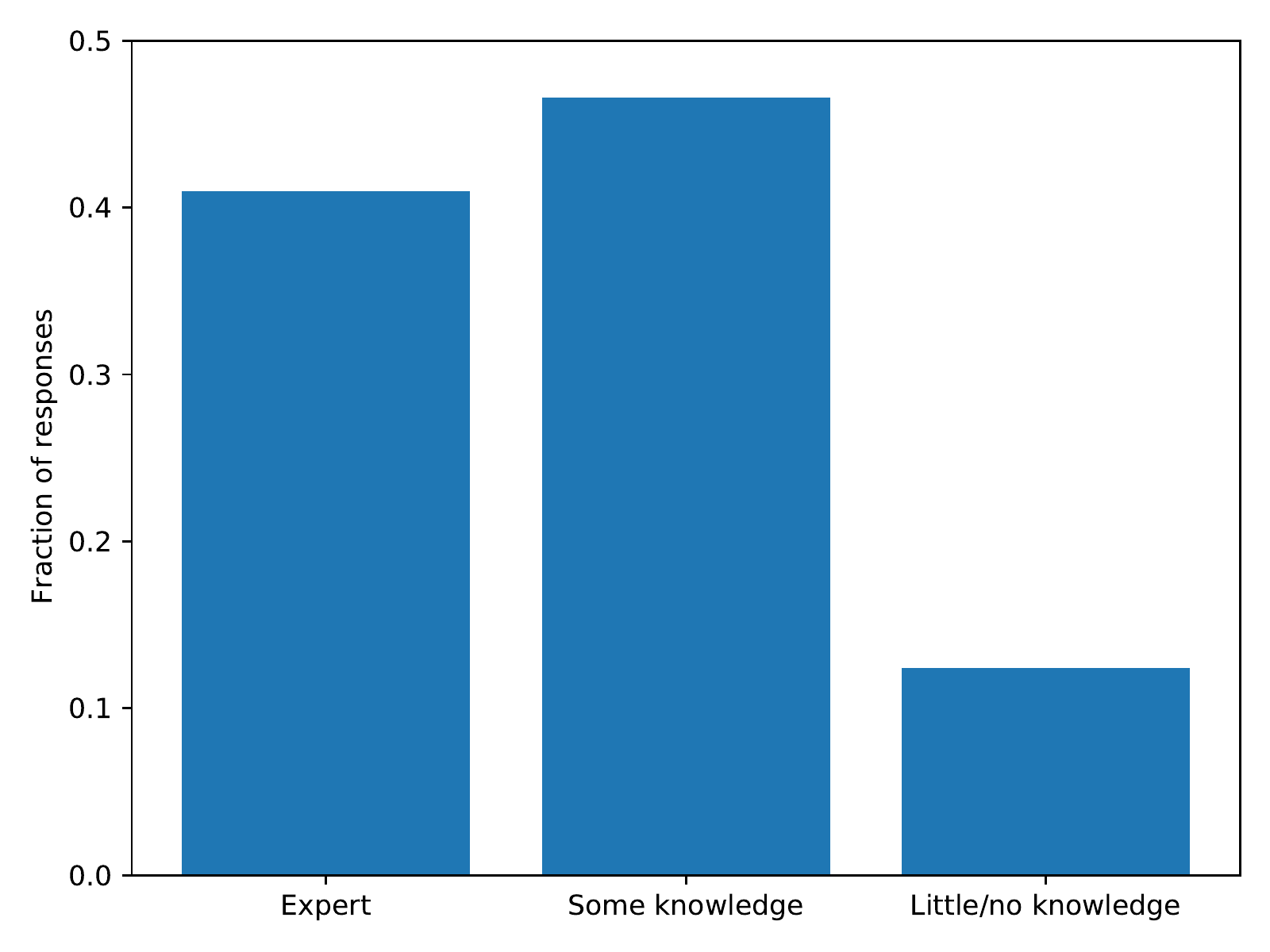}
\includegraphics[width=7in]{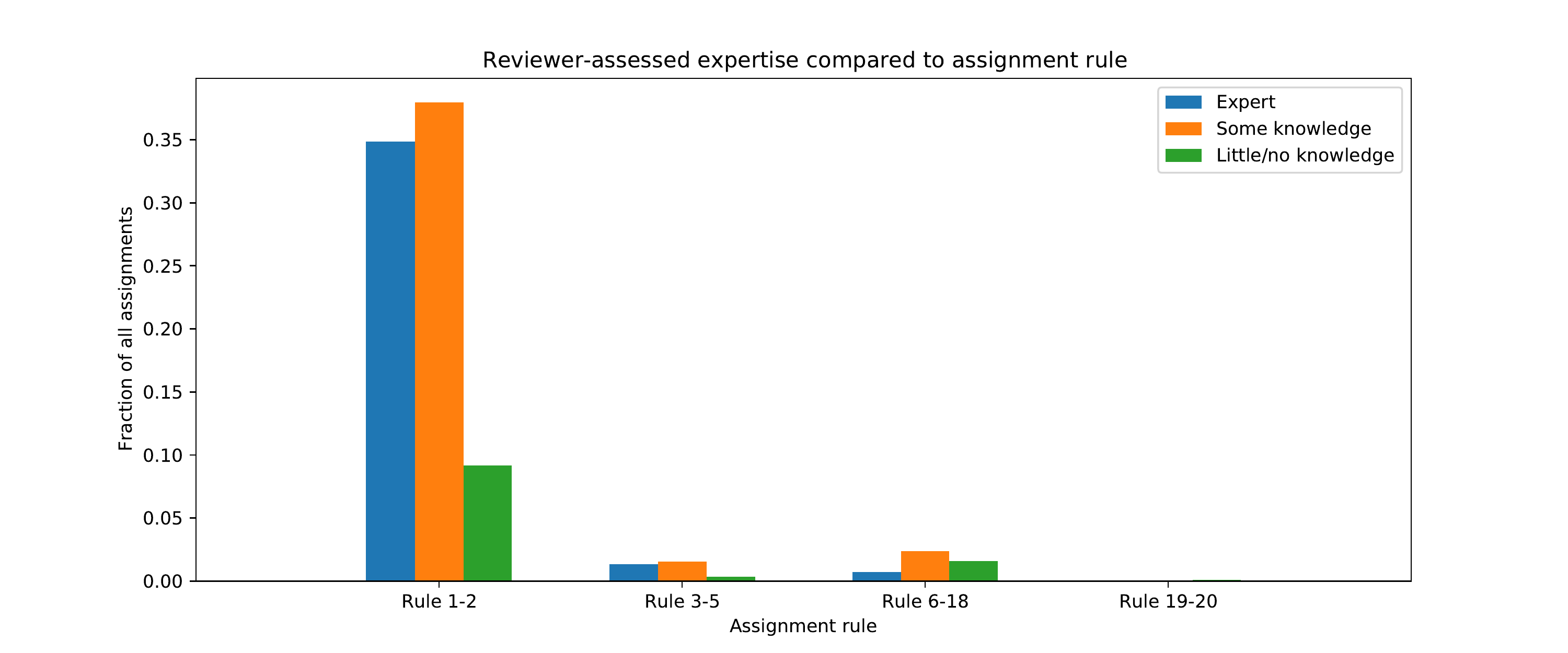}
\caption{(Top) Histogram of the reviewer's self-assessment of their expertise on individual proposal assignments. (Bottom) Fraction of all assignments per assignment rule group split by reviewer self-assessment.}
\label{fig:rev_expert}
\end{figure}

Reviewers provided a self-assessment of their expertise for each proposal into one of three levels: (1) This is my field of expertise; (2) I have some general knowledge of this field; and (3) I have little or no knowledge of the field. In a traditional panel review, selected panelists are expected to have expertise on all proposals in their panel at least at the level of ``some general knowledge", if not ``my field". The top panel of Figure~\ref{fig:rev_expert} shows a histogram of the reviewer responses, which shows that reviewers self-assessed themselves as experts or as having some knowledge on 88\% of the proposal assignments. This is comparable to the 94\% of the review assignments that were matched with the keywords specified by the  reviewers, indicating that the assignment algorithm made reasonable matches to the reviewer expertise as judged by the reviewers themselves. 

\begin{figure}
\centering
\includegraphics[width=\textwidth, trim=0in 0.5in 0in 0.5in, clip]{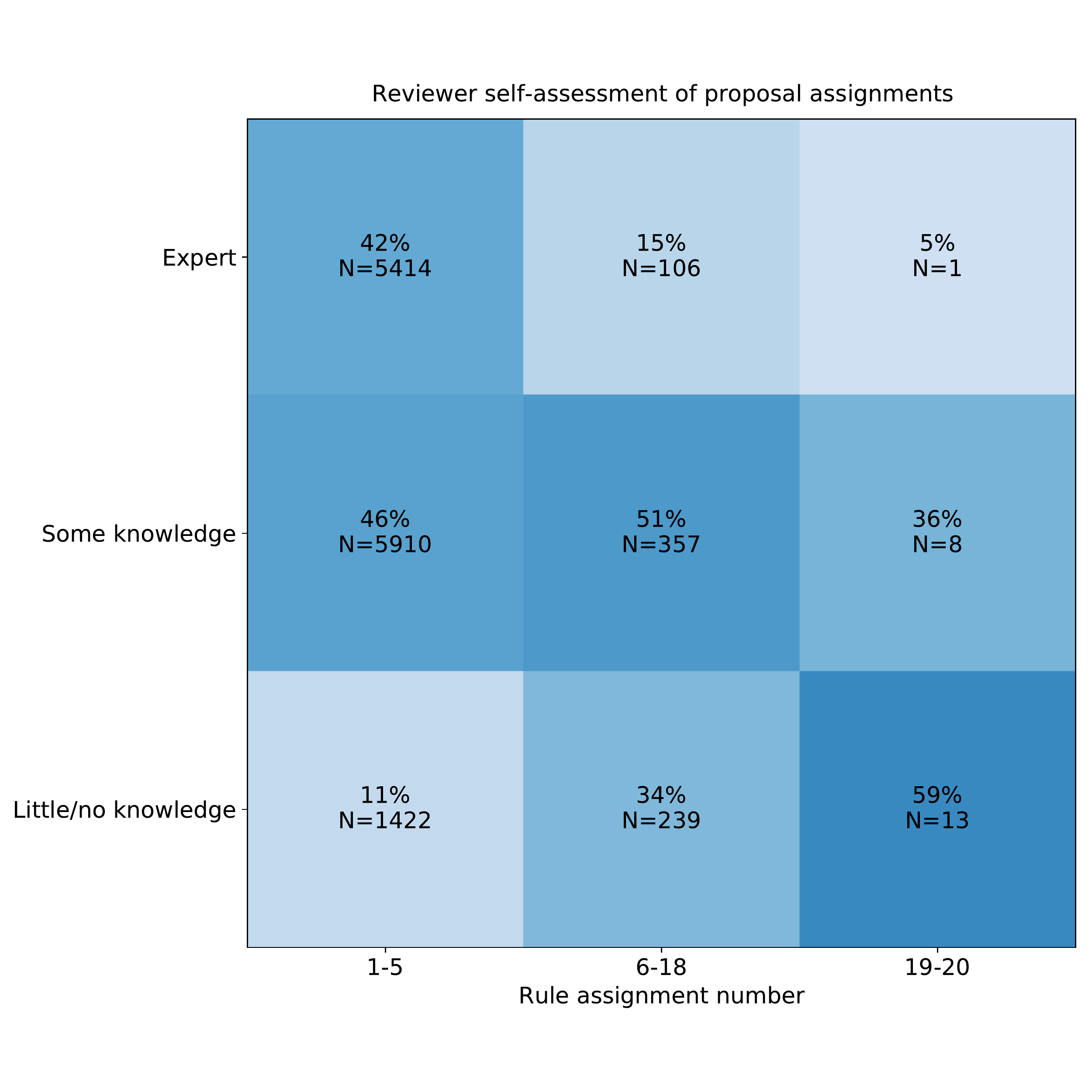}
\caption{Distribution of the reviewer self-assessment of their expertise on individual assigned proposals across assignment rule group. Rules 1-5 represent exact matches between reviewer expertise keywords and the assigned proposal keyword, Rules 6-18 represent assignments similar to the reviewer's expertise, and Rules 19-20 indicate no correspondence between the reviewer's expertise and assigned proposal. The responses are normalized separately for each rule assignment number.}
\label{fig:rev_expert_rules}
\end{figure}

To assess the efficacy of the proposal assignment rules, the bottom panel of Figure~\ref{fig:rev_expert} and Figure~\ref{fig:rev_expert_rules} show the distribution of reviewer expertise by the groups of rule numbers used to assign proposals (see Section~\ref{sec:assign}). As seen in Figure~\ref{fig:rev_expert_rules}, as the assignment rule increased, a greater percentage of reviewers declared they had little or no expertise, which is to be expected. While more than half of the reviewers with Rule 19 assignments said they had little or no knowledge in the proposal, only 24 of the 14970 proposal assignments (0.2\%) were assigned under Rule 19. Most proposals where the reviewer deemed that they little or no knowledge were assigned using the keywords specified by the reviewer. This suggests that the keywords used by ALMA are too broad or some reviewers are very selective in defining their expertise when it comes to individual proposals.

\begin{figure}
\centering
\includegraphics[width=\textwidth, trim=0 1in 0 1in, clip]{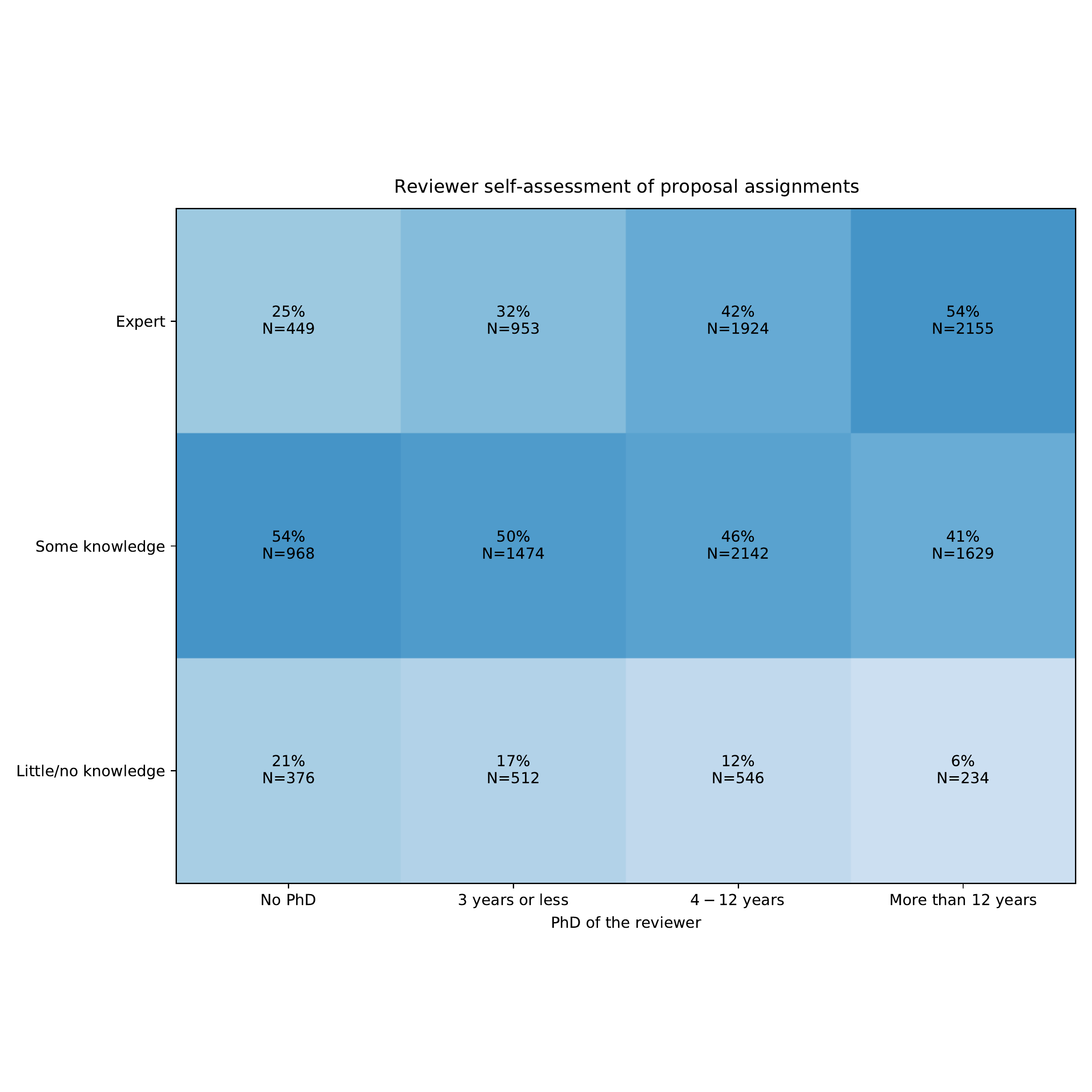}
\caption{Distribution of the reviewer self-assessment of their expertise on individual assigned proposals across career levels. The responses are normalized separately for each career level.}
\label{fig:rev_expert_career}
\end{figure}

The reviewer expertise on individual proposals was also cross-matched with the career status of the reviewer, which is grouped into no PhD, PhD within 3 years or less, PhD within 4-12 years, and PhD more than 12 years ago. Not surprisingly, more experienced researchers rated themselves as experts on a higher percentage, and non-experts on a lower percentage, of their assignments than early career researchers, as shown in Figure~\ref{fig:rev_expert_career}.

\subsubsection{Reviewer feedback}
\label{subsec:rfeedback}

\begin{figure}
\centering
\includegraphics[width=\textwidth]{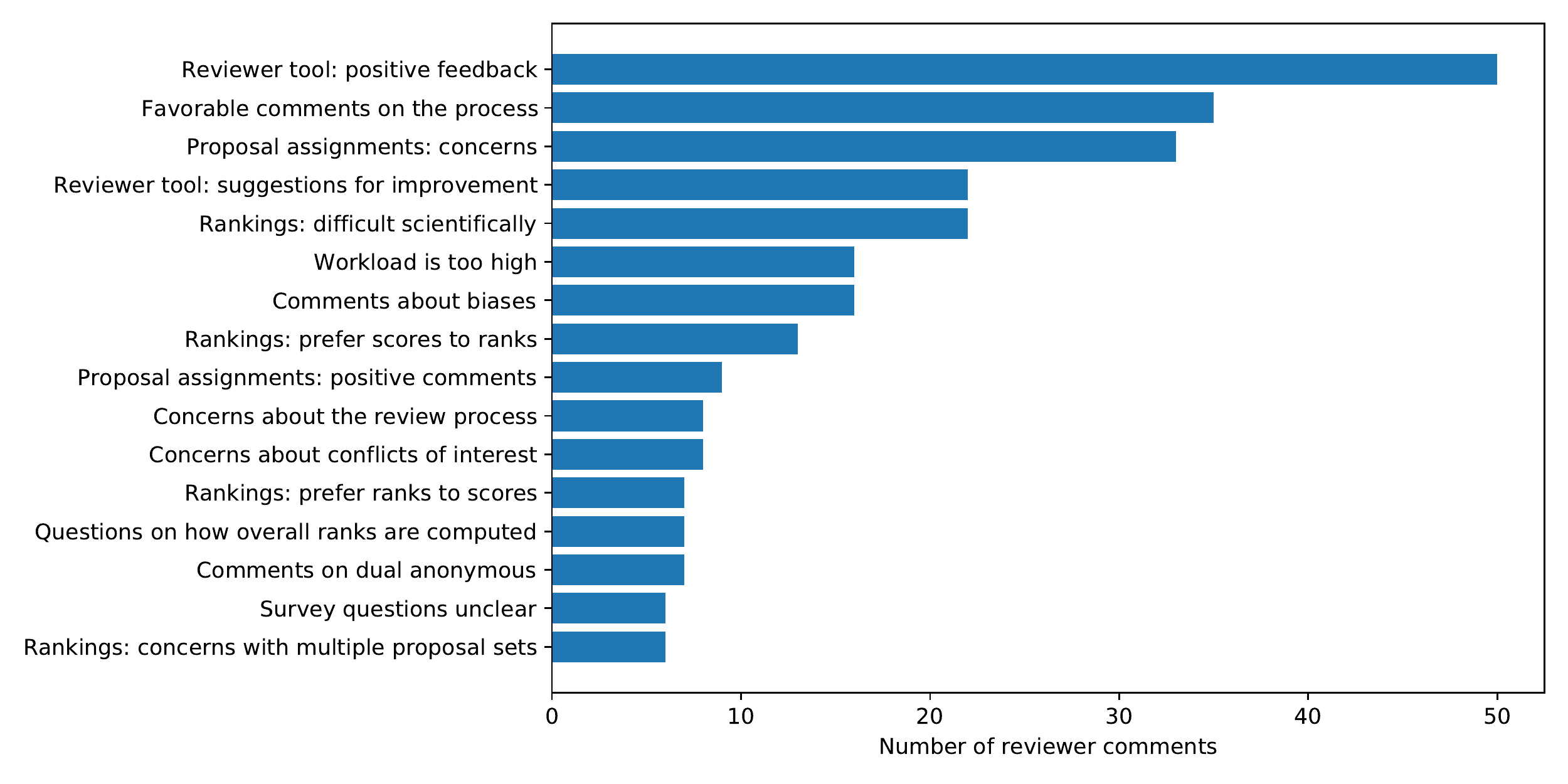}
\caption{The number of comments received via the reviewer survey regarding various topics in the free-form feedback. Only topics mentioned by five or more reviewers are indicated.}
\label{fig:feedback_reviewers}
\end{figure}

The PHT reviewed all 195 free-form comments submitted by the reviewers and categorized them into topics. While only 19\% of reviewers provided comments, it nonetheless provides an assessment of the experience (both positive and negative) reviewers had with the process. Figure~\ref{fig:feedback_reviewers} shows the frequency with which different topics were mentioned by reviewers.

The most frequent comment was general praise for the reviewer tool itself in terms of ease of use and reliability. The second most frequent comment was overall support of the distributed peer review process, including comments that it was a valuable learning experience and remarks from former panel members that they preferred the distributed peer system. The top concern raised by reviewers, which was the third most common comment, relates to the proposal assignments. Most reviewers with such concerns indicated that they were not experts in some of the assigned proposals, though a few reviewers expressed concerns that the proposals matched too well and they would have preferred a broader range of proposals. Several reviewers mentioned that it was difficult scientifically to rank the proposals in priority order given that there was not much distinction between the scientific merit of the proposals. Several reviewers also indicated that they felt the workload of 10 proposals was too high.

\subsection{Rank analysis by self-assessed reviewer expertise}

In Section~\ref{sec:output}, the analysis of individual ranks was performed using the assignment rule as a proxy for the quality of each assignment. In that section, it is shown that reviewers with matching expertise who were assigned proposals in a single category give ranks consistent with a uniform distribution, and reviewers assigned proposals in an additional category (even when they are experts) or similar but not matching expertise to their assignments tend to give better than average ranks for that assignment, in the 8.9\% of relevant assignments. Given the high response rate of the reviewer survey, the individual ranks submitted by reviewers were again analyzed for trends instead using their self-assessed expertise. An analysis is also presented of the 1370 Rule 1-2 review assignments whose reviewers claimed they had little or no expertise.

\begin{figure}
\includegraphics[width=\textwidth]{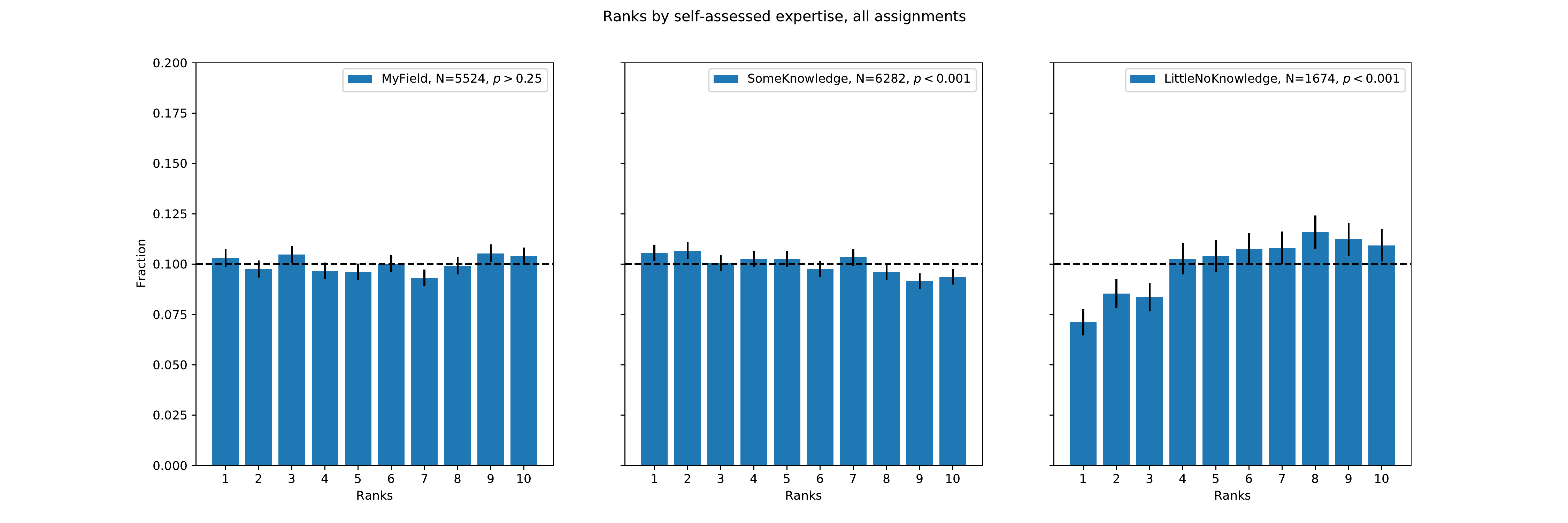}
\caption{Distribution of ranks separated by reviewer self-assessment of their expertise. The dotted line in each panel denotes the expectation of a uniform distribution among the ten rank options, and the $p$-value in each panel shows the result of the Anderson-Darling test. Error bars in each panel reflect confidence intervals ($\sqrt N$) given the observed numbers of counts.}
\label{fig:all_selfassess}
\end{figure}

The aggregate ranks per reviewer self-assessment are shown in Figure~\ref{fig:all_selfassess} along with the statistical significance of comparing each distribution to the expected uniform distribution, as in Section~\ref{sec:output}. Self-assessed experts rank proposals in a manner consistent with the uniform distribution, and those with some knowledge tend to give better than average ranks (similar to the trend observed with assignment rule). The most significant trend, however, is with reviewers claiming that they have little or no knowledge about an assignment; such reviewers tend to under-rank those proposals. Both trends are statistically significant, but the latter is more pronounced, as evident in Figure~\ref{fig:all_selfassess}. 

\begin{figure}
\includegraphics[width=\textwidth, trim=1in 0in 1in 0in, clip]{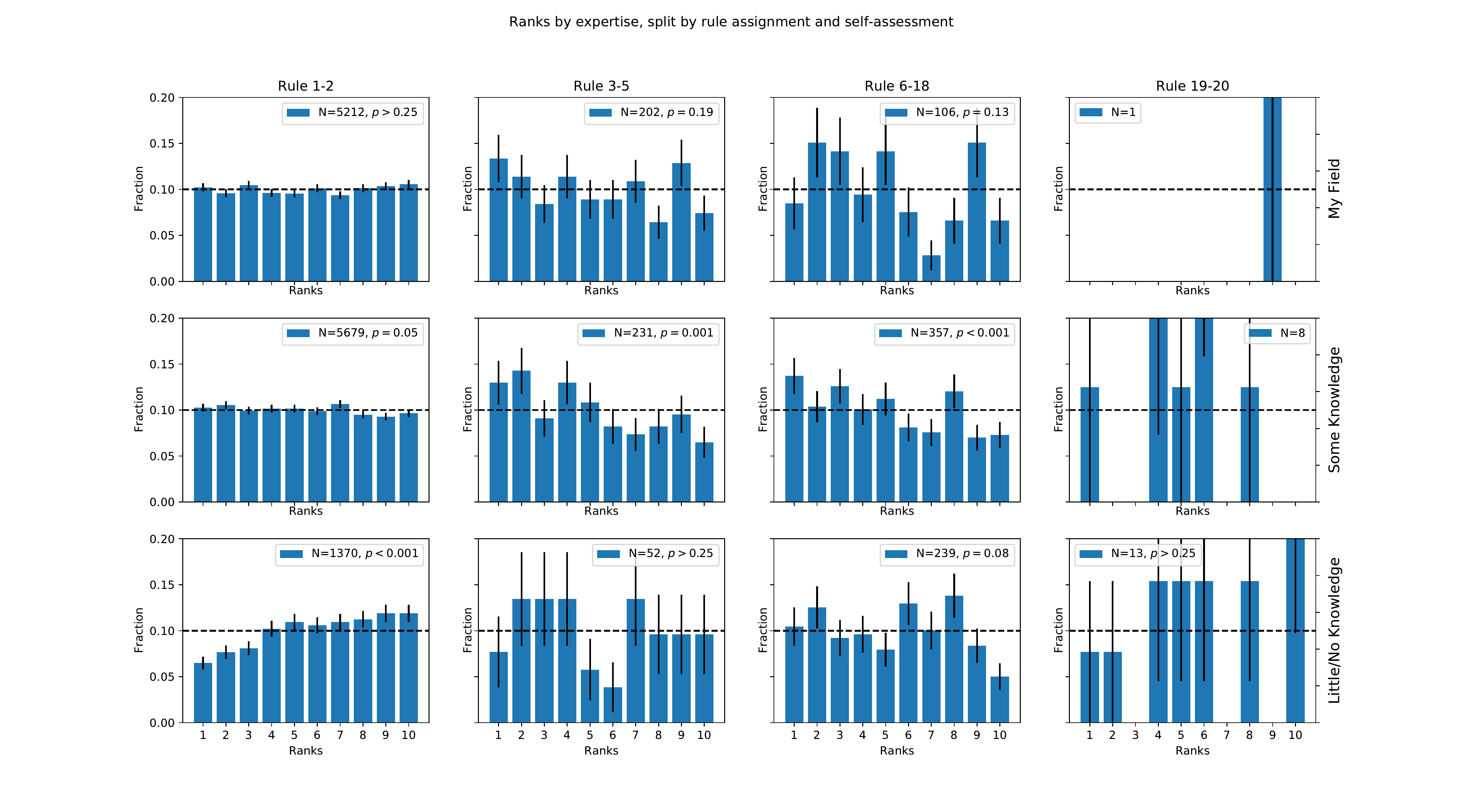}
\caption{Distribution of ranks with reviewer self-assessment, separated by assignment rule group. The dotted line in each panel denotes the expectation of a uniform distribution among the ten rank options, and the $p$-value in each panel shows the result of the Anderson-Darling test. Error bars in each panel reflect confidence intervals ($\sqrt N$) given the observed numbers of counts.}
\label{fig:all_selfassess_4panel}
\end{figure}

Going a step further, we examine the rank distributions in each combination of assignment rule and self-assessed expertise. These distributions are shown in Figure~\ref{fig:all_selfassess_4panel} separated by rule group (experts to non-experts go left to right) and self-assessment (experts to non-experts go top to bottom). Experts by both measures are expected in the top left and top left center panels, and complete non-experts are expected in the bottom right panel. No significant deviation from a uniform distribution is detected for any reviewer who indicated ``My field" (panels along the top row). All panels for reviewers indicating they had ``Some knowledge" tend to give better than average ranks for that assignment (as described in Section~\ref{sec:output}), though the trend is only significant in the Rule 3-5 and Rule 6-18 panels, consistent with the analysis previously reported. However, the reviewers in the bottom left panel of Figure~\ref{fig:all_selfassess_4panel}, with assignments matching their specified expertise but a self-assessment of ``Little/no knowledge", appear to drive the trend seen in the right panel of Figure~\ref{fig:all_selfassess}. 

\begin{figure}
\includegraphics[width=\textwidth, trim=1in 0in 1in 0in, clip]{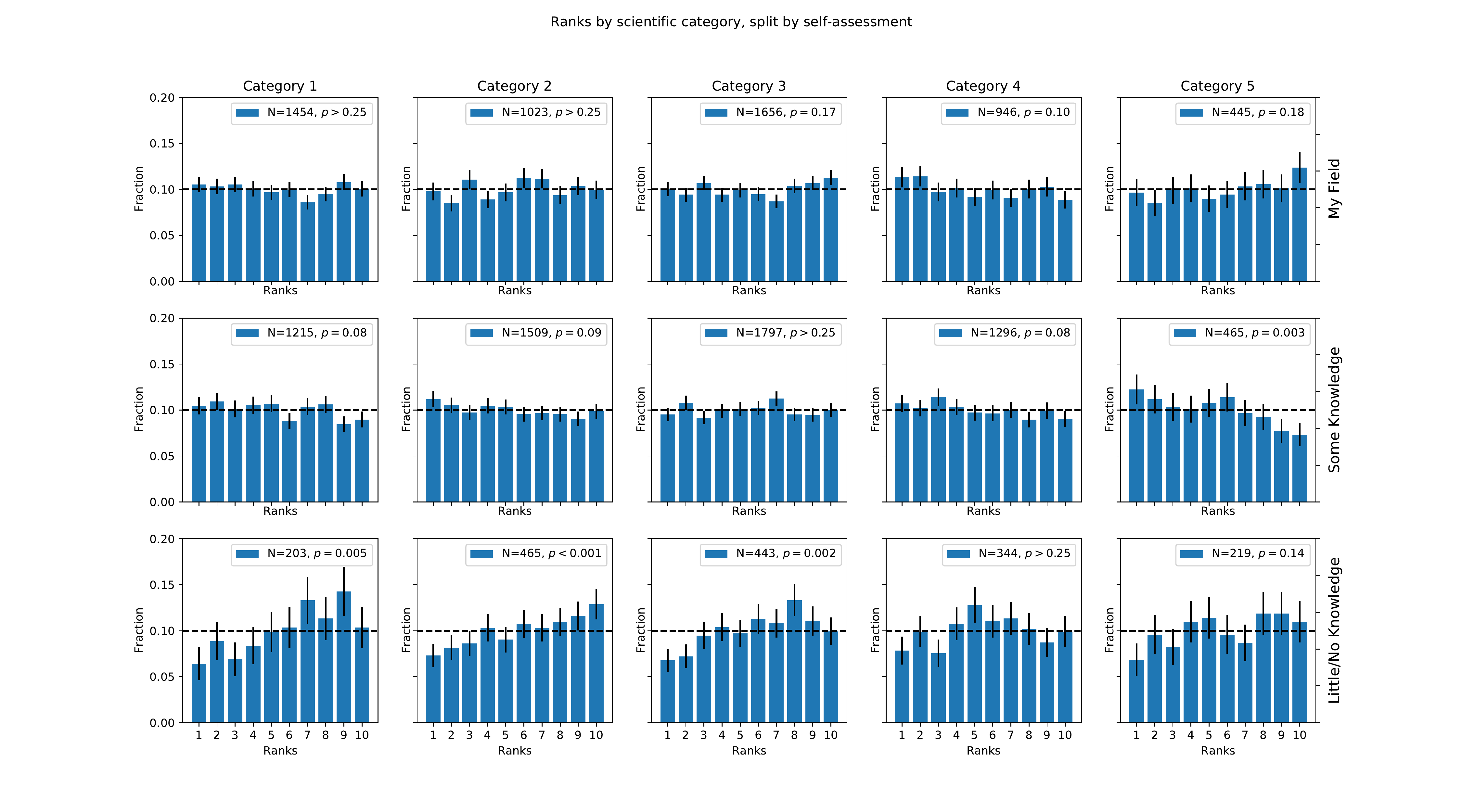}
\caption{Distribution of ranks with reviewer self-assessment, split by scientific category. The dotted line in each panel denotes the expectation of a uniform distribution among the ten rank options, and the $p$-value in each panel shows the result of the Anderson-Darling test. Error bars in each panel reflect confidence intervals ($\sqrt N$) given the observed numbers of counts.}
\label{fig:scicat_selfassess}
\end{figure}

Given the previous result that some evidence of trends in the ranks with assignment suitability over scientific category may exist at a low level (Section~\ref{subsec:multcats} and Figure~\ref{fig:totscicat}), we also examine the rank distributions separated by self-assessed expertise and category in Figure~\ref{fig:scicat_selfassess}. No significant trend with scientific category exists for reviewers who indicated ``My field" (panels along the top row). Reviewers indicating they had ``Some knowledge" submitted ranks consistent with a uniform distribution except in Category 5, where they tended to give better ranks to those assignments ($\it{p}$-value of 0.003). This matched the trend observed in Section~\ref{sec:output} and Figures~\ref{fig:all_selfassess} and \ref{fig:all_selfassess_4panel}. The signal seen in the group of reviewers claiming ``Little/no knowledge", where proposals are ranked below average, exists significantly in Categories 1, 2, and 3 but appears to be present in all categories. In Category 1 such assignments comprise 4\% of all assignments, compared to the other four categories (7-14\% of assignments), perhaps indicating that Category 1 keywords are better defined.

Reviewers with little knowledge regarding assignments that match their submitted expertise are a puzzle, and in addition to scientific category, we further examine this group along a number of axes: \\ 
\indent $\bullet$ The signal is present in reviewers from every region (except Chile) and for reviewers with and without PhDs. However, students/PhD reviewers comprise 25\%/75\% of this group, indicating that students are over-represented in this demographic when considering the overall balance of assignments made to students/PhD reviewers (14\%/86\%). \\
\indent $\bullet$ The trend is most significant for reviewers with a single proposal set ($p<0.001$ measured in 767 assignments, more than twice the number of assignments as the next assignment load bin). The trend is also present for reviewers with two proposal sets (N=340, $p<0.01$) and three proposal sets (N=140, $p<0.02$). \\ 
\indent $\bullet$ The trend is apparent for all numbers of submitted keywords, but it is detected with at least marginal significance only when the number of relevant assignments is more than $\sim$100. Reviewers who submitted 3 and 4-6 keywords are the most numerous along this axis and show the signal most significantly ($p<0.001$) over 466 and 478 assignments, respectively. Even though reviewers were required to select at least three keywords, potentially introducing a situation where they may feel forced to claim more areas of expertise than they would have preferred, this constraint cannot fully explain the trend that we see among reviewers with matched keyword assignments but little or no self-assessed knowledge. 

In reality, this puzzling set of reviewer feedback is likely not the result of any single factor, and the impact of these ranks is relatively low on the final rank-ordered list of proposals given that they are represented in every scientific category. As described in previous sections, these trends will be monitored by the JAO to determine whether future changes to the review process may be required.

\subsection{PI survey}
\label{subsec:survey_pi}

PIs of proposals that were reviewed in the distributed peer review process were invited to complete a survey designed to gauge their satisfaction in the process and the reviews. PIs were asked to provide their career status, answer seven general questions about the review process, and rate the helpfulness of each individual comment that they received. Of the 1011 unique PIs, 464 (46\% of all PIs) answered the survey questions, 442 PIs provided ratings of the review quality on behalf of 670 submitted proposals, and 187 PIs provided free-form comments. 

Appendix~\ref{app:psurvey} presents the PI survey results. Many of the questions were also asked in a PI survey conducted after the Cycle 7 Main Call, where the proposals were reviewed using traditional panels, and after the Cycle 7 Supplemental Call, which used distributed peer review. Where appropriate, the survey results from these previous review processes are also shown in Appendix~\ref{app:psurvey}.

\subsubsection{PI survey questions}

As detailed in Appendix~\ref{app:psurvey}, 80\% of PIs found their comments to be overall ``fully" or ``mostly" clear and understandable, 60\% found them ``fully" or ``mostly" accurate, and 41\% indicated that they would ``fully" or ``mostly" help to improve future proposals. Also, 99\% of PIs found that the comments were written in a respectful manner. These combined percentages are comparable to those from the PI survey for the Cycle 7 consensus reports produced by the panels, though PIs generally found the consensus reports to be ``fully" clear, understandable, and accurate 2-3 times more often than in the distributed peer review cycles. However, PIs were also several times more likely to find that their consensus reports would not help them improve future proposals compared to the distributed reviews. The best consensus reports appear to be better than the collective set of 10 comments from distributed peer review, but the distributed reviews are more likely to be at least somewhat helpful. The PI responses in the two distributed review cycles (Cycle 7 Supplemental Call and Cycle 8 Main Call) are found to be quite similar, which is noteworthy considering how many more reviewers were involved in Cycle 8.

PIs were also asked for their opinion on whether they were more or less concerned about confidentiality and biases in the proposal review process. The majority of PIs (77\%) thought any confidentiality concerns in distributed peer review were similar to or less than panel reviews, or had no strong opinion, while 23\% of PIs were more concerned about confidentiality with distributed peer review. Similarly, 65\% of PIs thought that distributed peer review is just as robust or more robust against biases compared to panel review or had no strong opinion, and 35\% thought distributed peer review was less robust against biases.

\subsubsection{PI assessment of individual reviews}

\begin{figure}
\centering
\includegraphics[width=\textwidth]{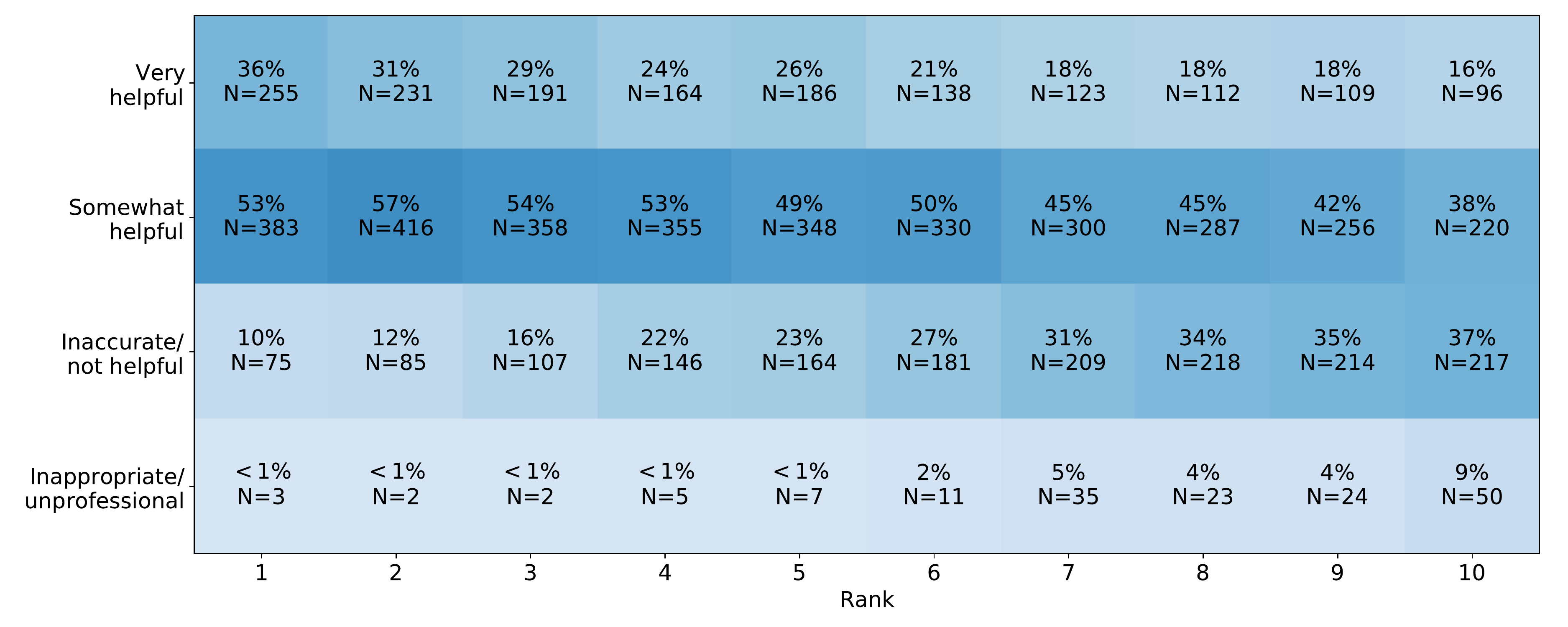}
\caption{Correlation between the helpfulness of a comment as assessed by the PI versus the proposal rank. The responses are normalized separately for each rank value.}
\label{fig:rank_help}
\end{figure}

PIs were asked to assess the helpfulness of each individual comment that they received for their submitted proposal(s). Similar to the Cycle 7 Supplemental Call, PIs rated 73\% of the individual comments to be ``very" or ``somewhat" helpful. On the other hand, 24\% of the comments were rated as inaccurate or not helpful, and 2.4\% were rated inappropriate or unprofessional. Figure~\ref{fig:rank_help} shows the correlation between the helpfulness of a comment as assessed by the PI versus the rank assigned by the reviewer. A clear correlation is observed in that comments associated with a good individual rank are rated more helpful than comments associated with a poor rank, suggesting that PIs were most favorably disposed to comments likely to be the most positive.

\subsubsection{PI feedback}

\begin{figure}
\centering
\includegraphics[width=\textwidth]{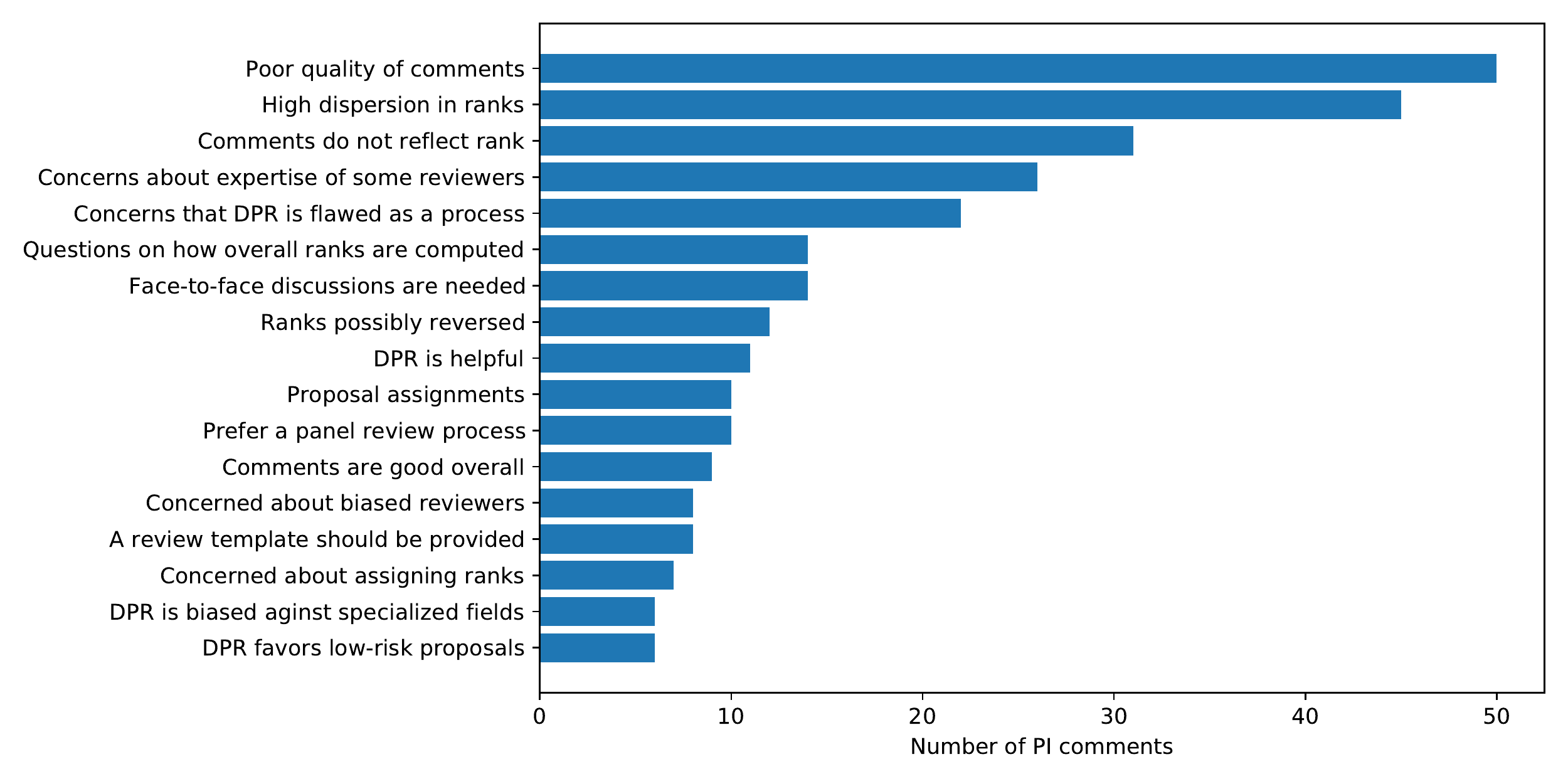}
\caption{The number of comments received from PIs in various topics in the free-form feedback. Only topics mentioned by five or more PIs are indicated.}
\label{fig:feedback_pis}
\end{figure}

Similar to the Cycle 8 reviewer survey (see Section~\ref{subsec:rfeedback}), we classified the free-form feedback received from PIs to establish the predominant concerns. Figure~\ref{fig:feedback_pis} shows the results of the analysis. Of the 187 comments received from 1011 PIs, the majority of the feedback expressed concerns about some aspect of the distributed peer review process. The top concern expressed by PIs was related to the poor quality of some of the comments. A related issue is that some comments did not adequately justify the assigned rank, leading to some PIs expressing concerns about the expertise of the reviewers. Many PIs also commented on the high dispersion of their assigned ranks.  

\subsection{Analysis of PI helpfulness ratings}
By correlating the PI rating of the helpfulness of individual reviews with other demographic data, we can begin to assess what factors contribute to a useful comment from the PI perspective. In the following sections, we evaluate how the quality of the comments as judged by the PIs depends on the reviewer expertise on the proposal, the career status of the reviewer, the length of the comments, and the workload of the reviewer.

\subsubsection{Reviewer expertise}

\begin{figure}
\centering
\includegraphics[width=0.45\textwidth]{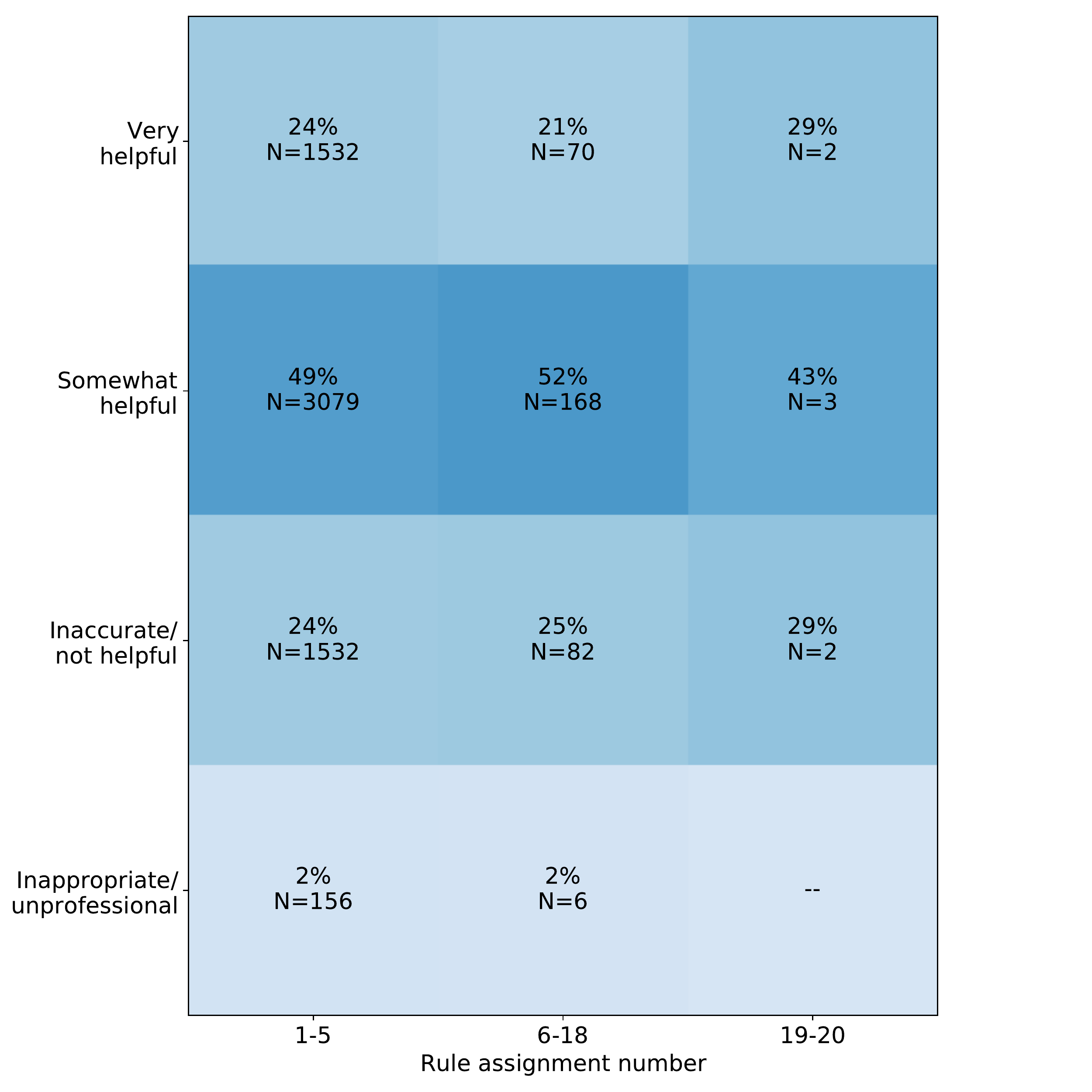}
\includegraphics[width=0.45\textwidth]{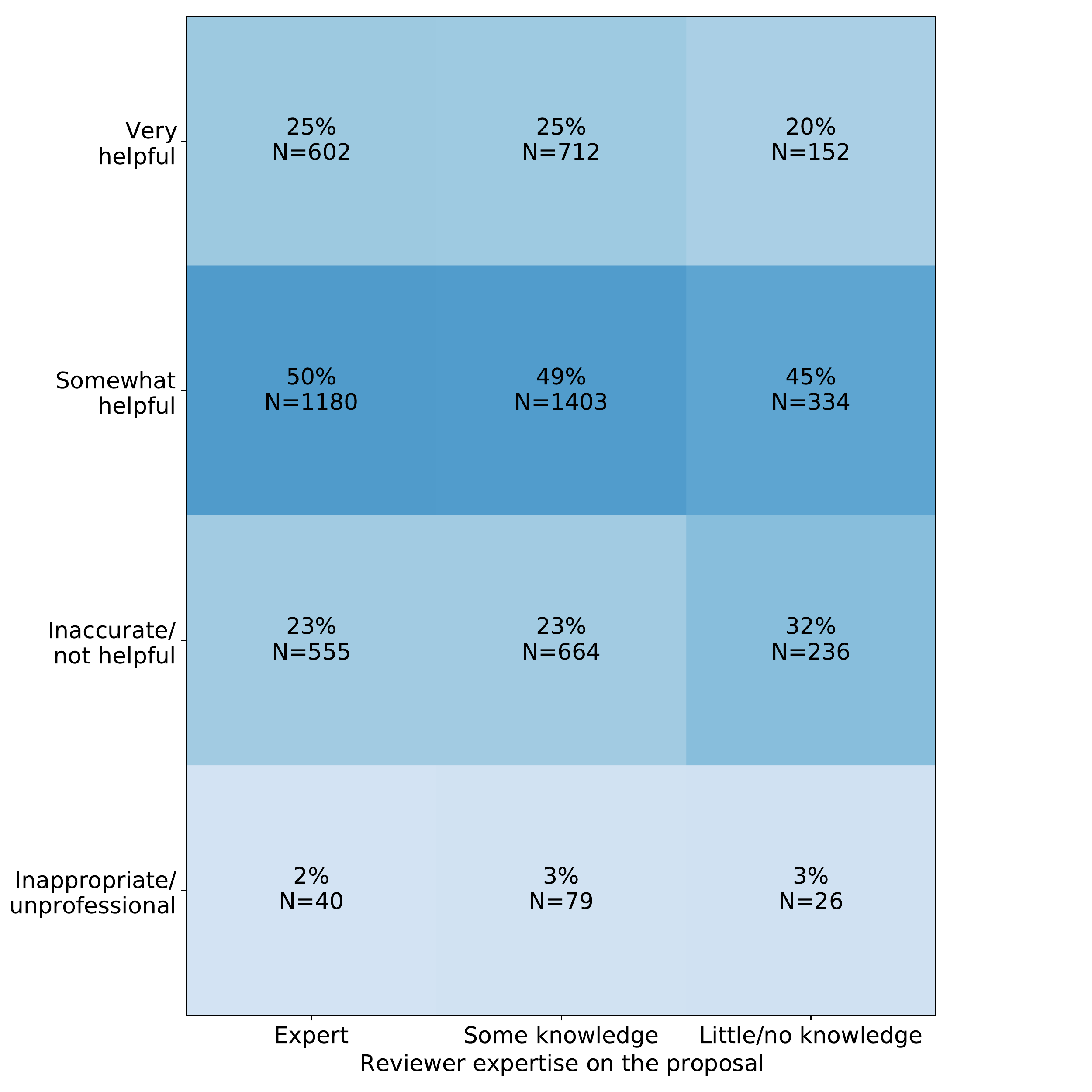}
\caption{Correlation between the helpfulness of the review as determined by the PI with the proposal assignment rule (left) and the reviewer self-assessment of their expertise (right). The responses are normalized separately for each expertise ``bin" column.}
\label{fig:rule_help}
\end{figure}

We first correlate the PIs' assessments of comment quality with how well the expertise of the reviewers matched the assigned proposal. We have two measures of the reviewer expertise: the keywords selected by the reviewer and the self-assessment by the reviewers of their expertise on a given proposal. Figure~\ref{fig:rule_help} shows the correlation between the helpfulness of the proposal comment as assessed by the PI with the rule assignment number (left panel) and the reviewer's self-assessment of their expertise (right panel). Comments submitted where the reviewer-specified expertise matched the assigned proposal keywords (Rules 1-5) are rated equally as helpful as comments submitted for Rule 6-18 assignments. In both cases, assignments where the reviewers are experts (by either rule number or self-assessment) garner similar helpfulness ratings compared to reviewers who have some knowledge. However, reviewers with little or no knowledge on a proposal by their own self-assessment are more likely to write an inaccurate comment than a helpful one. 

The fact that the comments rated ``inaccurate or unhelpful" were primarily written by reviewers assigned proposals with keywords that they selected suggests that the rule assignment algorithm itself is not the main issue. The keywords may be too broad and may need revising. ALMA may also need to provide more guidance to PIs and reviewers on how to select keywords. Machine-learning algorithms may also help improve proposal assignments \citep{Strolger17,Kerzendorf20}. However, such approaches also achieve $\sim$90\% success rates in matching proposals with reviewer  expertise, comparable to the success rate achieved by ALMA, so achieving a higher success rate may require fine-tuning any of these methods or combining different approaches. 

\subsubsection{Career status of the reviewer}

\begin{figure}
\centering
\includegraphics[width=0.6\textwidth]{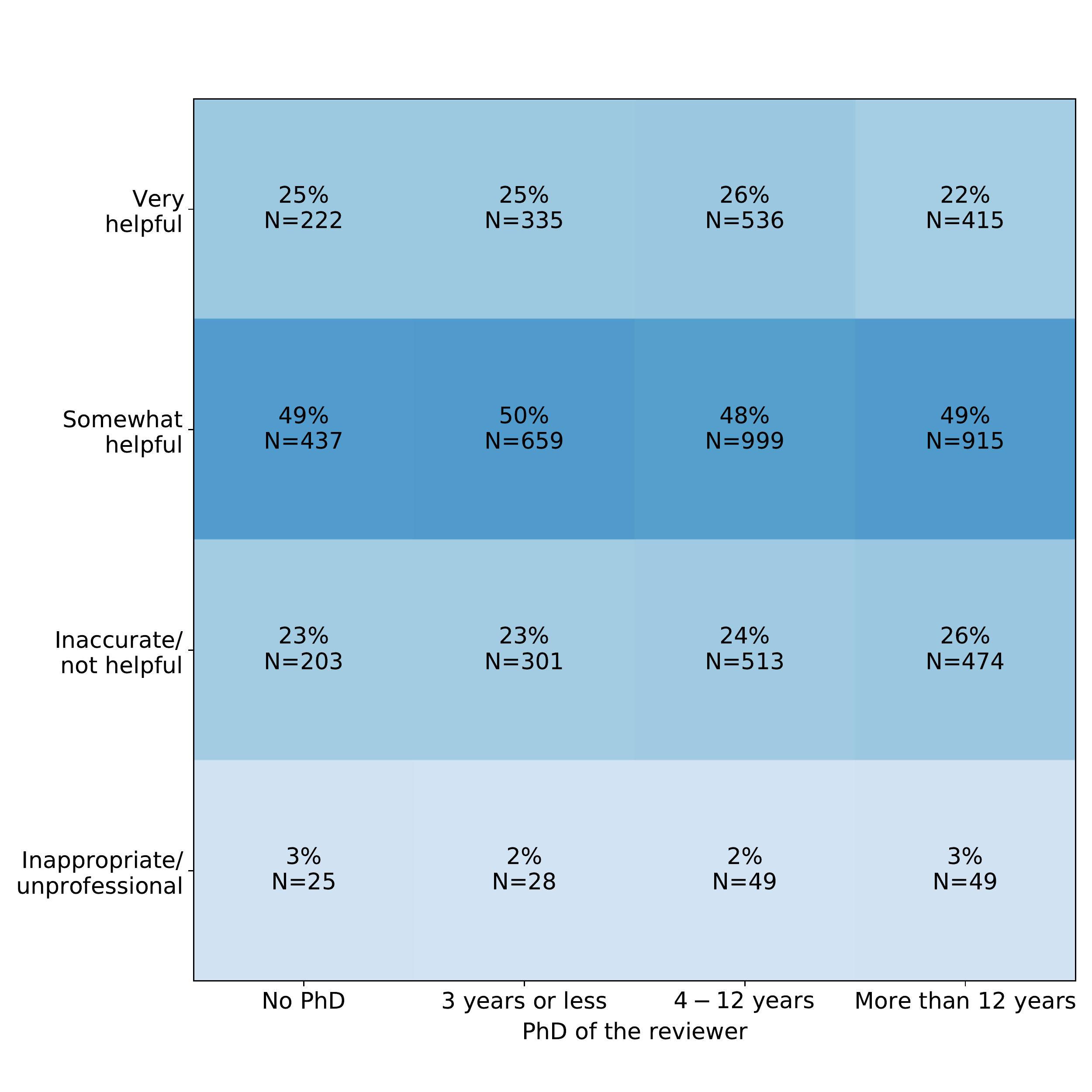}
\caption{Correlation between the helpfulness of the comment, as determined by the PI, with the career status of the reviewer. The responses are normalized separately for each career stage column.}
\label{fig:career_help}
\end{figure}

ALMA allows PIs without PhDs to be reviewers in the distributed peer review process as long as they specify a mentor to assist them with the proposal reviews. Such reviewers do not have as much scientific experience as more senior reviewers, and they likely have little or no experience in reviewing proposals. Reviewers were asked to provide their career status in terms of the number of years since they received their PhD (or no PhD). Figure~\ref{fig:career_help} shows the correlation between the career status of the reviewer and the helpfulness of their comments. Despite a relative lack of experience in the proposal review process, early career researchers (reviewers with no PhD or with a PhD only in the past 3 years) had the lowest rate of inaccurate and unhelpful comments. Conversely, the most senior reviewers who received their PhD more than 12 years ago have the lowest percentage of ``very helpful" comments and the highest percentage of ``inaccurate/not helpful" comments. Statistically the differences between the demographic groups are not significant. This indicates that the comments (though not necessarily the rankings) have similar quality regardless of the career status of the reviewer.

\subsubsection{Length of the comments}
\label{subsec:length}

\begin{figure}
\centering
\includegraphics[width=0.6\textwidth]{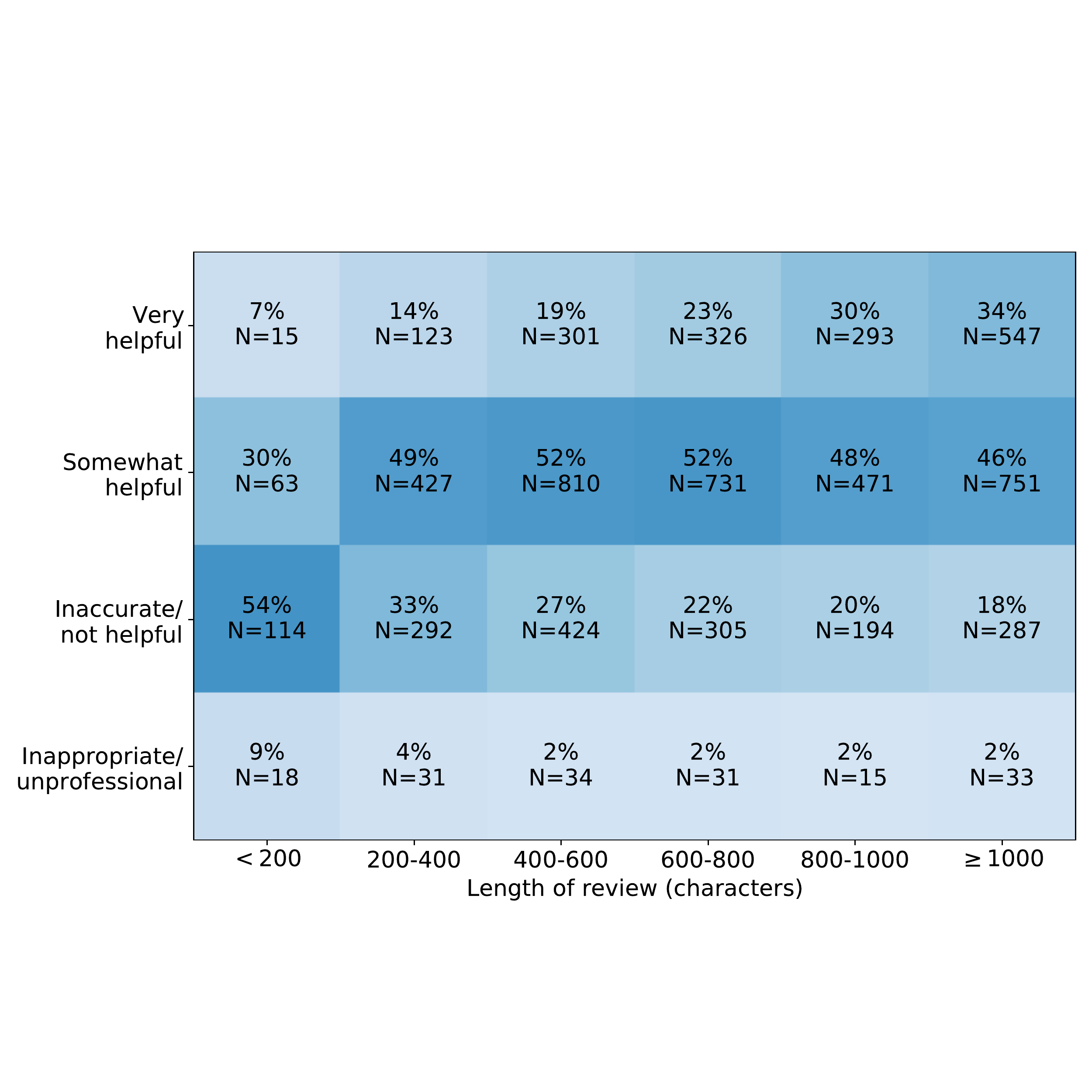}
\caption{Correlation between the helpfulness of the review as determined by the PI with the length of the review in characters. The responses are normalized separately for each column.}
\label{fig:length_help}
\end{figure}

The length of the review comments vary considerably from reviewer to reviewer, as described in Section~\ref{subsec:comments}. To evaluate if there is a threshold at which the comments become particularly helpful, Figure~\ref{fig:length_help} shows the correlation between the length of the comment and its helpfulness rating. Not surprisingly, longer comments have a clear tendency to be more helpful than shorter comments. A length of at least 200 characters is the threshold for at least half of the comments to be somewhat useful, as the majority of shorter comments are deemed to be inaccurate or not helpful. Longer comments become increasingly judged as very helpful, although even 18\% of the longest comments are judged to be inaccurate or not helpful. While the length of the comment does not necessarily reflect quality, the introduction of a minimum comment length may help reduce the number of unhelpful reviews, as would providing more overall guidance to the reviewers on writing proposal reviews.  

\subsubsection{Workload of the reviewer}

\begin{figure}
\centering
\includegraphics[width=0.75\textwidth, trim=0in 2in 0in 2in, clip]{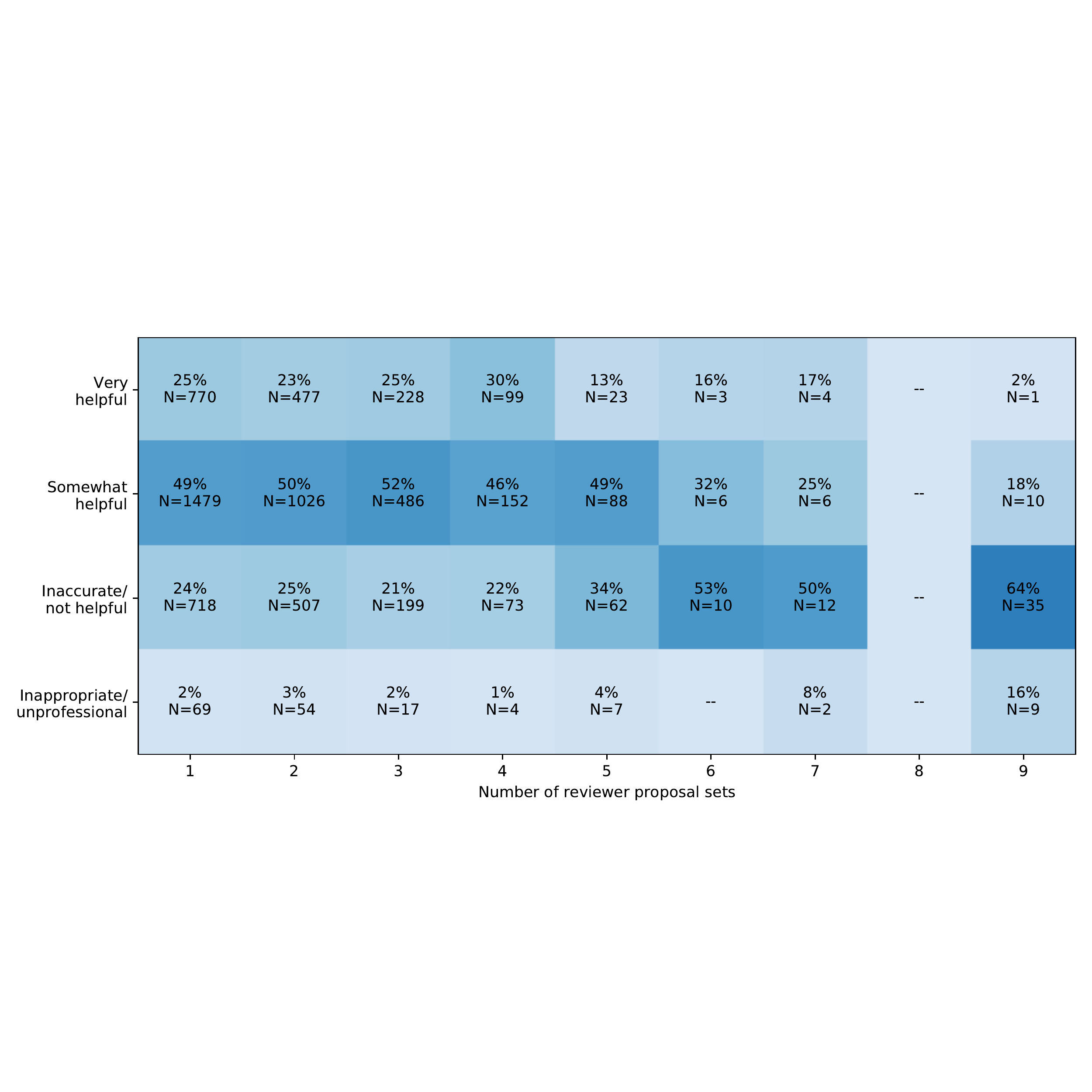}

\caption{Correlation between the helpfulness of the review as determined by the PI with the number of proposals sets assigned to the reviewer. The percentages are normalized separately for each number of proposal sets.}
\label{fig:nsets_help}
\end{figure}

No restrictions are placed on the number of proposal sets that may be assigned to a reviewer. As the assigned number of proposal sets increases, it might be expected for the quality of the comments to diminish as there may be less time available for a thorough evaluation of all assigned proposals. Figure~\ref{fig:nsets_help} shows the correlation between the helpfulness of the comments and the number of proposal sets assigned to a reviewer. Even with as many as 4 proposal sets, reviewers wrote comments deemed just as helpful as those written by reviewers with a single proposal set. Only when a reviewer was assigned 5 or more proposal sets (as was the case for 11 reviewers in total) was there a decline in the comment quality. 

%% file: summary.tex
\section{Summary}
\label{sec:summary}

In this paper, we have summarized the process and results of the ALMA Cycle 8 distributed peer review, which was implemented in response to the challenges presented by high reviewer workloads in traditional panel reviews and the increasing number of submitted proposals. Based on nearly 15,000 ranks and comments submitted by reviewers, the demographics of the reviewers and their expertise, and the feedback from reviewers and PIs through surveys, we assess the overall performance of the review process. Our main takeaways are as follows:

\begin{itemize}
\item ALMA implemented distributed peer review to assess 1497 proposals requesting less than 25 hours of 12-m array time or less than 150 hours on the 7-m array in ``standalone" mode in the Cycle 8 Main Call. Of the accepted proposals from both distributed peer review and the panels, 87\% of the highest priority proposals (Grade A and B) went through the distributed peer review process to account for 59\% of the available observing time on the 12-m Array. Of the 1497 proposals which underwent distributed peer review, 99.7\% of reviews were submitted on time. A grace period was provided to the remaining reviewers, and in the end only 1 proposal was rejected for a lack of submitted reviews. Based on the relatively low instance of requests for user support during the review process, the tools and process worked well. \\

\item Several updates to the distributed peer review process were implemented based on feedback from the pilot program in the Cycle 7 Supplemental Call. These changes included adding the ability for reviewers to specify their expertise to improve the matching between reviewers and proposals, the addition of an optional Stage 2 where reviewers can read the anonymized comments of other reviewers assigned the same proposals and adjust their own ranks and comments,  improvements to modernize the maneuverability of the rankings within the user-facing tool, and adding similar keywords to the internal assignment matching algorithm to further improve the assignment quality. \\

\item Over 1000 unique reviewers participated in the process, including 164 reviewers without PhDs, and 90\% of reviewers (97\% of reviewers without PhDs) assessed 10 or 20 proposals. The majority (83\%) submitted scientific keywords representing their expertise to use in the assignment matching process, while the remaining reviewers were assumed to be expert in the keywords of their submitted proposals. The ``supply" of reviewers specifying expertise in every keyword was in all cases greater than the ``demand" given by submitted proposals in each keyword, but some keywords were only minimally ``over-supplied". The majority of reviewers who submitted expertise keywords did so in more than one scientific category. \\

\item Review assignments were created to prioritize matches between a reviewer's expertise and the content of the assigned proposals. Of the 14,970 assignments created, exact matches between a reviewer's submitted expertise keyword and the keyword on the submitted proposal comprised 94\% of assignments. The remaining assignments consisted of similar or related content relative to a reviewer's expertise in 5\% of cases and unrelated content in less than 1\% of cases. \\

\item Most reviewers participated in the optional Stage 2 process, during which they were able to read the anonymized comments of other reviewers assigned the same proposals and make changes to their own ranks and comments. However, only a minority (15\%) of ranks or comments to the PI were modified during Stage 2. \\

\item In the 5.3\% of review assignments consisting of similar or related content relative to a reviewer's specified expertise, reviewers tend to rank those proposals more strongly than average. Further, the same trend is detected in the 3.4\% of assignments where exact matches were made with the declared expertise of the reviewer (i.e., they were experts on those assignments), but those reviewers were assigned proposals in more than one category. Given the low fraction of review assignments which fall into these demographics, the overall impact on the aggregate ranks per scientific category are small, as no scientific category shows a statistically significant deviation from the expected distribution of submitted ranks. 
\\

\item The distribution of comment lengths written to PIs were found to correlate with three quantities: the reviewer's submitted rank for the proposal, the assignment load carried by the reviewer, and the reviewer's region. Reviewers wrote the shortest comments for their strongest-ranked proposals and the longest comments for their weakest-ranked proposals. Those with 5 or more proposal sets wrote significantly shorter comments than reviewers with smaller assignment loads, and East Asian reviewers wrote the shortest comments of all regions. \\

\item The final ranked list of proposals was created by averaging the submitted reviewer ranks and sorting by the average rank. Two methods of outlier rejection were considered but not used, and a pairwise ranking method for combining the ranks called Plackett-Luce was explored as described in Appendix~\ref{app:ranks}. A variety of simulations were run comparing the dispersion in ranks for individual proposals to those we measure from the Cycle 8 distributed peer review data, as well as the Cycle 7 panel review data. The dispersion in Cycle 8 ranks is most similar to the equivalent ranks that Cycle 7 panel reviewers produced before the panel meetings, both of which are much larger than can be explained by the fact that reviewers rank only a subset of proposals. This finding indicates that the dispersion is dominated by differences of opinion between reviewers on the relative scientific merit of proposals, which is qualitatively similar to the result found by \citet{Patat18} in a large study of pre-meeting scores awarded during ESO panel reviews. \\

\item A survey of the reviewers revealed valuable information concerning their self-assessed expertise on each assignment they received, and 88\% of the responses indicated that they were expert or had some knowledge of the proposals they were asked to review; this fraction is similar to the fraction (94\%) of assignments made with matching keywords. The majority of assignments where reviewers rated themselves as having ``little/no knowledge" on an assignment were actually in cases where those assignments matched the expertise keywords provided by the reviewer. This result suggests that the keywords currently in use may be too broad or that reviewers are very selective in defining their expertise. Reviewers tend to assign better-than-average ranks to proposals in which they assess that they have ``some knowledge" (most significantly in Category 5), while they tend to give below-average ranks to proposals where they feel they have ``little/no knowledge." \\

\item A survey of the PIs revealed more valuable demographic data regarding the participants of the distributed peer review process. Similar to the results of the Cycle 7 PI survey, which was based on a traditional panel review, PIs found 80\% of their comments to be ``fully" or ``mostly" clear and understandable, 60\% found them to be accurate, and 41\% felt that they would help to improve future proposals. PIs expressed concerns about the dispersions observed in their ranks as well as the quality of some comments that they received. They also submitted helpfulness ratings for each comment they received, which show a strong correlation with the accompanying rank, such that strong ranks garnered helpful ratings, and weak ranks garnered unhelpful ratings. Correlating the results of the two surveys revealed that reviewer career status had no bearing on whether the resulting comments were found to be helpful. The length of the comment and assignment load of the reviewer were both found to correlate with PI satisfaction with the comments. Comments with less than 200 characters were found to be broadly inaccurate or unhelpful, and reviewers with assignment loads of 5 or more proposal sets were also found to produce comments that were more likely to be unhelpful.
\\
\end{itemize}

Distributed peer review creates a new dynamic in the proposal review process in both explicit and implicit ways. Fundamentally, many astronomers are asked to review a relatively small number of proposals, instead of inviting relatively few people to read and discuss many proposals. In addition to the subsequent change in assignment load, this change explicitly translates to a more inclusive review process and a much broader range of available expertise. Nonetheless, any review process is subject to potential systematics and biases. As ALMA continues to use distributed peer review in future cycles, the PHT will continue to improve the process by examining the results of each cycle carefully and incorporating feedback from the community in order to ensure fairness in the review process.

\begin{acknowledgments}

JDM would like to thank the anonymous referee whose suggestions helped to strengthen the conclusions of this paper. The PHT would like to acknowledge the support of the ALMA Regional Centers, whose members provided testing efforts and invaluable feedback on the Reviewer Tool, for their support of the international ALMA community through the distributed review process. We would also like to thank the ALMA community for their contributions to the process through their feedback and involvement over the last four years; the process has evolved, and will continue to evolve, as a result of these interactions with a community committed to fairness and inclusion in proposal review.

\end{acknowledgments}

\software{SciPy \citep{Jones01}, kSamples \citep{Scholz19}, choix \citep{Maystre18}}

%% file: appendices/rules.tex
\appendix
\section{Rules for Participation in Distributed Peer Review} 
\label{app:rules}
The following rules of the distributed review process were posted to the ALMA Science Portal.\footnote{https://almascience.nrao.edu/proposing/alma-proposal-review/distributed-peer-review} \\
\indent 1. All participants in the review process are expected to behave in an ethical manner. If it is found that a reviewer has not behaved in an ethical manner, the proposal(s) on which the reviewer is acting as the designated reviewer may be rejected. \\
\indent 2. Each proposal must designate one reviewer to participate in the review process. The designated reviewer may be the PI of the proposal or one of the co-Is. \\
\indent 3. The reviewer must be specified in the Observing Tool (OT) at the time of proposal submission and cannot be changed after the proposal deadline.  \\
\indent 4. PIs who do not have a PhD may be selected as the designated reviewer. In such cases, a mentor must be specified who will assist the PI in the review process. The mentor does not need to be part of the proposal team, but must have a PhD in astronomy or a related field, and must be specified in the OT at the time of proposal submission. Co-Is who do not have a PhD are not eligible to be selected as reviewers. \\
\indent 5. Proposals will be assigned to reviewers based on the expertise of the reviewer as specified on the reviewer's user profile. If a reviewer has not registered their expertise, then the assignment algorithm will use the keywords of their submitted proposals. \\
\indent 6. Reviewers must declare any major conflicts of interest of their assigned proposals. Any proposals with a major conflict of interest will be replaced by another proposal. \\
\indent 7. During Stage 1, each designated reviewer will be assigned to review ten proposals. The reviewer must rank the proposals relative to each other in order of scientific priority from 1 to 10 (1 being the strongest, and 10 being the weakest), and write a review for each one of them. If a person is the designated reviewer on multiple proposals, they will receive ten unique review assignments per submitted proposal. \\
\indent 8. If a reviewer does not submit their reviews and ranks by the Stage 1 deadline (3 June 2021 15:00 UT), the proposal for which they were identified as the designated reviewer will be rejected. \\
\indent 9. During Stage 2, the Stage 1 reviews will be shown anonymously to the other (9) reviewers of each proposal. The reviewers will then be able to re-rank their assigned proposals and edit their reviews. \\
\indent 10. Reviews and ranks will be sent anonymously to PIs without any editing by the JAO. If reviewers participate on Stage 2, then their edited reviews and modified ranks will be sent to the PIs, otherwise the Stage 1 reviews and ranks will be sent. \\
\indent 11. All participants in the review process agree to keep the materials confidential and will not use the materials for any other means other than the proposal review. Participants will delete any proposals and any other review materials after they have completed their assessments. \\
\indent 12. All communications between the JAO Proposal Handling Team (PHT) and the reviewers will be done by email. The PHT will use the email address associated with the reviewers’ ALMA user account. Please make sure to keep your user profile updated so you do not lose important information. \\

%% file: appendices/guidelines.tex
\section{Guidelines for Reviewers} 
\label{app:guidelines}
The following guidelines for reviewers were posted to the ALMA Science Portal.\footnote{https://almascience.nrao.edu/proposing/alma-proposal-review/distributed-peer-review} \\
\indent 1. The overall scientific merit of the proposed investigation and its potential contribution to the advancement of scientific knowledge. 
\begin{itemize}
\item Does the proposal clearly indicate which important, outstanding questions will be addressed? 
\item Will the proposed observations have a high scientific impact on this particular field and address the specific science goals of the proposal? ALMA encourages reviewers to give full consideration to well-designed high-risk/high-impact proposals even if there is no guarantee of a positive outcome or definite detection. 
\item Does the proposal present a clear and appropriate data analysis plan? 
\end{itemize}

\indent 2. The suitability of the observations to achieve the scientific goals. 
\begin{itemize}
\item Is the choice of target (or targets) clearly described and well justified? 
\item Are the requested signal-to-noise ratio, angular resolution, spectral setup, and u-v coverage sufficient to achieve the science goals? 
\end{itemize}
 
The following guidance, as well as an example review, was provided to assist reviewers in writing helpful comments to the PI: \\

\indent 1. Summarize both the strengths and weaknesses of the proposal. 
\begin{itemize}
\item A summary of both the strengths and weaknesses can help PIs understand what aspects of the project are strong, and which aspects need to be improved in any future proposal. 
\item Reviews should focus on the major strengths and major weaknesses. Avoid giving the impression that a minor weakness was the cause of a poor ranking. Many proposals do not have obvious weaknesses but are just less compelling than others; in such a case, acknowledge that the considered proposal is good but that there were others that were more compelling. 
\item Take care to ensure that the strengths and weaknesses do not contradict each other.
\end{itemize}

\indent 2. Be objective.
\begin{itemize}
\item Be as specific as possible when commenting on the proposal. Avoid generic statements that could apply to most proposals.
\item If necessary, provide references to support your critique.
\item All reviews should be impersonal, critiquing the proposal and not the proposal team. For example, do not write ``The PI did not adequately describe recent observations of this source.", but instead write ``The proposal did not adequately describe recent observations of this source.”.
\item Reviewers cannot be sure at the time of writing reviews whether the proposed observations will be scheduled for execution. The reviews should be phrased in such a way that they are sensible and meaningful regardless of the final outcome.
\end{itemize}

\indent 3. Be concise.
\begin{itemize}
\item It is not necessary to write a lengthy review. An informative review can be only a few sentences in length if it is concise and informative. But, please avoid writing only a single sentence that does not address specific strengths and weaknesses.
\end{itemize}

\indent 4. Be professional and constructive.
\begin{itemize}
\item It is never appropriate to write inflammatory or inappropriate comments, even if you think a proposal could be greatly improved.
\item Use complete sentences when writing your reviews. We understand that many reviewers are not native English speakers, but please try to use correct grammar, spelling, and punctuation.
\end{itemize}

\indent 5. Be aware of unconscious bias.
\begin{itemize}
\item We all have biases and we need to make special efforts to review the proposals objectively. 
\end{itemize}

\indent 6. Be anonymous.
\begin{itemize}
\item Do not identify yourself in the reviews to the PIs. In case of distributed peer review, these reviews will not be checked and edited by the JAO. They will be sent verbatim to the PIs, and they will also be shared with other reviewers during Stage 2.
\item Do not spend time trying to guess who is the proposal team behind the proposal you are reviewing. Your review should be based solely on the scientific merit of the proposal. The identity of the team behind it has no relevance for your review. 
\end{itemize}

\indent 7. Other best practices.
\begin{itemize}
\item Do not summarize the proposal: The purpose of the review is to evaluate the scientific merits of the proposal, not to summarize it. While you may provide a concise overview of the proposal, it should not constitute the bulk of the reviews.
\item Do not include statements about scheduling feasibility. If there are any scheduling feasibility issues with the proposal, the JAO will address them directly with the PI.
\item Do not include explicit references to other proposals that you are reviewing, such as project codes.
\item Do not ask questions. A question is usually an indirect way to indicate there is a weakness in the proposal, but the weakness should be stated explicitly. For example, instead of ``Why was a sample size of 10 chosen?" write ``The proposal did not provide a strong justification for the sample size of 10." 
\item Do not use sarcasm or any insulting language.
\end{itemize}

\indent 8. Re-read your reviews and scientific rankings.
\begin{itemize}
\item Once you have completed your assessments, re-read your reviews and ask how you would react if you received them. If you feel that the reviews would upset you, revise them. 
\item Check to see if the strengths and weaknesses in the reviews are consistent with the scientific rankings. If not, consider revising the reviews or the rankings.
\end{itemize}

%% file: appendices/keywords.tex
\section{Scientific Category/Keyword Pairs}
\label{app:keywords}
The following list presents the scientific categories, keywords, and their shorthands which were available to PIs in Cycle 8. \\

Category 1 – Cosmology and the high redshift universe \\
\indent a. Lyman Alpha Emitters/Blobs (LAE/LAB) \\
\indent b. Lyman Break Galaxies (LBG) \\
\indent c. Starburst galaxies \\
\indent d. Sub-mm Galaxies (SMG) \\
\indent e. High-z Active Galactic Nuclei (AGN) \\
\indent f. Gravitational lenses \\
\indent g. Damped Lyman Alpha (DLA) systems \\
\indent h. Cosmic Microwave Background (CMB)/Sunyaev-Zel'dovich Effect (SZE) \\
\indent i. Galaxy structure \& evolution \\
\indent j. Gamma Ray Bursts (GRB) \\
\indent k. Galaxy Clusters \\
 
Category 2 – Galaxies and galactic nuclei \\
\indent a. Starbursts, star formation \\
\indent b. Active Galactic Nuclei (AGN)/Quasars (QSO) \\
\indent c. Spiral galaxies \\
\indent d. Merging and interacting galaxies \\
\indent e. Surveys of galaxies \\
\indent f. Outflows, jets, feedback \\
\indent g. Early-type galaxies \\
\indent h. Galaxy groups and clusters \\
\indent i. Galaxy chemistry \\
\indent j. Galactic Centers/nuclei \\
\indent k. Dwarf/metal-poor galaxies \\
\indent l. Luminous and Ultra-Luminous Infra-Red Galaxies (LIRG \& ULIRG) \\
\indent m. Giant Molecular Clouds (GMC) properties \\

Category 3 – ISM, star formation and astrochemistry \\
\indent a. Outflows, jets and ionized winds \\
\indent b. High-mass star formation \\
\indent c. Intermediate-mass star formation \\
\indent d. Low-mass star formation \\
\indent e. Pre-stellar cores, Infra-Red Dark Clouds (IRDC) \\
\indent f. Astrochemistry \\
\indent g. Inter-Stellar Medium (ISM)/Molecular clouds \\
\indent h. Photon-Dominated Regions (PDR)/X-Ray Dominated Regions (XDR) \\
\indent i. HII regions \\
\indent j. Magellanic Clouds \\

Category 4 – Circumstellar disks, exoplanets and the solar system \\
\indent a. Debris disks \\
\indent b. Disks around low-mass stars \\
\indent c. Disks around high-mass stars \\
\indent d. Exoplanets \\
\indent e. Solar system: Comets \\
\indent f. Solar system: Planetary atmospheres \\
\indent g. Solar system: Planetary surfaces \\
\indent h. Solar system: Trans-Neptunian Objects (TNOs) \\
\indent i. Solar system: Asteroids \\

Category 5 – Stellar evolution and the Sun \\
\indent a. The Sun \\
\indent b. Main sequence stars \\
\indent c. Asymptotic Giant Branch (AGB) stars \\
\indent d. Post-AGB stars \\ 
\indent e. Hypergiants \\
\indent f. Evolved stars: Shaping/physical structure \\
\indent g. Evolved stars: Chemistry \\
\indent h. Cataclysmic stars \\
\indent i. Luminous Blue Variables (LBV) \\
\indent j. White dwarfs \\ 
\indent k. Brown dwarfs \\
\indent l. Supernovae (SN) ejecta \\
\indent m. Pulsars and neutron stars \\
\indent n. Black holes \\ 
\indent o. Transients \\

%% file: appendices/survey_pi.tex
\section{PI survey results}
\label{app:psurvey}

This appendix lists the questions and responses for the Cycle 8 PI survey. Many of the survey questions were adopted from a similar ESO survey in their pilot review using distributed peer review \citep{Patat19}. ALMA PIs were also asked similar questions in the surveys conducted after the Cycle 7 Main Call which used panel reviews, and the Cycle 7 Supplemental Call with distributed peer review \citep[see][]{Carpenter20b}. The response rate for the Cycle 7 surveys were 23\% and 70\% for the Main and Supplemental calls, respectively. Where appropriate, we show the survey results from Cycle 8 with these previous ALMA survey results.
 
\subsection{Are the individual comments on your proposal clear and understandable?}

\begin{itemize}
\item Fully: The comments are clear, whether I agree with the comments scientifically or not. (96 responses)
\item Mostly: Most of the comments are clear. (281 responses)
\item Somewhat: Some of the comments are clear. (87 responses)
\item No: Few, if any, of the comments are clear. (9 responses)
\end{itemize}

\begin{figure}[ht]
\centering
\includegraphics[width=\textwidth]{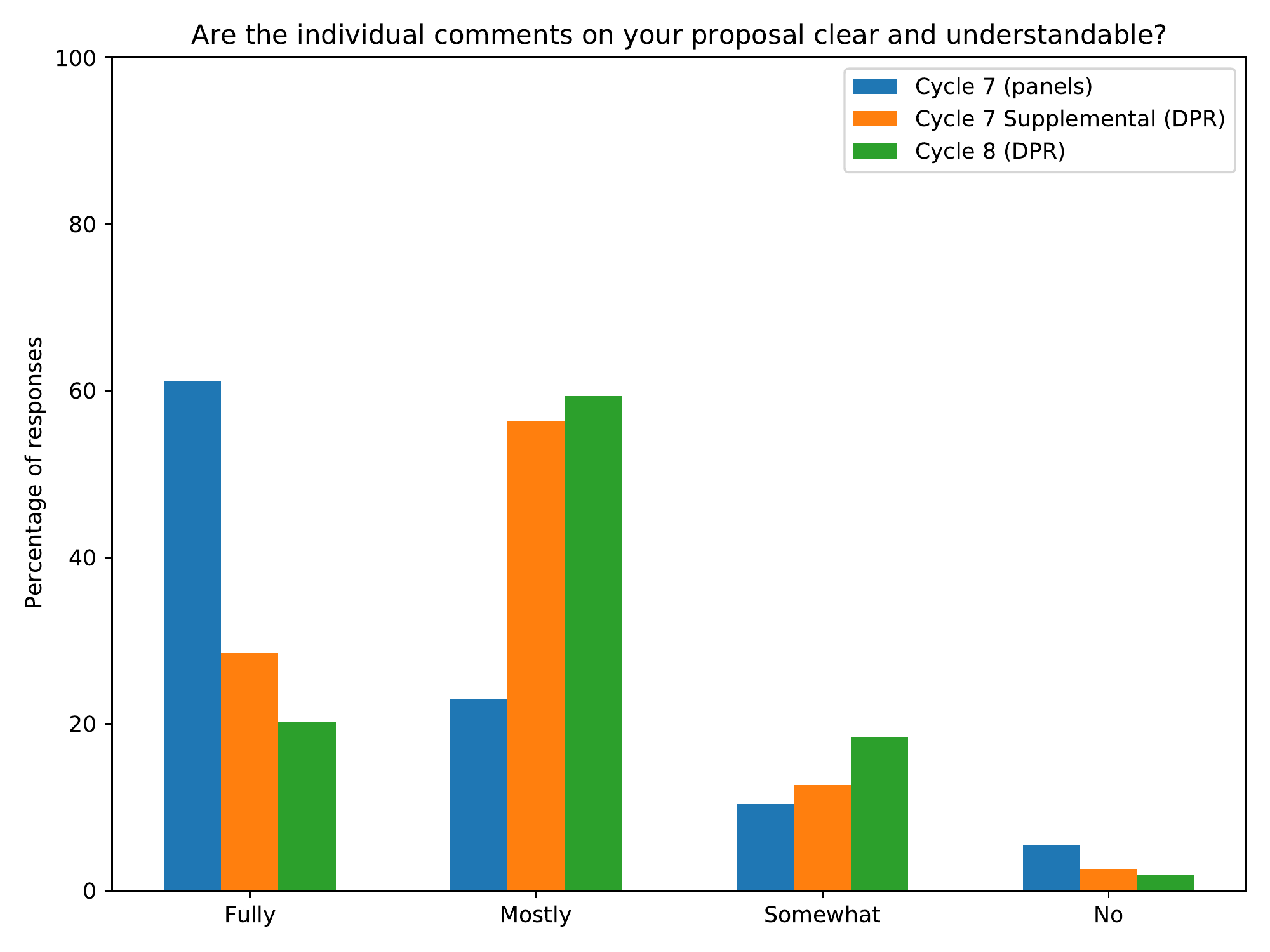}
\caption{Results of the PI survey question asking if the overall comments are clear and understandable.}
\label{fig:psurv_clear}
\end{figure}
\clearpage

\subsection{Are the comments scientifically accurate?}

\begin{itemize}
\item Fully: The comments are scientifically accurate. (39 responses)
\item Mostly: Most of the comments are scientifically accurate. (239 responses)
\item Somewhat: Some of the comments are scientifically accurate. (169 responses)
\item No: Few, if any, of the comments are scientifically accurate. (16 responses)
\end{itemize}

\begin{figure}[h]
\centering
\includegraphics[width=\textwidth]{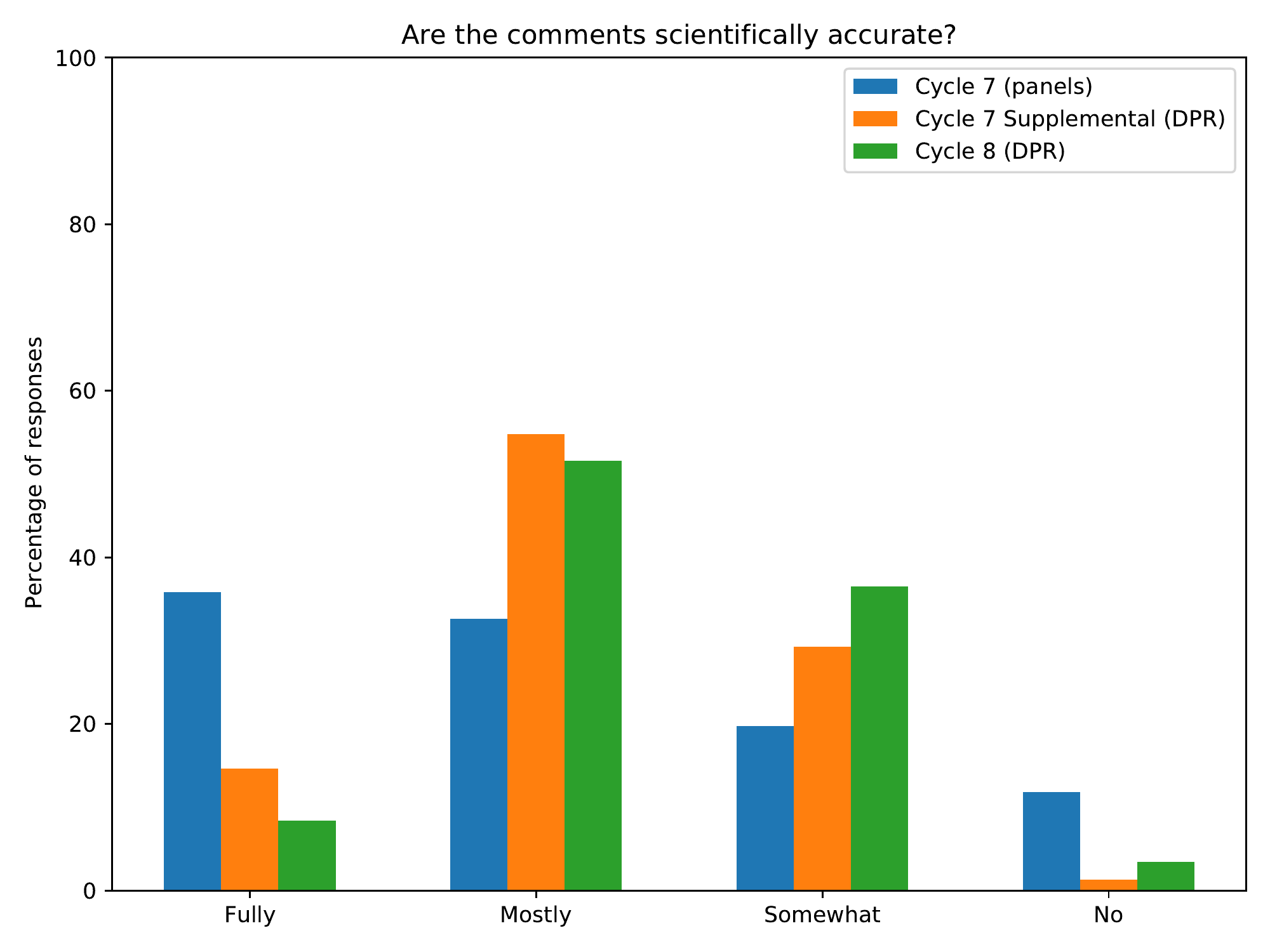}
\caption{Results of the PI survey question asking if the overall comments are scientifically accurate.}
\label{fig:psurv_accurate}
\end{figure}
\clearpage

\subsection{Will the comments help you to improve future ALMA proposals?}
\begin{itemize}
\item Fully: The comments will allow me to improve future proposals. (48 responses)
\item Mostly: Most of the comments will allow me to improve future proposals. (143 responses)
\item Somewhat: Some of the comments might help me strengthen future proposals. (237 responses)
\item No: The comments will not help me improve future proposals. (34 responses)
\end{itemize}

\begin{figure}[h]
\centering
\includegraphics[width=\textwidth]{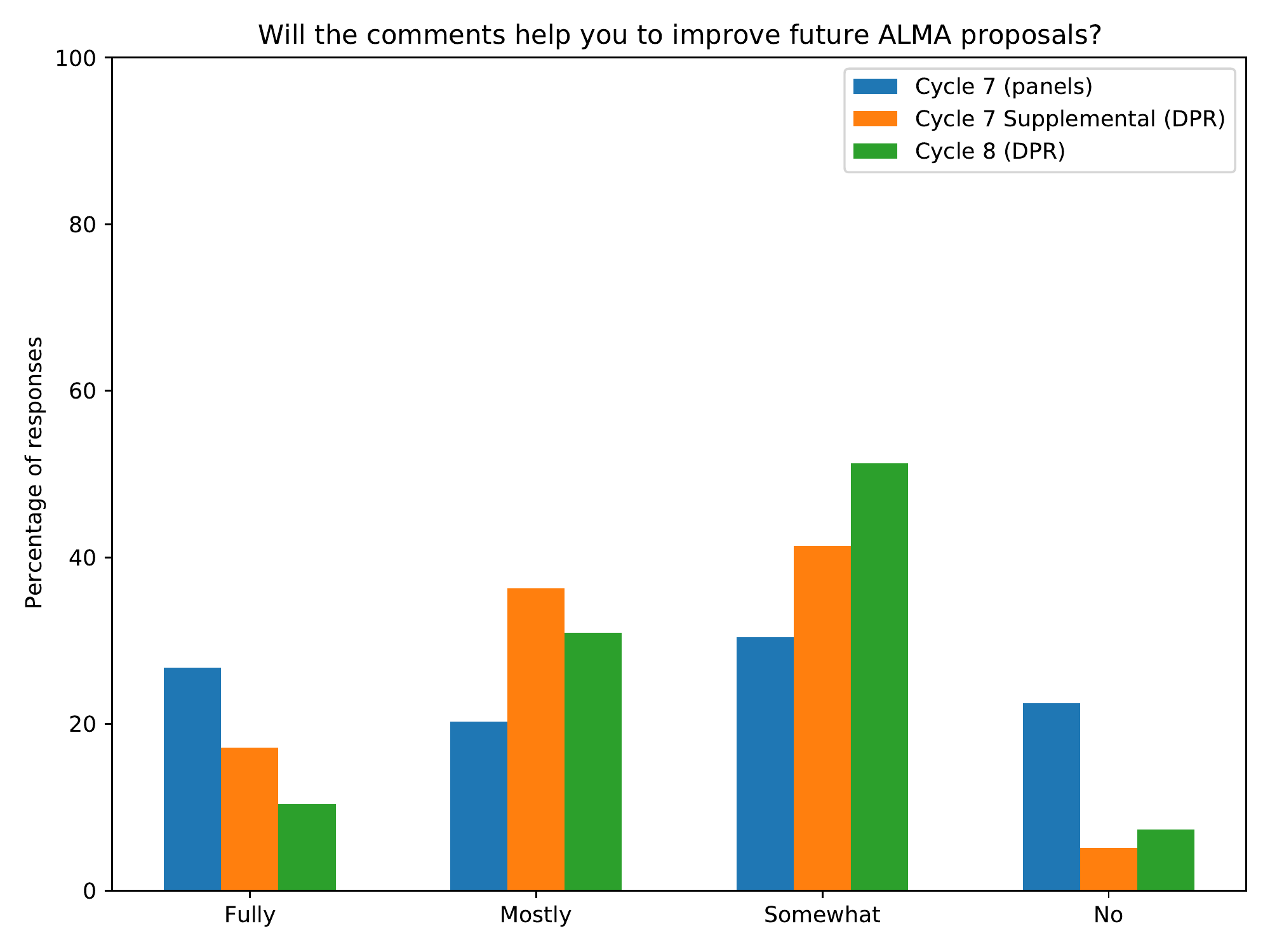}
\caption{Results of the PI survey question asking if the overall comments will help improve future proposals.}
\label{fig:psurv_improve}
\end{figure}
\clearpage

\subsection{Were the comments written in a respectful and professional manner?}
\begin{itemize}
\item Yes: All of the comments were respectful even if I do not agree with the comments scientifically. (354 responses)
\item Somewhat: Most of the comments were respectful. (103 responses)
\item No: Many of the comments contain unprofessional remarks. (3 responses)
\end{itemize}

\begin{figure}[h]
\centering
\includegraphics[width=\textwidth]{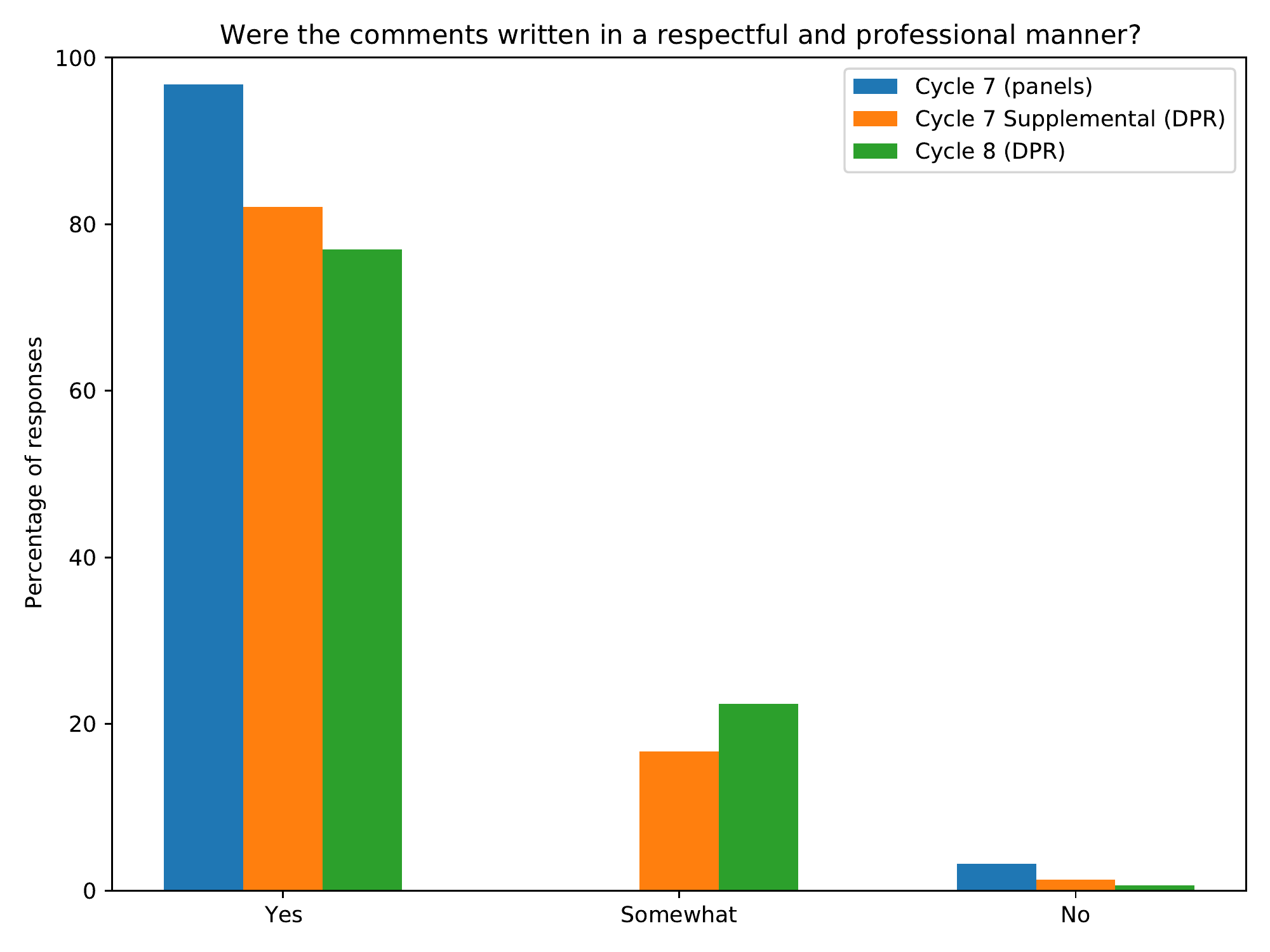}
\caption{Results of the PI survey question asking if overall comments were written in a respectful and professional manner. In the PI survey for the Cycle 7 main call, the only options were ``yes" and ``no".}
\label{fig:psurv_respect}
\end{figure}
\clearpage

\subsection{ Are you concerned about confidentiality in ALMA review processes?}
\begin{itemize}
\item I am neither more nor less concerned about confidentiality issues in the Distributed Peer Review process compared to the Panel Review process. (158 responses)
\item I am more concerned about confidentiality in the Distributed Peer Review process. (108 responses)
\item I am less concerned about confidentiality in the Distributed Peer Review process. (65 responses)
\item I have no strong opinion on this point. (131 responses)
\end{itemize}

\begin{figure}[h]
\centering
\includegraphics[width=\textwidth]{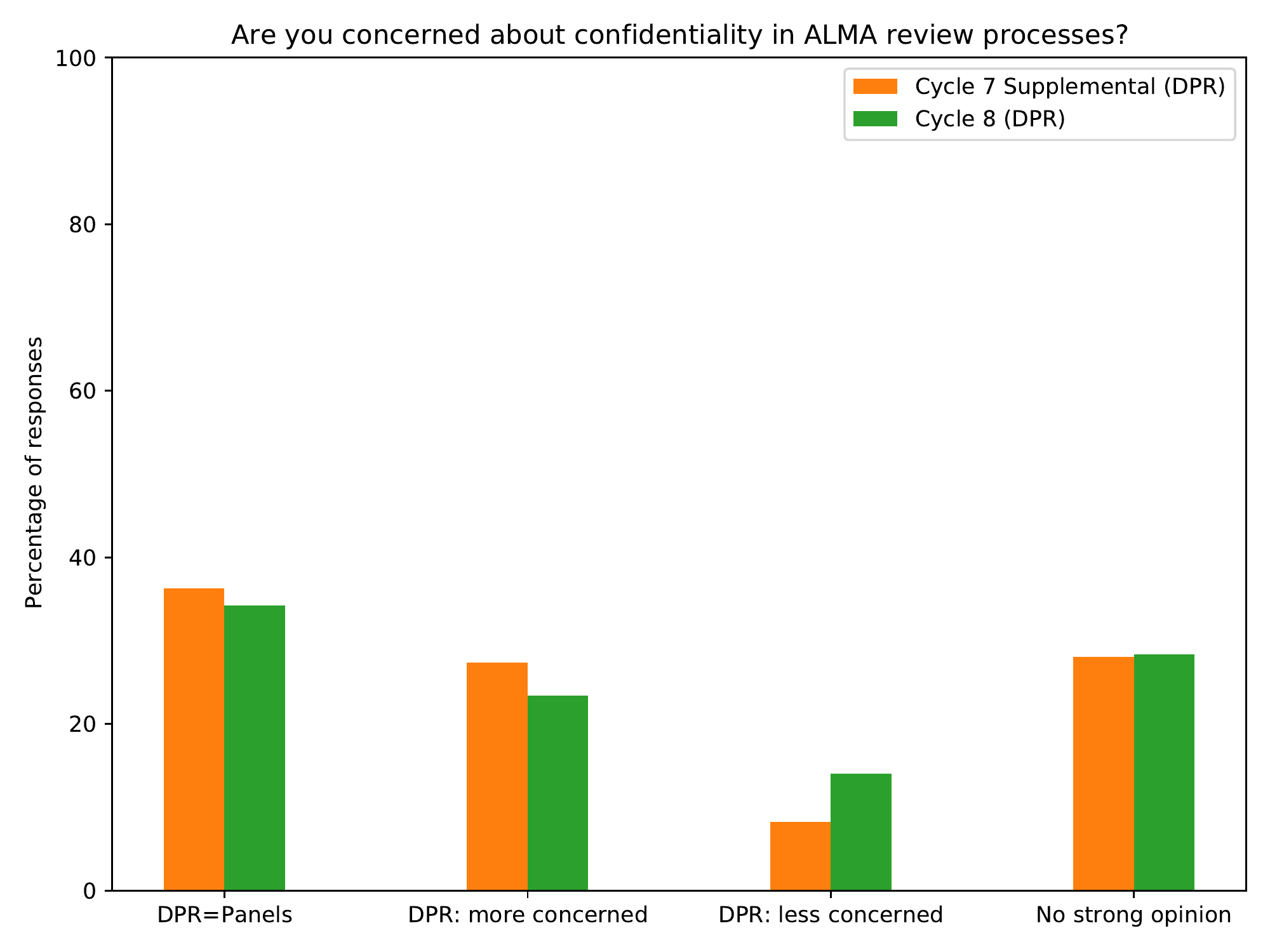}
\caption{Results of the PI survey question asking if they are more concerned or less concerned about confidentiality in distributed peer review versus panel reviews.}
\label{fig:psurv_confidentiality}
\end{figure}
\clearpage

\subsection{Are you concerned about the robustness of ALMA review processes against any biases?}
\begin{itemize}
\item I think that the Distributed Peer Review process is as robust against biases as the Panel Review process. (90 responses)
\item I think that the Distributed Peer Review process is more robust against biases compared to the Panel Review process. (113 responses)
\item I think that the Distributed Peer Review process is less robust against biases compared to the Panel Review process. (161 responses)
\item I have no strong opinion on this point. (98 responses)
\end{itemize}

\begin{figure}[h]
\centering
\includegraphics[width=\textwidth]{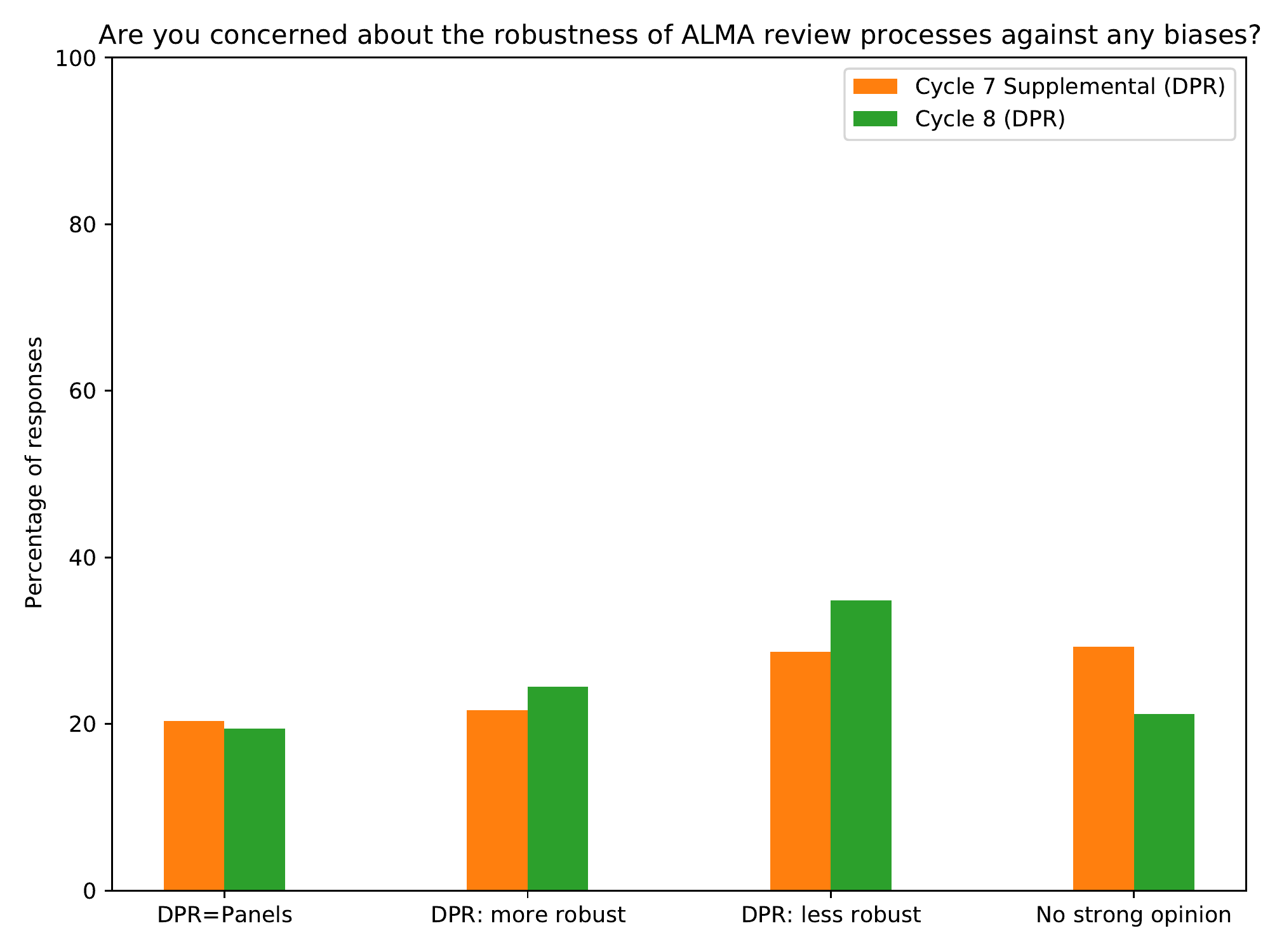}
\caption{Results of the PI survey question asking if they are more concerned or less concerned about biases in distributed peer review versus panel reviews.}
\label{fig:psurv_biases}
\end{figure}
\clearpage

\subsection{How many years has it been since you obtained your PhD?}
\begin{itemize}
\item I do not have a PhD yet. (80 responses)
\item 3 years or fewer. (101 responses)
\item Between 4 and 12 years. (147 responses)
\item More than 12 years. (133 responses)
\end{itemize}

\begin{figure}[h]
\centering
\includegraphics[width=\textwidth]{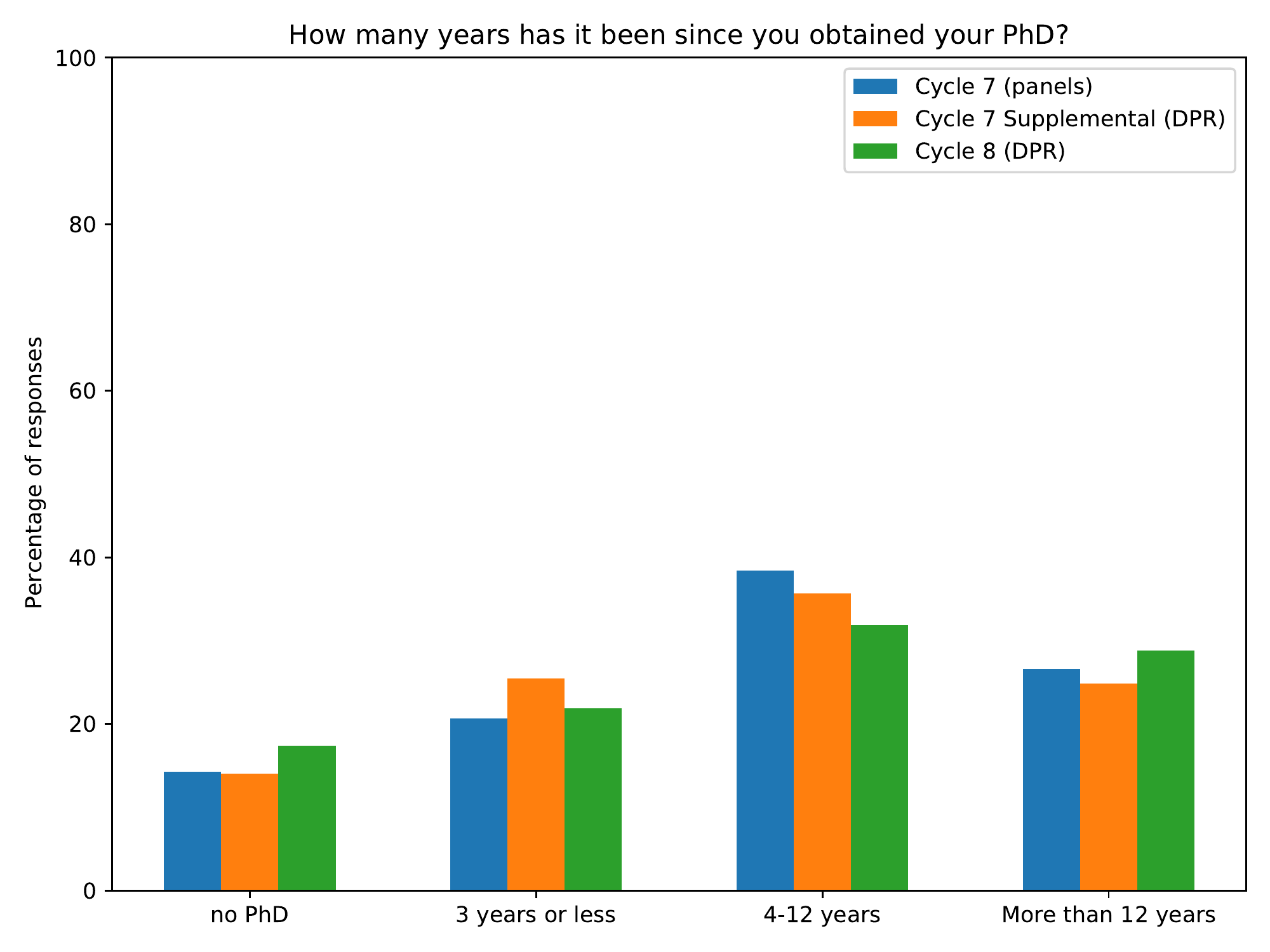}
\caption{Results of the PI survey asking  the number of years since they obtained their PhD.}
\label{fig:psurv_phd}
\end{figure}
\clearpage

\subsection{Please take a few minutes to rate the helpfulness of each review that you received, indicating the extent to which this comment will help to improve your proposal in the future. Positive comments like ``best proposal I ever read” can be ranked as not helpful as it does not improve the proposal further.}
\begin{itemize}
\item This review is very helpful. (1605)
\item This review is somewhat helpful. (3253 responses)
\item This review is inaccurate or otherwise not helpful. (1616 responses)
\item This review is inappropriate or unprofessional. (162)
\end{itemize}

\begin{figure}[h]
\centering
\includegraphics[width=\textwidth]{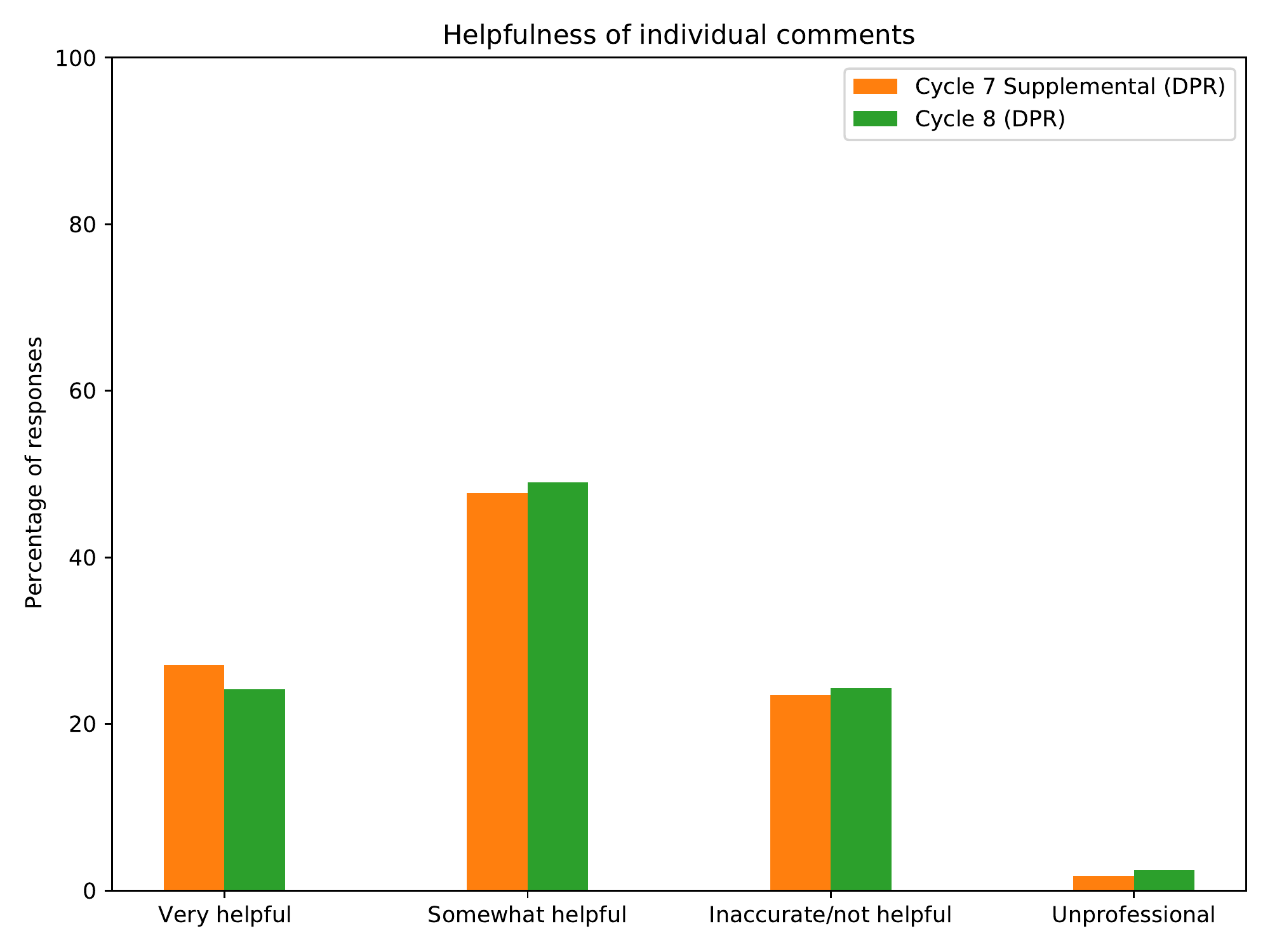}
\caption{Results of the PI survey question asking to rate the helpfulness of the individual reviews of their proposal.}
\label{fig:psurv_individual}
\end{figure}
\clearpage

%% file: appendices/ranks_comp.tex
\section{Computation of the overall ranks}
\label{app:ranks}

Each reviewer ranks the proposals in their set from 1 to 10. Nominally each proposal then will receive 10 ranks. The individual ranks need to be combined to form an overall ranked list. Two approaches have been investigated to form this ranked list: (i) averaging the individual ranks and (ii) modeling the ranks using the Plackett-Luce algorithm. The advantages and disadvantages of these two approaches are discussed below.  For the Cycle 8 main call, the overall ranks were computed by averaging the individual ranks without outlier rejection.

\subsection{Averaging individual ranks}

The overall rank of the proposal can be determined by averaging the ranks from the individual reviewers,\footnote{In practice, the ranks are normalized in the handful of cases where reviewers rank less than 10 proposals.} and then sorting by the average rank. The main advantage of this approach is the simplicity and robustness of the calculation. The main disadvantage is that an individual rank from a reviewer reflects the relative quality of the proposals in their proposal set. In the case where, for example, a reviewer receives 10 excellent proposals, one of the proposals will be assigned rank 10 even if the reviewer found all proposals to be very strong. Similarly, if the reviewer is assigned 10 relatively poor proposals, one will need to be assigned a rank of 1. Therefore, from the PI perspective, interpreting a rank is ambiguous without knowing the strength of the other proposals.

\subsection{Plackett-Luce model}

The Plackett-Luce model \citep{Plackett75} overcomes the main disadvantage of straight averaging by considering the relative strength of the proposal within a proposal set. Following \citet{Turner19} and \citet{Maystre18}, consider a reviewer that has $n$ proposals to review in their proposal set:
\begin{equation}
    S = \{i_1, i_2, ..., i_n\}.
\end{equation}
Following Luce's choice axiom \citep{Luce59}, we assume that a reviewer's preference for proposal $i_j$ over proposal $i_k$ is independent of the remaining proposals in the set $S$. Therefore, the probability of selecting proposal $i_j$ as the top proposal from $S$ is given by
\begin{equation}
    P(i_j|S) = {e^{\theta_{i_j}} \over \sum_{i\in S}\,\,e^{\theta_i}},
\end{equation}
where $\theta_{i_j}$ represents the ``strength" or ``worth" of proposal $i_j$. Larger differences in $\theta_{i_j}$ then correspond to larger differences in the perceived scientific merit.  The rankings then can be viewed as a sequence of choices. The top proposal is selected from all proposals in the set, the second ranked proposal is selected from the remaining proposals, and so forth until all proposals are ranked. The joint probability of the specific ranking $i_1 \succ ... \succ i_n$ is then
\begin{eqnarray}
    \mathbb{P}(i_1\succ i_2\succ...\succ i_n|S) &\,\,=\,\,&
    {e^{\theta_{i_1}}\over{e^{\theta_{i_1}}+e^{\theta_{i_2}}+...+ e^{\theta_{i_n}}}}\,\,
    {e^{\theta_{i_2}}\over{e^{\theta_{i_2}}+...+ e^{\theta_{i_n}}}}\,\,
    ...\,\,
    {e^{\theta_{i_{n-1}}}\over{e^{\theta_{i_{n-1}}} +e^{\theta_{i_n}}}}\\
     &\,\,=\,\,& \prod_{j=1}^n {e^{\theta_{i_j}} \over {\sum_{i\in A_j} e^{\theta_i}}},
\end{eqnarray}
where $A_j$ is the set of proposals $\{i_j, i_{j+1}, ..., i_n\}$ from which $i_j$ is chosen. From the independent rankings from each of the reviewers, maximum likelihood techniques can be used to solve for the values of $\theta_i$, and therefore the rank-ordered list of proposals. Priors on the parameters $\theta_i$ can be placed to regulate the solution. In practice, each parameter was solved with a Gaussian prior with variance $\alpha^{-1}$. Unless otherwise stated, we set $\alpha=10^{-6}$ to apply loose constraints on the model parameters. The solutions were obtained with the python code {\tt choix} version 0.3.4 \citep{Maystre18}. 

\subsection{Simulations}
\label{subsec:ranks_sim}

Monte Carlo simulations were used to determine the efficacy of using the average ranks and the Plackett-Luce model. The basic assumption is that there is an intrinsic relative scientific merit to the proposals. A ``true rank" was assigned randomly to each proposal. Ten proposals are then assigned to a reviewer in the simulation following the same proposal assignments used in the actual Cycle 8 distributed peer review process. The simulation encompasses the 1496 proposals that successfully completed Stage 1. The simulated overall ranks were then computed by averaging the individual ranks or applying the Plackett-Luce algorithm.

In conducting the simulations, a reviewer's ability to order the proposals by the true rank was set by $\sigma_\mathrm{ranks}$. For a given reviewer, the ranks were randomized using a normal probability distribution with dispersion $\sigma_\mathrm{ranks}$. The randomized ranks were then sorted, which defined the rank ordered list of proposals for that reviewer. For $\sigma_\mathrm{ranks}=0$, reviewers can perfectly recover the true proposal rankings. The individual ranks from the reviewers were then averaged without outlier rejection to derive the overall ranked list of proposals.

\begin{figure}
\centering
\plotone{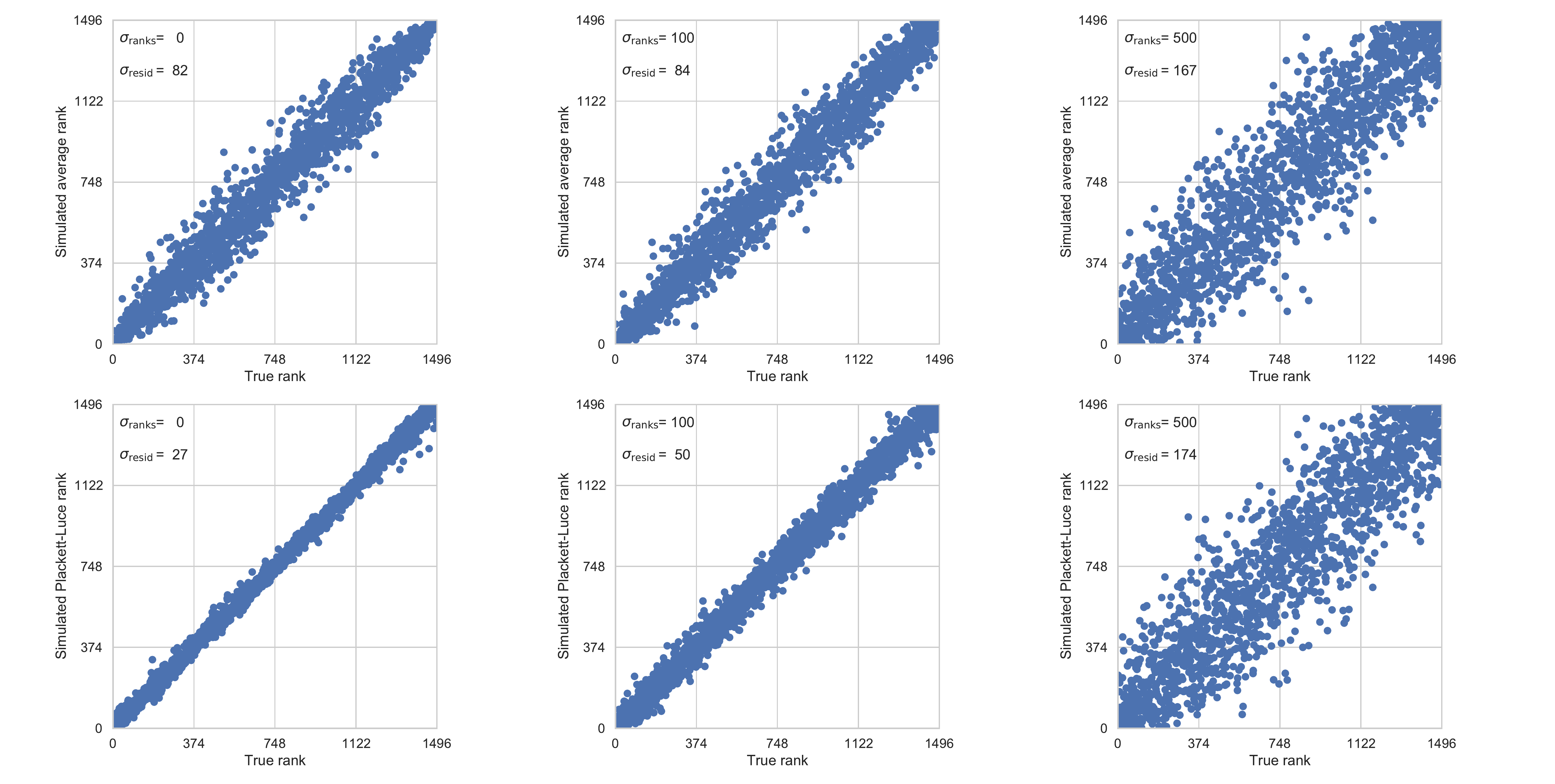}
\caption{Monte Carlo simulations that explore the efficiency of averaging and the Plackett-Luce model to recover the ``true rankings" (see text). The top row shows the results for averaging individual ranks and the bottom row shows the Plackett-Luce model. The panels from left to right illustrate different effectiveness in recovering the true proposal ranks as parameterized by the gaussian dispersion $\sigma_\mathrm{ranks}$.} 
\label{fig:sim3}
\end{figure}

Figure~\ref{fig:sim3} shows the results of the simulations. The abscissa in each plot shows the true proposal rank and the ordinate shows the derived overall rank. The top row shows the recovered ranks computed by averaging the individual ranks, and the bottom row shows the ranks recovered from the Plackett-Luce model. The panels from left to right show the results for $\sigma_\mathrm{ranks}= 0$, 100, and 500. The residuals after fitting a linear relation between the recovered and true ranks are indicated by $\sigma_\mathrm{resid}$. 

As shown in the top left panel of Figure~\ref{fig:sim3}, when reviewers have the perfect ability to recover the proposal ranks, the true ranks can be recovered with an rms of $\sim$5\% (=82/1496) when averaging the ranks. This is because a given reviewer ranks a small subset of all proposals and the true ranks can only be recovered in the limit of a very large number of reviewers or reviewing all proposals. For the same simulation, the Plackett-Luce model  can recover the true rankings with an rms of 1.8\%, which is three times smaller than straight averaging, since it can take into account the competition within a proposal set. However, as the randomness of the reviewer ranks increases (i.e., as $\sigma_\mathrm{ranks}$ increases), the difference between averaging and Plackett-Luce decreases. For $\sigma_\mathrm{ranks}=100$, Plackett-Luce can recover the true ranks with 1.7 times smaller rms, while for $\sigma_\mathrm{ranks}=500$, the rms of the residuals are essentially the same.

\subsection{Application to Cycle 8}
\label{subsec:app_cycle8}

\begin{figure}
\centering
\plotone{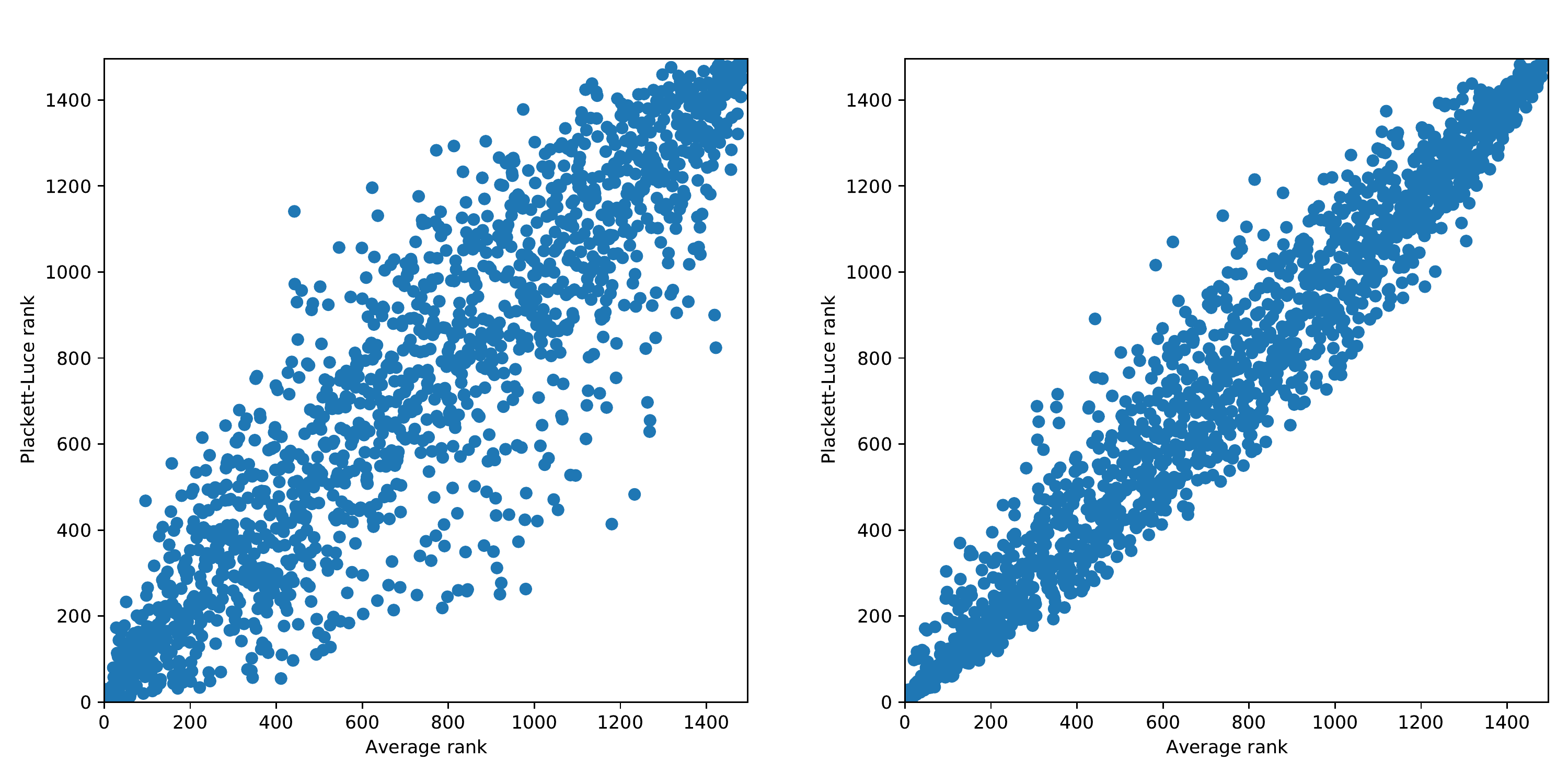}
\caption{Correlation in the overall ranks for Cycle 8 distributed peer review computing using the average ranks without outlier rejection and the Plackett-Luce model. The Plackett-Luce solutions were obtained with the regularization parameter $\alpha=10^{-6}$  (left panel) and $\alpha=10$ (right panel).}
\label{fig:average_pl}
\end{figure}

Figure~\ref{fig:average_pl} compares the overall ranks derived from averaging the individual ranks and the Plackett-Luce model. The Plackett-Luce ranks were derived by applying weak constraints in the model parameters (left panel) and strong constraints (right panel). A general correlation between the ranks is seen as expected. However, if no constraints are applied to the model parameters, individual proposals in the bottom quartile of the average ranks are found in the second quartile by the Plackett-Luce model. These extreme outliers are removed by applying model constraints on the parameters as shown in the right panel. Identifying the specific reason for the differences in the ranks for individual proposals is not straightforward since in the Plackett-Luce model, the rankings depend on the how all of the proposals are rated relative to each other across all of the proposal sets.

As shown in Section~\ref{subsec:scatter}, the scatter in the individual ranks is dominated by the individual opinions of the reviewers. In this regime, simply averaging the ranks recovers the true ranks with similar precision as the Plackett-Luce model when $\sigma_\mathrm{ranks}$ is large (see Section~\ref{subsec:ranks_sim}). Given the scatter in the ranks, and the sensitivity of the Plackett-Luce results in this regime to the regularization parameter (see Figure~\ref{fig:average_pl}), we used the simple averaging of individual ranks to compute the overall ranks in Cycle 8. Further investigation of the Plackett-Luce algorithm is needed to understand how best to apply this model in practice.

%% file: main.bbl
\begin{thebibliography}{}
\expandafter\ifx\csname natexlab\endcsname\relax\def\natexlab#1{#1}\fi
\providecommand{\url}[1]{\href{#1}{#1}}
\providecommand{\dodoi}[1]{doi:~\href{http://doi.org/#1}{\nolinkurl{#1}}}
\providecommand{\doeprint}[1]{\href{http://ascl.net/#1}{\nolinkurl{http://ascl.net/#1}}}
\providecommand{\doarXiv}[1]{\href{https://arxiv.org/abs/#1}{\nolinkurl{https://arxiv.org/abs/#1}}}

\bibitem[{{Andersen} {et~al.}(2019){Andersen}, {Chiboucas}, {Geball},
  {Salinas}, {Scharwaechter}, {Eckersley}, {Kim}, {Taylor}, {Dupuy}, \&
  {Sanmartim}}]{Andersen19}
{Andersen}, M., {Chiboucas}, K., {Geball}, T., {et~al.} 2019, in American
  Astronomical Society Meeting Abstracts, Vol. 233, American Astronomical
  Society Meeting Abstracts \#233, 455.03

\bibitem[{{Braatz} {et~al.}(2021){Braatz}, {Biggs}, , {Sanhueza}, \&
  {Corvill\'on}}]{Braatz21}
{Braatz}, J., {Biggs}, A., , {Sanhueza}, P., \& {Corvill\'on}, A. 2021, {ALMA
  Cycle 8 2021 Proposer’s Guide}.
\newblock
  \url{https://almascience.nrao.edu/documents-and-tools/cycle8/alma-proposers-guide}

\bibitem[{{Carpenter}(2020)}]{Carpenter20a}
{Carpenter}, J. 2020, \pasp, 132, 024503, \dodoi{10.1088/1538-3873/ab3e18}

\bibitem[{{Carpenter} {et~al.}(2020){Carpenter}, {Donovan Meyer},
  {Corvill\'on}, {Impellizzeri}, {Kurowski}, \& {Chalevin}}]{Carpenter20b}
{Carpenter}, J., {Donovan Meyer}, J., {Corvill\'on}, A., {et~al.} 2020, ALMA
  Memo 616

\bibitem[{{Carpenter} {et~al.}(2022){Carpenter}, {Corvillon}, {Donovan Meyer},
  {Plunkett}, {Kurowski}, {Chalevin}, \& {Macias}}]{Carpenter22}
{Carpenter}, J.~M., {Corvillon}, A., {Donovan Meyer}, J., {et~al.} 2022, arXiv
  e-prints, arXiv:2203.11334.
\newblock \doarXiv{2203.11334}

\bibitem[{Jones {et~al.}(2001)Jones, Oliphant, Peterson, {et~al.}}]{Jones01}
Jones, E., Oliphant, T., Peterson, P., {et~al.} 2001, {SciPy}: Open source
  scientific tools for {Python}.
\newblock \url{http://www.scipy.org/}

\bibitem[{{Kerzendorf} {et~al.}(2020){Kerzendorf}, {Patat}, {Bordelon}, {van de
  Ven}, \& {Pritchard}}]{Kerzendorf20}
{Kerzendorf}, W.~E., {Patat}, F., {Bordelon}, D., {van de Ven}, G., \&
  {Pritchard}, T.~A. 2020, Nature Astronomy, 4, 711,
  \dodoi{10.1038/s41550-020-1038-y}

\bibitem[{Luce(1959)}]{Luce59}
Luce, R.~D. 1959, Individual Choice Behavior: A Theoretical analysis (New York,
  NY, USA: Wiley)

\bibitem[{Maystre(2018)}]{Maystre18}
Maystre, L. 2018, PhD thesis, École Polytechnique Fédérale de Lausanne

\bibitem[{{Merrifield} \& {Saari}(2009)}]{Merrifield09}
{Merrifield}, M.~R., \& {Saari}, D.~G. 2009, Astronomy and Geophysics, 50,
  4.16, \dodoi{10.1111/j.1468-4004.2009.50416.x}

\bibitem[{{Patat}(2018)}]{Patat18}
{Patat}, F. 2018, \pasp, 130, 084501, \dodoi{10.1088/1538-3873/aac463}

\bibitem[{{Patat} {et~al.}(2019){Patat}, {Kerzendorf}, {Bordelon}, {Van de
  Ven}, \& {Pritchard}}]{Patat19}
{Patat}, F., {Kerzendorf}, W., {Bordelon}, D., {Van de Ven}, G., \&
  {Pritchard}, T. 2019, The Messenger, 177, 3, \dodoi{10.18727/0722-6691/5147}

\bibitem[{Plackett(1975)}]{Plackett75}
Plackett, R.~L. 1975, Journal of the Royal Statistical Society. Series C
  (Applied Statistics), 24, 193.
\newblock \url{http://www.jstor.org/stable/2346567}

\bibitem[{Scholz \& Zhu(2019)}]{Scholz19}
Scholz, F., \& Zhu, A. 2019, kSamples: K-Sample Rank Tests and their
  Combinations.
\newblock \url{https://CRAN.R-project.org/package=kSamples}

\bibitem[{Scholz \& Stephens(1987)}]{Scholz87}
Scholz, F.~W., \& Stephens, M.~A. 1987, Journal of the American Statistical
  Association, 82, 918.
\newblock \url{http://www.jstor.org/stable/2288805}

\bibitem[{{Sobkowicz} {et~al.}(2013){Sobkowicz}, {Thelwall}, {Buckley},
  {Paltoglou}, \& {Sobkowicz}}]{Sobkowicz13}
{Sobkowicz}, P., {Thelwall}, M., {Buckley}, K., {Paltoglou}, G., \&
  {Sobkowicz}, A. 2013, EPJ Data Sci., 2, 2, \dodoi{10.1140/epjds14}

\bibitem[{{Strolger} {et~al.}(2017){Strolger}, {Porter}, {Lagerstrom},
  {Weissman}, {Reid}, \& {Garcia}}]{Strolger17}
{Strolger}, L.-G., {Porter}, S., {Lagerstrom}, J., {et~al.} 2017, \aj, 153,
  181, \dodoi{10.3847/1538-3881/aa6112}

\bibitem[{Turner {et~al.}(2019)Turner, van Etten, Firth, \&
  Kosmidis}]{Turner19}
Turner, H.~L., van Etten, J., Firth, D., \& Kosmidis, I. 2019, Modelling
  rankings in R: the PlackettLuce package.
\newblock \doarXiv{1810.12068}

\end{thebibliography}
